\documentclass[a4paper,12pt]{article}
\pdfoutput=1
\usepackage{jheppub}
\usepackage{graphicx}
\usepackage{amssymb}
\usepackage{amsmath}
\usepackage{amsfonts}
\usepackage{hyperref}
\usepackage{xcolor}

	\def\bea{\begin{eqnarray}}
	\def\eea{\end{eqnarray}}
	\def\be{\begin{equation}}
	\def\ee{\end{equation}}
	\def\nn{\nonumber\\}

	\def\mm{\mathcal}
	
	\def\pa{\partial}

	\definecolor{db}{rgb}{0,0.08,0.45}
	\definecolor{brick}{rgb}{0.6,0.1,0.3}
	
	\definecolor{zz}{rgb}{1,0,0}
	\definecolor{zz2}{rgb}{0.7,0.1,0.1}
	\definecolor{yy}{rgb}{0.05,0.9,0.05}
	\definecolor{ww}{rgb}{0.6,0.1,0.3}
	
	\definecolor{rr}{cmyk}{0,0,0,1}
	\definecolor{vv}{rgb}{0.5,0,0.5}
	\definecolor{ss}{cmyk}{0,0,0,1}

	\definecolor{brick}{rgb}{0.5,0,0.5}

	\def\a{\alpha} \def\b{\beta} \def\g{\gamma} \def\d{\delta}    \def\n{\nu}  \def\s{\sigma}    \def\l{\lambda}  
	\def\G{\Gamma}  \def\D{\Delta}   \def\L{\Lambda}   \def\O{\Omega}

\title{Loops in AdS: From the Spectral Representation to Position Space}

\author[\Psi]{Dean Carmi}
\affiliation[\Psi]{Institute of Physics, \'Ecole Polytechnique F\'ed\'erale de Lausanne (EPFL), Rte de la Sorge, BSP 728, CH-1015 Lausanne, Switzerland}
\emailAdd{deancarmi1@gmail.com}

\abstract{
We compute a family of scalar loop diagrams in $AdS$. We use the spectral representation to derive various bulk vertex/propagator identities, and these identities enable to reduce certain loop bubble diagrams to lower loop diagrams, and often to tree-level exchange or contact diagrams. An important example is the computation of the finite coupling 4-point function of the large-$N$ conformal $O(N)$ model on $AdS_3$. Remarkably, the re-summation of bubble diagrams is equal to a certain contact diagram: the $\bar{D}_{1,1,\frac{3}{2},\frac{3}{2}} (z,\bar z)$ function. Another example is a scalar with $\phi^4$ or $\phi^3$ coupling in $AdS_3$: we compute various 4-point (and higher point) loop bubble diagrams with alternating integer and half-integer scaling dimensions in terms of a finite sum of contact diagrams and tree-level exchange diagrams. The 4-point function with external scaling dimensions differences obeying $\D_{12}=0$ and $\D_{34}=1$ enjoys significant simplicity which enables us to compute in quite generality. For integer or half-integer scaling dimensions, we show that the $M$-loop bubble diagram can be written in terms of Lerch transcendent functions of the cross-ratios $z$ and $\bar z$. Finally, we compute 2-point bulk bubble diagrams with endpoints in the bulk, and the result can be written in terms of Lerch transcendent functions of the AdS chordal distance. We show that the similarity of the latter two computations is not a coincidence, but arises from a vertex identity between the bulk 2-point function and the double-discontinuity of the boundary 4-point function.
}

\begin{document} 
\maketitle
\flushbottom

\section{Introduction}
\label{sec:1}

There are several reasons why one would want to study quantum field theory on AdS space \cite{Callan:1989em,Breitenlohner:1982jf,Aharony:1999ti,Aharony:2010ay,Aharony:2012jf,Burges:1985qq,Witten:1998zw,Burgess:1984ti,Inami:1985dj}. AdS is a curved background which is maximally symmetric, thus enabling to perform various calculations which are not possible in less symmetric spaces. AdS acts like a box with an infinite volume, and it has a natural scale, the AdS radius,  which enables to study RG flows by varying the AdS radius. AdS is conformally equivalent to flat space with a flat boundary, thus studying QFTs on AdS can teach us about boundary conformal theories in flat space, \cite{McAvity:1995zd}. Most importantly, AdS is a great laboratory for studying quantum gravity, and it plays a unique role in the gauge gravity duality \cite{Maldacena:1997re,Gubser:1998bc,Witten:1998qj}.

In the AdS/CFT correspondence, the generating function for CFT correlators is equal to the path integral in AdS with boundary sources. Thus the perturbative AdS diagram expansion is equal to the $1/N$ expansion of CFT correlators. There has been a large effort in computing scattering amplitudes in AdS, and much of this work has been concerned with computing tree-level diagrams. AdS tree-level diagrams have been computed in position space \cite{Liu:1998th,Liu:1998ty,Dolan:2000ut,Freedman:1998tz,DHoker:1998ecp,Freedman:1998bj,DHoker:1998bqu,DHoker:1999mqo,Zhou:2018sfz,Hijano:2015zsa}, and in momentum space \cite{Raju:2010by,Raju:2011mp,Albayrak:2019asr,Albayrak:2018tam}. There have been computations in Mellin space \cite{Penedones:2010ue,Fitzpatrick:2011ia,Paulos:2011ie,Rastelli:2017udc,Rastelli:2016nze,Cardona:2017tsw,Yuan:2017vgp,Yuan:2018qva}, embedding space \cite{Costa:2014kfa,Costa:2011mg}, and computations employing the crossing and/or super-symmetry of the boundary theory  \cite{Aharony:2016dwx,Henriksson:2017eej,Alday:2017xua,Alday:2017gde,Mazac:2018mdx,Caron-Huot:2018kta,Alday:2017vkk,Aprile:2017bgs,Aprile:2017qoy}. Loop diagrams in AdS are technically challenging, much more so than flat-space amplitudes. This is partly due to the bulk-to-bulk propagators being complicated hypergeometric functions of the AdS geodesic distance. Additionally, momentum space is not as useful because of the absence of translation invariance. As a result of this, there have been relatively very few computations of loop amplitudes in AdS, and much fewer beyond 1-loop. For these reasons we believe that obtaining analytic expressions for loop diagrams would be very interesting and useful. 

Let us discuss some of the literature on loop diagrams in AdS. \cite{Fitzpatrick:2010zm,Fitzpatrick:2011hu} showed that the 1-loop bubble diagram (with no external legs) in AdS can be written as an infinite sum of bulk-to-bulk correlators.\cite{Aharony:2016dwx} considered $\phi^3$ and $\phi^4$ on AdS at 1-loop, and computed double trace anomalous dimensions $\g^{(2)}_{n,l}$ and the Mellin amplitude. \cite{Mazac:2018mdx} considered $\phi^4$ in $AdS_2$, using crossing symmetry and the extremal functional method they computed the 1 and 2-loop 4-point function.\footnote{Note that \cite{Aharony:2016dwx,Mazac:2018mdx} obtain the crossing symmetric sum of diagrams, and not individual diagrams.}. Brute force position space computations of individual loop diagrams were done in \cite{Bertan:2018khc,Bertan:2018afl,Beccaria:2019stp}. Cutting rules and unitarity methods in AdS are discussed in \cite{Fitzpatrick:2011dm,Ponomarev:2019ofr,bb}. \cite{Liu:2018jhs} considered 1-loop polygon diagrams in AdS, e.g the 3-point triangle diagram, the 4-point box diagram, etc. They used spectral/split representation methods to compute the pre-amplitude\footnote{The pre-amplitude is the integrand of the spectral integrals.} in terms of the cross-kernel 6J symbol. \cite{Yuan:2017vgp,Yuan:2018qva} used the spectral representation to derive results about the Mellin space pre-amplitude and it's singularity structure. The computations in that work were rather systematic, which enabled to go to higher loops. \cite{Giombi:2017hpr} computed the $AdS_3$ 1-loop 2-point bubble diagram in the spectral representation by performing the spectral representation integrals. This result gives the correction to anomalous dimension, and some of their results are generalized to spinning bubbles. In \cite{Carmi:2018qzm} we derived the 1-loop scalar bubble diagram in the spectral representation\footnote{For a fermion, the bubble is derived in $d=1,2$.} in any $d$. More importantly, it was shown that the spectral representation, analogously to momentum space for flat-space amplitudes, is ideal for re-summing bubble diagrams. This observation enabled to derive the spectral representation of the exact 4-point function for the large-$N$ $O(N)$ model and Gross-Neveu model on AdS.

We will compute diagrams in position space, and we will focus on scalar fields in $AdS$. Some of the computations hold in any dimension $d$, but most of them are done for $AdS_3$. In section~\ref{sec:spec} we review the spectral representation and write formulas for the spectral representations for Witten diagrams, which will be used frequently in the following sections.  In section~\ref{sec:relswitten1} we derive several bulk vertex/propagator identities. These identities will be used to obtain relations between different classes of Witten diagrams, and also to compute various loop Witten diagrams in terms of contact and exchange diagrams. In section~\ref{sec:3} we compute the finite coupling 4-point function in position space for the conformal large-$N$ $O(N)$ model on $AdS_3$. We first perform a brute force computation of the conformal block expansion sum, which gives us an explicit analytic result for the double-discontinuity of the 4-point function. Then using a bubble identity, we compute the full 4-point function in terms of a contact diagram ($\bar{D}_{1,1,\frac{3}{2},\frac{3}{2}} (z,\bar z)$ function)!  In section~\ref{sec:4} we consider a scalar field on $AdS_3$ with $\phi^4$ interaction, for the case of 4-point function with external scaling dimensions $\D_{12}=\D_{34}=0$. We write down the conformal block expansion in terms of sums of LegendreQ functions, and show that in very special cases one can perform the summation. In section~\ref{sec:5} we likewise consider a scalar field in $AdS_3$, but here for the class of 4-point functions with external scaling dimensions $\D_{12}=0$ and $\D_{34}=1$. Significant simplification occurs for this class of 4-point functions, which enables us to compute in quite generality. The general contact diagram is a ${}_2 F_1$ function, the exchange diagram is a ${}_3 F_2$, and the ($\D=1$) 1-loop bubble diagram is a ${}_4 F_3$. More generally, for integer or half-integer scaling dimensions, we show that the $M$-loop bubble diagram can be expanded in a basis of Lerch transcendent functions (i.e generalizations of polylogarithms) of the cross-ratios $z$ and $\bar z$. In section~\ref{sec:6} we compute 2-point bulk bubble diagrams with endpoints in the bulk. Once again, for integer or half-integer scaling dimensions in the bubble, the $M$-loop bubble diagram can be expanded in a basis of Lerch transcendent functions of the AdS chordal distance. We show that the similarity of the latter two computations is not a coincidence, but arises from a vertex identity between the bulk 2-point function and the double-discontinuity of the boundary 4-point function. In section~\ref{sec:7} we discuss the results and future directions. In Appendix~\ref{sec:A} we derive a few more useful vertex/propagator identities. In appendices \ref{sec:appb} - \ref{sec:Mellinp} we provide details of computations which were not shown in the main text.

\section{The Spectral representation}
\label{sec:spec}
The spectral representation of 2-points in AdS encapsulates the symmetries of AdS, and it is analogous to momentum space of Feynman diagrams, see e.g. \cite{Carmi:2018qzm,Penedones:2007ns,Costa:2014kfa,Costa:2012cb}. One advantage of the spectral/split representation is that it enables (at least in principle) to compute various classes of bubble diagrams \cite{Yuan:2017vgp,Yuan:2018qva,Carmi:2018qzm}. \cite{Yuan:2018qva} called such diagrams "generalized bubble diagrams". In this work we will explicitly compute bubble diagrams.
\begin{figure}[t]
\centering
\includegraphics[clip,height=3cm]{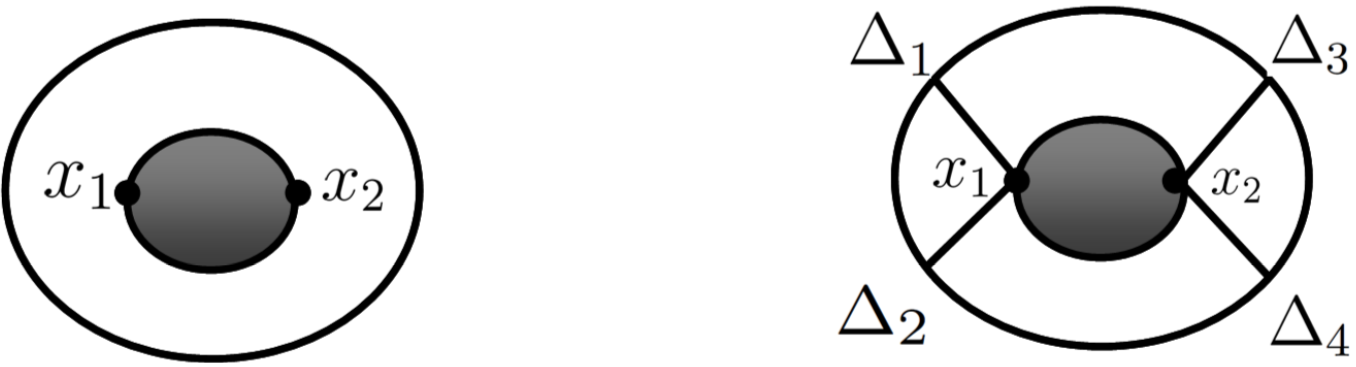}
\caption{\textbf{Left:} The grey blob represents a general bulk two point function $F(x_1,x_2)$, with points $x_1$ and $x_2$. Other than that, the blob is completely general. \textbf{Right:} Attaching 4 legs to the boundary gives a CFT 4-point function $g_4(z, \bar z)$. This class of diagrams has 2 bulk-to-boundary propagators attached to $x_1$, and 2 attached to $x_2$.}
\label{fig:bubblerelations16}
\end{figure}
Consider a function $F(x_1,x_2)$ of two bulk points $x_1$ and $x_2$ in AdS, see Fig.~\ref{fig:bubblerelations16}-Left. One can expand this function in AdS harmonic functions\footnote{$\n$ space in AdS is analogous to momentum space in flat space.} $\O_\n(x_1,x_2)$:
\begin{align} 
\label{eq:nseeeeb}
F(x_1,x_2)=  \int_{-\infty}^\infty d\n  \tilde{F}_\n  \Omega _{\nu}(x_1,x_2) \,,
\end{align}
where $ \tilde{F}_\n$ is the spectral transform of $F(x_1,x_2)$, and $\n$ is the spectral parameter.
The AdS harmonic function is:
\begin{align} 
\label{eq:ksjnf2}
&\O_\n (x_1,x_2) =  \frac{i\n}{2\pi} \Big( G_{\frac{d}{2}+i\n}(x_1,x_2) -G_{\frac{d}{2}-i\n}(x_1,x_2) \Big)
\nn
&= \frac{h-\frac{d}{2}}{2\pi} \Big( G_{h}(x_1,x_2) -G_{d-h}(x_1,x_2) \Big)
\end{align} 
where $G_{\frac{d}{2}+i\n}(x_1,x_2)$ are bulk-to bulk propagators, $h \equiv \frac{d}{2}+i\n$, and we always put the AdS radius to be $L_{AdS}=1$.
The bulk-to-bulk propagator of a scalar field with mass $M^2=\D(\D-d)$ in $AdS_{d+1}$ in position space is:
\begin{align} 
G_\D(x_1,x_2) =  \frac{\G_\D}{2\pi^{\frac{d}{2}} \G_{\D-\frac{d}{2}+1}} \zeta^{-\D} {}_2F_1 (\D,\D-\frac{d-1}{2},2\D-d+1,-4\zeta^{-1})
\end{align} 
where $\zeta= \frac{(z_1-z_2)^2+(\vec{x}_1-\vec{x}_2)^2}{z_1z_2}$ is the square of the chordal distance between the two points $x_1$ and $x_2$. The spectral representation of the bulk-to-bulk propagator is:
\begin{align} 
\label{eq:bulkspect4}
G_\D (x_1,x_2) = \int_{-\infty}^\infty \frac{ d\nu}{\n^2+(\D-\frac{d}{2})^2}  \Omega _{\nu}(x_1,x_2) \ \ \ \ \ , \ \ \ \ \ \tilde{F}_\n =\frac{1}{\n^2+(\D-\frac{d}{2})^2}
\end{align} 
So the spectral representation in this case is a simple pole in the $\n$ plane, plus it's shadow pole at $i\n \to - i\n$.
The spectral representation has the very useful property that convolutions in position space become products in spectral space:
\begin{align} 
\int d^{d+1}x f(x_1,x)F(x,x_2) = \int_{-\infty}^\infty d\n \tilde{f}_\n \tilde{F}_\n \O_\n (x_1,x_2)
\end{align} 
where $\tilde{f}_\n$ is the spectral representation of $f(x_1,x_2)$:  $f(x_1,x_2) =  \int_{-\infty}^\infty d\n \tilde{f}_\n \O_\n (x_1,x_2)$, and likewise for $\tilde{F}_\n$.

\subsection{The spectral representation of the 1-loop bubble}
The 1-loop bubble spectral function $\tilde{B}_\n$ was computed in general $d$ in \cite{Carmi:2018qzm}, in terms of a ${}_5 F_4$ function. We will focus in this paper on $AdS_3$, i.e $d=2$, in which case the bubble spectral representation is:
 \begin{align}
 \label{eq:bubble4}
 \tilde{B}_\n = \frac{\psi(\D-\frac{1}{2}+\frac{i\n}{2})-\psi(\D-\frac{1}{2}-\frac{i\n}{2})}{8\pi (i\n)} \ \ \ \ \ \ , \ \ \ \ \ for \ \ d=2
 \end{align}
where $\psi$ the digamma function. The poles $ \tilde{B}_\n$ are:
\begin{align}
\label{eq:poles6}
 \tilde{B}_\nu \overset{1+i\nu\, \sim  \,2\Delta+2n}{\sim}-\frac{1}{4\pi}\frac{1}{1+i\nu-(2\Delta+2n)} \times \frac{1}{(2\D+2n-1)}
\end{align}
The bubble function simplifies when $\D$ is an integer or half integer. When $\D=1$: 
\begin{align} 
\label{eq:b1}
\tilde{B}^{(\D=1)}_\nu   = -\frac{1}{8} \frac{ \cot ( \frac{ \pi (1+i\n) }{2})  }{(i\n)} = \frac{1}{16}\frac{ \G_{\frac{1}{2}+\frac{i\n}{2}}\G_{\frac{1}{2}-\frac{i\n}{2}}}{\G_{1+\frac{i\n}{2}}\G_{1-\frac{i\n}{2}} }
\end{align} 
where in the second equality we simply wrote it in terms of gamma functions. Likewise the bubbles for $\D=\frac{1}{2}$ and for $\D=\frac{3}{2}$ are:
\begin{align} 
\label{eq:b2}
\tilde{B}^{(\D=\frac{1}{2})}_\nu   = \frac{ \tan ( \frac{ \pi (1+i\n) }{2}) }{  8i\n}-\frac{1}{4\pi (i\n)^2} = \frac{-1}{4(i\n)^2} \frac{\G_{1+\frac{i\n}{2}}\G_{1-\frac{i\n}{2}} }{ \G_{\frac{1}{2}+\frac{i\n}{2}}\G_{\frac{1}{2}-\frac{i\n}{2}}} - \frac{1}{4\pi(i\n)^2}
\end{align} 
and
\begin{align} 
\label{eq:b3}
\tilde{B}^{(\D=\frac{3}{2})}_\nu   =\frac{ \tan ( \frac{ \pi (1+i\n) }{2}) }{  8i\n}+\frac{1}{4\pi (i\n)^2} = \frac{-1}{4(i\n)^2} \frac{\G_{1+\frac{i\n}{2}}\G_{1-\frac{i\n}{2}} }{ \G_{\frac{1}{2}+\frac{i\n}{2}}\G_{\frac{1}{2}-\frac{i\n}{2}}} + \frac{1}{4\pi(i\n)^2}
\end{align} 

\subsection{Attaching bulk-to-boundary propagators}
Consider the bulk 2-point  function Eq.~\ref{eq:nseeeeb} and Fig.~\ref{fig:bubblerelations16}-left, and then attach two bulk-to-boundary propagators to $x_1$ and one bulk-to-boundary propagator to $x_2$ in order to create a CFT 3 point function. The CFT 3 point function has a space time dependence which is fixed by the conformal symmetry. It is easy to show that it is proportional to the spectral function $\tilde{F}_{\n}$. Now consider instead attaching 4 bulk-to-boundary propagators with scaling dimensions $(\D_1$,  $\D_2$, $\D_3$, $\D_4)$ to an internal 2-point arbitrary blob $F(x_1,x_2)$, as in Fig.~\ref{fig:bubblerelations16}-Right. This diagram is given by:
\begin{align} 
\label{eq:4pointspec}
\widehat{g}_4  =\int d x_1 dx_2 F(x_1,x_2) 
 K_{\Delta_1}(P_1,x_1)K_{\Delta_2}(P_2,x_1) K_{\Delta_3}(P_3,x_2)K_{\Delta_4}(P_4,x_2)
\end{align} 
where the integrals are over $AdS_{d+1}$. In Eq.~\ref{eq:AAA1} of Appendix~\ref{sec:appbb} we show the derivation of the spectral representation of $g_4(z, \bar z)$:
\begin{align} 
\label{eq:sdbsmd7}
&g_4(z, \bar z)= \int_{-\infty}^\infty d\nu  \tilde{F}_\n 
\times \frac{  \G_{\frac{\Delta_1+\D_2}{2} +\frac{i \nu-\frac{d}{2}}{2}} \G_{\frac{\Delta_1+\D_2}{2} -\frac{i \nu+\frac{d}{2}}{2}} \G_{\frac{\Delta_3+\D_4}{2} +\frac{i \nu-\frac{d}{2}}{2}} \G_{\frac{\Delta_3+\D_4}{2} -\frac{i \nu+\frac{d}{2}}{2}}  }{\Gamma_{i\nu} \Gamma_{\frac{d}{2}+i\n} }
\nn
& \Big( \G_{\frac{\D_2-\Delta_1}{2}+\frac{i \nu+\frac{d}{2}}{2}} \G_{\frac{\Delta_1-\D_2}{2}+\frac{i \nu+\frac{d}{2}}{2}} \G_{\frac{\Delta_3-\D_4}{2}+\frac{i \nu+\frac{d}{2}}{2}} \G_{\frac{\D_4-\Delta_3 }{2}+\frac{i \nu+\frac{d}{2}}{2}} \Big) \mathcal{K}^{\D_i}_{\frac{d}{2}+i\nu} (z,\bar z)
\end{align}
See also \cite{Ponomarev:2019ofr}. Where $\tilde{F}_\n $ is the spectral representation of $F(x_1,x_2) $, $\mathcal{K}^{\D_i}_{\frac{d}{2}+i\nu}(z,\bar z)$ is the conformal block. We will use Eq.~\ref{eq:sdbsmd7} repeatedly in this paper. 
Let us stress that $g_4$ is the stripped correlator, obtained by stripping off the factor $A_{\D_i}$ in Eq.~\ref{eq:AAA1} and \ref{eq:prefactor}:
\begin{align}
\label{eq:lkjh7}
g_4 (z,\bar z)\equiv \frac{\widehat{g}_4}{A_{\D_i}}
\end{align}
Let us re-write the 4-point function by using the identity $\G_x \G_{1-x}= \frac{\pi}{\sin \pi (x)}$:
\begin{align} 
\label{eq:rt1}
&g_4(z, \bar z)=\pi^2 \int d\nu  \tilde{F}_\n 
\frac{\Bigg[ \frac{  \G_{\frac{\Delta_1+\D_2}{2} -\frac{d}{4}+\frac{i \nu}{2}}  \G_{\frac{\Delta_3+\D_4}{2} -\frac{d}{4}+\frac{i \nu}{2}}  }{\G_{1-\frac{\Delta_1+\D_2}{2} +\frac{d}{4}+\frac{i \nu}{2}}  \G_{1-\frac{\Delta_3+\D_4}{2} +\frac{d}{4}+\frac{i \nu}{2}}  }  \Bigg]}{\Gamma_{i\nu} \Gamma_{\frac{d}{2}+i\n} \sin \pi (\frac{\Delta_1+\D_2}{2} -\frac{d}{4}-\frac{i \nu}{2}) \sin \pi (\frac{\Delta_3+\D_4}{2} -\frac{d}{4}-\frac{i \nu}{2})}  
\nn
& \Big( \G_{\frac{\D_2-\Delta_1}{2}+\frac{i \nu+\frac{d}{2}}{2}} \G_{\frac{\Delta_1-\D_2}{2}+\frac{i \nu+\frac{d}{2}}{2}} \G_{\frac{\Delta_3-\D_4}{2}+\frac{i \nu+\frac{d}{2}}{2}} \G_{\frac{\D_4-\Delta_3 }{2}+\frac{i \nu+\frac{d}{2}}{2}} \Big) \mathcal{K}^{\D_i}_{\frac{d}{2}+i\nu} (z,\bar z)
\end{align}
The zeros of the $\sin$ denominators correspond to the double-trace poles. Now we consider taking the s-channel double-discontinuity of the 4-point function\footnote{We will later see instances in which the double-discontinuity is easier to compute than the 4-point function. One can then recover the 4-point function by using the dispersion relation of \cite{Carmi:2019cub}.}. The action of the s-channel double-discontinuity on the s-channel conformal block is\footnote{Note that the proper way to take the s-channel double-discontinuity, is to first re-include the pre-factor $A_{\D_i}$ of Eq.~\ref{eq:prefactor}.  In Eq.~\ref{eq:ddisc} we are simply dropping this pre-factor.}:
\begin{align}
\label{eq:ddisc}
&dDisc_s \big[ \mathcal{K}^{\D_i}_{\frac{d}{2}+i\nu} (z,\bar z) \big] 
\nn
&=\sin \pi \big(\frac{\Delta_1+\D_2}{2} -\frac{i \nu+\frac{d}{2}}{2}\big) \sin \pi \big(\frac{\Delta_3+\D_4}{2}  -\frac{i \nu+\frac{d}{2}}{2}\big) \mathcal{K}^{\D_i}_{\frac{d}{2}+i\nu} (z,\bar z)
\end{align}
Thus, the effect of taking the double-discontinuity of the 4-point function is to cancel the $\sin$ denominators:
\begin{align} 
\label{eq:sdbsmd7q}
&dDisc_s \big[ g_4 (z,\bar z ) \big]=  \pi^2 \int  d\nu  \tilde{F}_\n 
\frac{1}{\Gamma_{i\nu} \Gamma_{\frac{d}{2}+i\n} } \Bigg[ \frac{  \G_{\frac{\Delta_1+\D_2}{2} -\frac{d}{4}+\frac{i \nu}{2}}  \G_{\frac{\Delta_3+\D_4}{2} -\frac{d}{4}+\frac{i \nu}{2}}  }{\G_{1-\frac{\Delta_1+\D_2}{2} +\frac{d}{4}+\frac{i \nu}{2}}  \G_{1-\frac{\Delta_3+\D_4}{2} +\frac{d}{4}+\frac{i \nu}{2}}  }  \Bigg]
\nn
& \Big(  \G_{\frac{\D_2-\Delta_1}{2}+\frac{i \nu+\frac{d}{2}}{2}} \G_{\frac{\Delta_1-\D_2}{2}+\frac{i \nu+\frac{d}{2}}{2}} \G_{\frac{\Delta_3-\D_4}{2}+\frac{i \nu+\frac{d}{2}}{2}} \G_{\frac{\D_4-\Delta_3 }{2}+\frac{i \nu+\frac{d}{2}}{2}} \Big) \mathcal{K}^{\D_i}_{\frac{d}{2}+i\nu} (z,\bar z)
\end{align}
In the following subsections it will be useful to focus on a specific factor\footnote{We will see in Eq.~\ref{eq:importantn} that the same such factors arise more generally whenever we have a diagram in which 2 external legs are connected to a single bulk vertex.} in the RHS of Eq.~\ref{eq:sdbsmd7}, namely we define:
\begin{align} 
\label{eq:sdbsmd78}
&U^{(\D_1,\D_2)}_F \equiv \tilde{F}_\n  \times \G_{\frac{\Delta_1+\D_2}{2}-\frac{d}{4}+\frac{i \nu}{2}} \G_{\frac{\Delta_1+\D_2}{2}-\frac{d}{4}-\frac{i \nu}{2}} 
\end{align} 
$U^{(\D_1,\D_2)}_F $ depends on $\D_1+\D_2$ and on the spectral function $\tilde{F}_\n$. For conciseness we will not write the $\n$ dependence of $U$. Let us write the special case when $\tilde{F}_\n=1$, i.e just an interaction vertex with the bulk propagator amputated: 
\begin{align} 
\label{eq:AV}
U^{(\D_1,\D_2)}_{A.V} \equiv     \G_{\frac{\Delta_1+\D_2}{2}-\frac{d}{4}+\frac{i \nu}{2}} \G_{\frac{\Delta_1+\D_2}{2}-\frac{d}{4}-\frac{i \nu}{2}} 
\end{align} 
where "$A.V$" stands for "amputated vertex". Likewise the factor $U$ for a vertex with a bulk-to-bulk propagator of scaling dimension $\D$ is given by:
\begin{align} 
\label{eq:V}
U^{(\D_1,\D_2,\D)}_{V} \equiv     \G_{\frac{\Delta_1+\D_2}{2}-\frac{d}{4}+\frac{i \nu}{2}} \G_{\frac{\Delta_1+\D_2}{2}-\frac{d}{4}-\frac{i \nu}{2}} \times \frac{1}{\n^2+(\D-\frac{d}{2})^2}
\end{align} 
The factor $U$ for a 1-loop bubble with dimension scaling $\D$ is given by:
\begin{align} 
\label{eq:B}
U^{(\D_1,\D_2,\D)}_{B} \equiv     \G_{\frac{\Delta_1+\D_2}{2}-\frac{d}{4}+\frac{i \nu}{2}} \G_{\frac{\Delta_1+\D_2}{2}-\frac{d}{4}-\frac{i \nu}{2}} \times \tilde{B}^{(\D)}_\n 
\end{align} 

\subsection{More general diagrams}
\label{sec:moregeneral}
\begin{figure}[t]
\centering
\includegraphics[clip,height=3.9cm]{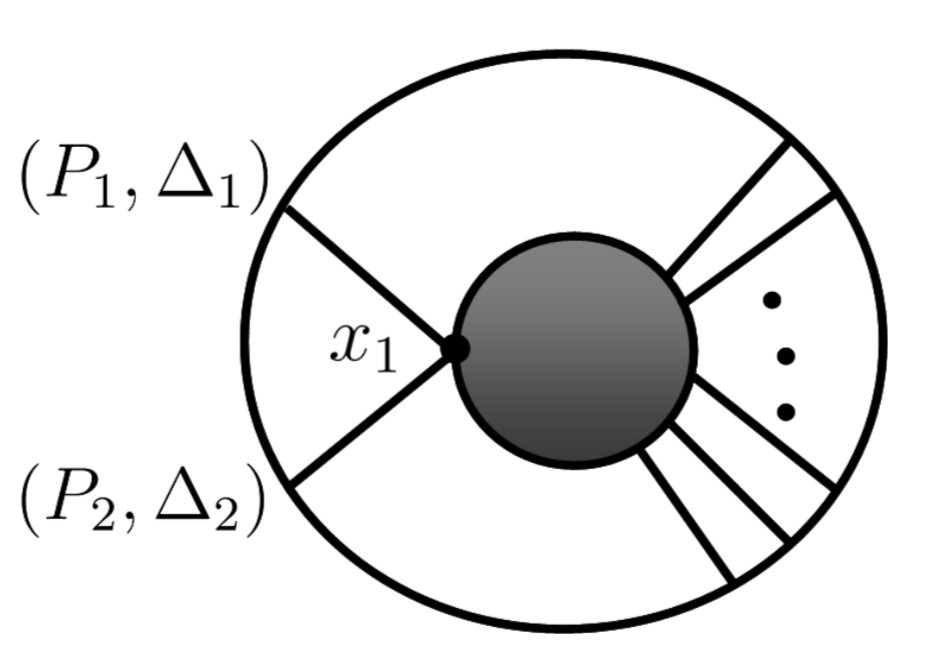}
\caption{An $N$-point function with a pair of external lines originating from the same vertex $x_1$, and ending on the boundary points $P_1$ and $P_2$. The grey blob is completely general.  }\label{fig:bubblerelations20}
\end{figure}
In Eq.~\ref{eq:sdbsmd7} we obtained the spectral representation of a scalar 4-point function, in the case when 2 external legs are attached to the bulk point $x_1$ and 2 external legs are attached to the point $x_2$, as in Fig.~\ref{fig:bubblerelations16}-Right. We will now show that one can say something interesting in more general cases. Consider a general Witten diagram $N$-point function for a scalar field in AdS. Now assume that there is a pair of eternal legs which connect to the same vertex, as in Fig.~\ref{fig:bubblerelations20}. In Fig.~\ref{fig:bubblerelations20} the boundary points $P_1$ and $P_2$ are attached to the vertex at $x_1$. The rest of the diagram, labeled as "Blob", is completely general. The expression for this diagram is given by\footnote{In Eq.~\ref{eq:lsd2} it is convenient to make $x_1$ a cubic vertex by including the bulk-to-bulk propagator $G_\D(x_1,x_2)$ attached to the vertex $x_1$. One can consider e.g quartic vertices by simply taking $\D \to \infty$.}:
\begin{align} 
\label{eq:lsd2}
\widehat{g}_N(P_1,P_2,\{ P_i\}) = \int \prod_{i=2} d^d x_i (Blob )\times \int d^d x_1  G_{\D}(x_1,x_2) K_{\D_1}(P_1,x_1) K_{\D_2}(P_2,x_1)
\end{align} 
where we singled out the $x_1$ vertex. In Eq.~\ref{eq:blobie1} of Appendix~\ref{sec:proofblob1} we obtain:
\begin{align} 
\label{eq:importantn}
&g_N(P_1,P_2,\{ P_i\}) =  \int \prod_{i=2} d^d x_i (Blob ) \int_{-\infty}^\infty d\nu  \frac{ \nu^2\sqrt{\mathcal{C}_{\frac{d}{2}+i\nu}\mathcal{C}_{\frac{d}{2}-i\nu}}  }{\n^2+(\D-\frac{d}{2})^2}
\nn
&\times \Gamma_{\frac{\D_1+\D_2}{2}-\frac{d}{4}+\frac{i\nu}{2}} \Gamma_{\frac{\D_1+\D_2}{2}-\frac{d}{4}-\frac{i\nu}{2}} \bigg[  \int dP_0  K_{\frac{d}{2}-i\nu}(P_0,x_2)\times U_2(P_0,\n) \bigg]
\end{align}
where in order to get rid of an overall factor we defined $g_N \equiv \frac{\pi}{A_0} \widehat{g}_N$, and $A_0$ is defined in Eq.~\ref{eq:a0nbvf}. The top line of Eq.~\ref{eq:importantn} does not depend on $\D_1$ and $\D_2$. The factor $U_2(P_0,\n)$ in the bottom line is defined in Eq.~\ref{eq:u2p}, and it depends only on the difference $(\D_1-\D_2)$ but does not depend on $(\D_1+\D_2)$. Thus the only dependence on $(\D_1+\D_2)$ comes from the gamma functions in the second line, and this dependence is exactly the same as in the top line of Eq.~\ref{eq:sdbsmd7}, which is just the factor $U^{(\D_1,\D_2)}_{A.V}$ Eq.~\ref{eq:AV}.

\section{Identities for Witten diagrams}
\label{sec:relswitten1}
In this section we derive various bulk vertex/propagator identities. These identities will enable to relate different Witten diagrams, as well as to compute various loop bubble diagrams in terms of contact diagrams and exchange diagrams. For the purpose of being self-contained, we also include 3 identities (identity 2',3' and 4) which were already derived in previous works. Table \ref{tbl:13} collects the identities and lists the number of space-time dimensions in which they apply. 

\begin{table}[h!]
  \begin{center}
    \caption{Listing all identities.}
    \label{tab:table1}
    \begin{tabular}{l|c|r} 
      \textbf{Identity} & \textbf{Dimensions} & \textbf{Comments} \\
      \hline
      1 & any $d$ & CFT 4-point function\\
      2 & any $d$ &CFT 4-point function\\
      3 & any $d$ & Propagator/vertex identity\\
      4 & $d=2$ & Bubble relation\\
      5 & $d=2$ & Bubble relation\\
      6 & $d=2$ & Bubble relation\\
      7 & $d=2$ & Bubble relation\\
      8 & any $d$ & CFT 4-point function\\
      9 & $d=2$ & Bubble relation\\
      10 & $d=2$ & Relation between 4-point\\ 
             & & and 2-point function\\
                   \hline
    \end{tabular}
    \label{tbl:13}
  \end{center}
\end{table}

Let us comment on the UV divergences of the diagrams that we are considering. Since UV divergences come from short distances, the UV divergences of diagrams in AdS match those of the same diagrams in flat space. Some of the identities below can possibly contain UV divergences (e.g identities 1,2,3), and in this case the divergences simply match between the left and right hand side of the identity. A sequence of 1-loop bubbles of the form in e.g Fig.~\ref{fig:bubblerelations2}, are non divergent in $AdS_3$\footnote{Such a 1-loop bubble has a naive divergence which goes like $\int^{\L} \frac{d^{d+1}k}{k^2(k+p)^2}\sim \L^{d+1-4}$, where $\L$ is a hard cutoff with dimension of mass. Above $d \geq 4$ this diagram become divergent. }. This fact makes, e.g, identities 5 and 6 possible.

\subsection{Identity 1}
\begin{figure}[t]
\centering
\includegraphics[clip,height=3.9cm]{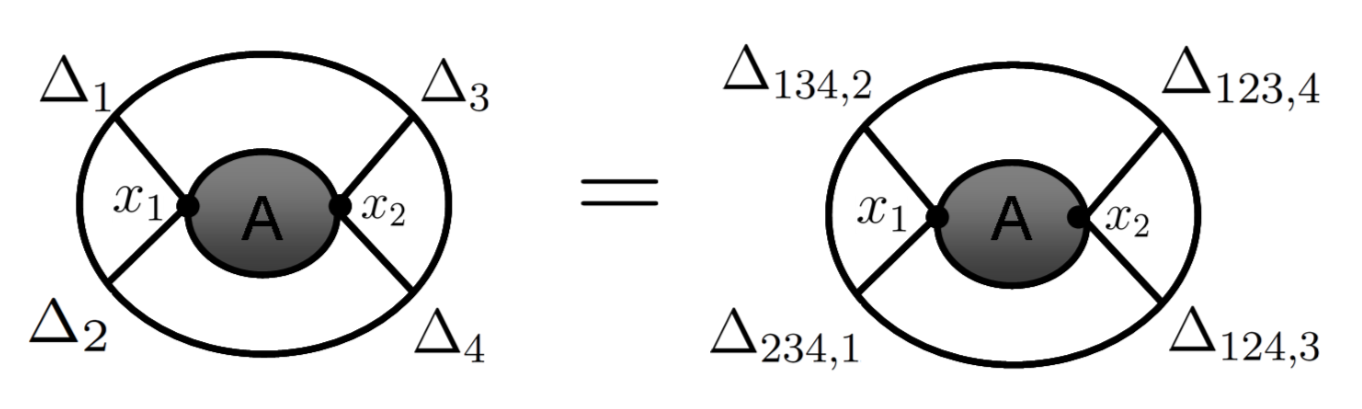}
\caption{Identity 1. For every choice of external scaling dimensions $\D_i$, the $4$-point function on the left equals the $4$-point function on the right, where we define $\D_{123,4}\equiv \frac{\D_1+\D_2+\D_3-\D_4}{2}$. These Witten diagrams are defined to be stripped of the factor $A_{\D_i}$ of Eq.~\ref{eq:lkjh7}. The grey blob 2-point function is general.   }
\label{fig:bubblerelations17}
\end{figure}

As a warm up, we will derive a simple identity for the 4-point function with external scaling dimensions $(\D_1,  \D_2, \D_3, \D_4)$ arising from a bulk 2-point function, Fig~\ref{fig:bubblerelations17}-Left. From Eq.~\ref{eq:sdbsmd7}, it is easy to see that the spectral representation is a product of 3 functions:
\begin{align} 
\label{eq:plangggd}
H_\n(\D_1+\D_2) \times H_\n(\D_3+\D_4) \times h_\n(\D_1-\D_2,\D_3-\D_4)
\end{align} 
where $H_\n$ is a function of the sum $(\D_1+\D_2)$ only, and $h_\n$ is a function of the differences of scaling dimensions only. Clearly Eq.~\ref{eq:plangggd} is invariant under the transformation $\D_i \to \D_i '$ , such that:
\begin{align} 
\label{eq:kjlsjnnd3}
\D_1'-\D_2' =\D_1-\D_2 \ \ \ \ \ \ \ \ \ \ ,\ \ \ \ \ \ \   \ \ \ \D_3'-\D_4' =\D_3-\D_4
\nn
\D_1'+\D_2' =\D_3+\D_4 \ \ \ \ \ \ \ \ \ \ ,\ \ \ \ \ \ \   \ \ \ \D_3'+\D_4' =\D_1+\D_2
\end{align}  
The solution to these equations is:
\begin{align} 
\label{eq:sdbsmd7s}
\D_1'= \D_{134,2} \ \ \ ,\ \ \ \D_2'= \D_{234,1} \ \ \ ,\ \ \ \D_3'= \D_{123,4} \ \ \ ,\ \ \ \D_4'= \D_{124,3}
\end{align} 
where we defined $\D_{123,4}\equiv \frac{\D_1+\D_2+\D_3-\D_4}{2}$ , and likewise for the other combinations. Thus we get that the 4-point function obeys the following identity, Fig.~\ref{fig:bubblerelations17}:
\begin{align} 
 \textbf{Identity\ 1:} \ \ \ \ \ \ \ \ \ \ \ \  \boxed{  g_4^{(\D_i')}(z,\bar z) =   g_4^{(\D_i)} (z, \bar z) }
\end{align} 
where $\D_i= (\D_1,\D_2,\D_3,\D_4)$ and $\D_i'$ is given by Eq.~\ref{eq:sdbsmd7s}. Note that when $\D_2=\D_1$ and $\D_3=\D_4$, Identity 1 becomes trivial, i.e $\{ \D_i' \} =\{ \D_i \}$. Identity 1 is true for any $d$.

\subsection{Identity 2}
\begin{figure}[t]
\centering
\includegraphics[clip,height=7.8cm]{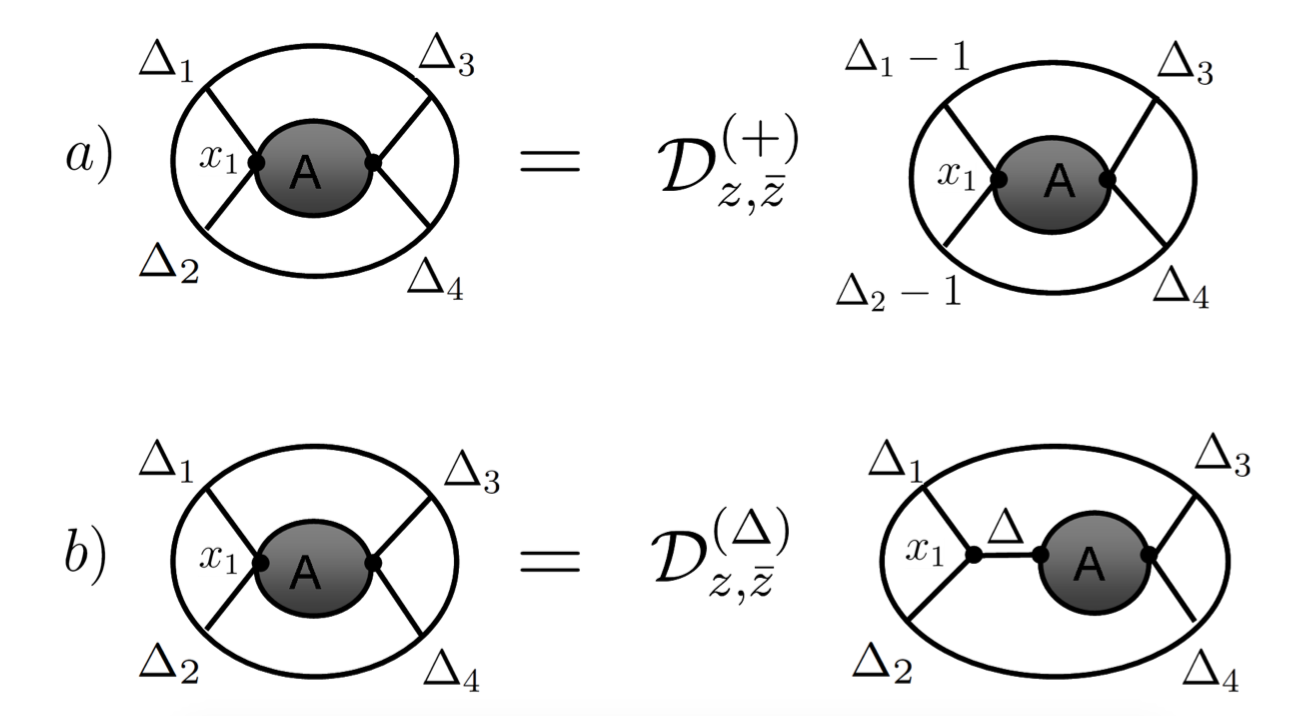}
\caption{\textbf{a)} Identity 2: The derivative operator raises the scaling dimension of the external legs. \textbf{b)} Identity 2':  The derivative operator kills a bulk-to-bulk propagator. The derivative operators are defined in Eqs.~\ref{eq:deriv12} and \ref{eq:deriv13}. We are suppressing numerical coefficients in these plots, and dropping the pre-factor $A_{\D_i}$ defined in Eq.~\ref{eq:prefactor}. \label{fig:bubblerelations15}}
\end{figure}
We derive here Identity 2, shown in Fig.~\ref{fig:bubblerelations15}. The scalar conformal block obeys the quadratic Casimir eigenvalue equation \cite{Dolan:2011dv}:
\begin{align}
\label{eq:der6}
2D_{z,\bar z}\ \mm{K}_{\frac{d}{2}+i\n} (z,\bar z) = -\big(\n^2+\frac{d^2}{4}\big)\mm{K}_{\frac{d}{2}+i\n} (z,\bar z) 
\end{align}
where the differential operator is:
\begin{align}
\label{eq:diff56}
D_{z,\bar z}  \equiv d_z + d_{\bar z} +(d-2) \frac{z\bar z}{z-\bar z} \Big( (1-z)\pa_z -  (1-\bar z)\pa_{\bar z}\Big)
\end{align}
and
\begin{align}
\label{eq:derivop}
d_z \equiv z^2 (1-z) \frac{d^2}{dz^2} - (a+b+1)z^2\frac{d}{dz} -abz
\end{align}
Where $a=\frac{\D_2-\D_1}{2}$ and $\frac{\D_3-\D_4}{2}$. Let us now combine Eq.~\ref{eq:der6} with Eq.~\ref{eq:AV}:
\begin{align}
&U_{A.V}^{(\D_1-1,\D_2-1)}\times \Big( \frac{-1}{4}\Big)\Big[ 2D_{z,\bar z} +\frac{d^2}{4}-  (\D_1+\D_2-3)^2 \Big] \mm{K}_{\frac{d}{2}+i\n} (z,\bar z)
\nn
&=\G_{\frac{\D_1+\D_2}{2}-\frac{3}{2}+\frac{i\n}{2}} \G_{\frac{\D_1 +\D_2}{2}-\frac{3}{2}-\frac{i\n}{2}}\bigg[  \frac{\n^2+\frac{d^2}{4}}{4} +\frac{(\D_1+\D_2-3)^2-\frac{d^2}{4}}{4} \bigg]\mm{K}_{\frac{d}{2}+i\n} (z,\bar z)
\nn
& =\G_{\frac{\D_1+\D_2}{2}-\frac{1}{2}+\frac{i\n}{2}} \G_{\frac{\D_1 +\D_2}{2}-\frac{1}{2}-\frac{i\n}{2}} \mm{K}_{\frac{d}{2}+i\n} (z,\bar z)=U_{A.V}^{(\D_1,\D_2)}  \mm{K}_{\frac{d}{2}+i\n} (z,\bar z)
\end{align}
We can write this more neatly in terms of the 4-point function $g_4$, which gives 
\begin{align} 
\label{eq:relation22}
\textbf{Identity\ 2:}  \ \ \ \ \ \ \boxed{g_4^{(\D_1,\D_2,\D_3,\D_4)}(z, \bar z) = \mm{D}_{z,\bar z}^{(+)} \ \ g_4^{(\D_1-1,\D_2-1,\D_3,\D_4)}(z, \bar z)  }
\end{align}
where we defined the raising operator:
\begin{align} 
\label{eq:deriv12}
\mm{D}_{z,\bar z}^{(+)}\equiv  - \frac{1}{4} \Big[2 D_{z,\bar z} +\frac{d^2}{4}-  (\D_1+\D_2-3)^2 \Big]
\end{align} 
In a similar way, we can combine Eq.~\ref{eq:der6} with Eq.~\ref{eq:V} to get: 
\begin{align} 
\textbf{Identity\ 2':} \ \ \ \ \  \ \boxed{U_{V}^{(\D_1,\D_2,\D)}  \times \mm{D}^{(\D)}_{z,\bar z}    \mm{K}_{\frac{d}{2}+i\n} (z,\bar z)= U_{A.V}^{(\D_1,\D_2)}  \mm{K}_{\frac{d}{2}+i\n} (z,\bar z) }
\end{align}
where we defined the following operator:
\begin{align}
\label{eq:deriv13}
\mm{D}_{z,\bar z}^{(\D)}\equiv   - 2D_{z,\bar z} + (\D-\frac{d}{2})^2-\frac{d^2}{4} 
\end{align}
Acting with this operator eliminates a bulk-to-bulk propagator. Identity 2' was previously derived\footnote{\label{footnote:56} We thank Xinan Zhou for pointing this out to us.} in \cite{Zhou:2018sfz}, and was used to relate exchange diagrams to contact diagrams.  Identities 2 and 2' are shown in Fig.~\ref{fig:bubblerelations15}.

\subsection{Identity 3}
\begin{figure}[t]
\centering
\includegraphics[clip,height=4.2cm]{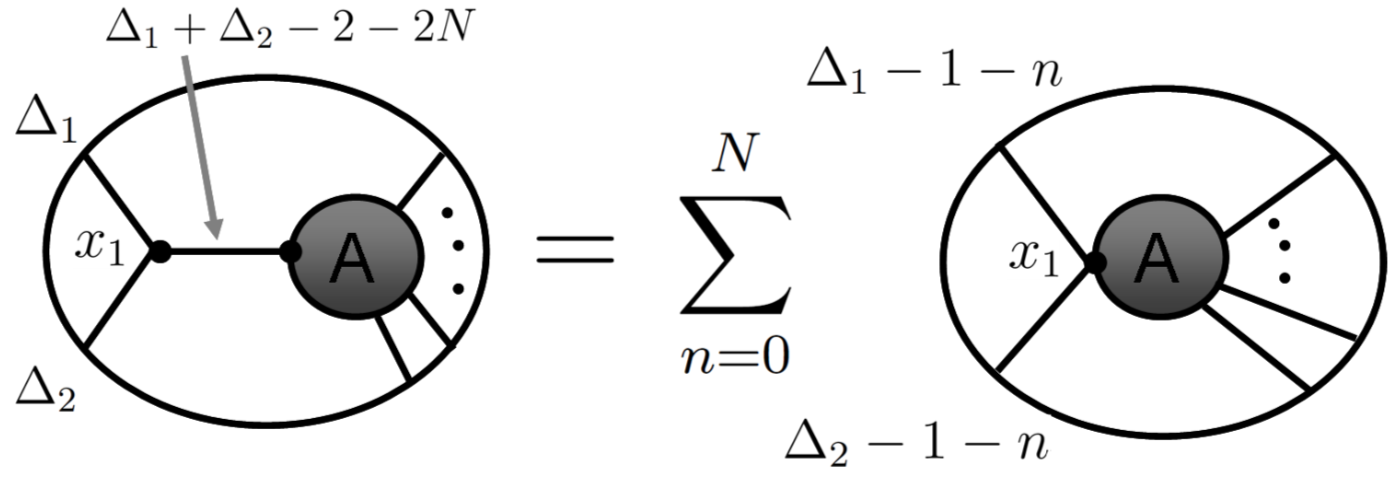}
\caption{ Identity 3'.  This identity generalizes the well known relations between tree level exchange 4-point diagrams and 4-point contact diagrams. We are suppressing numerical coefficients in these plots. The grey blob is general, and the diagrams correspond to stripped correlators.}
\label{fig:bubblerelations3}
\end{figure}
Plugging $\D=\D_1+\D_2-2$ in Eq.~\ref{eq:V} gives:
\begin{align} 
&U^{(\D_1,\D_2,\D)}_{V} =   \G_{\frac{\Delta_1+\D_2}{2}-\frac{d}{4}+\frac{i \nu}{2}} \G_{\frac{\Delta_1+\D_2}{2}-\frac{d}{4}-\frac{i \nu}{2}}  \frac{1}{\n^2+(\D_1+\D_2-2-\frac{d}{2})^2}
\nn
& =  \G_{\frac{\Delta_1+\D_2}{2}-1-\frac{d}{4}+\frac{i \nu}{2}} \G_{\frac{\Delta_1+\D_2}{2}-1-\frac{d}{4}-\frac{i \nu}{2}}  =  \frac{1}{4} U_{A.V}^{(\D_1-1,\D_2-1)}
\end{align} 
were we simply used the properties of the gamma functions, and used the definition in Eq.~\ref{eq:AV}. So we have:
\begin{align} 
\label{eq:145}
 \textbf{Identity\ 3:} \ \ \ \ \ \ \ \ \boxed{U_{V}^{(\D_1,\D_2,\D=\D_1+\D_2-2)}   =\frac{1}{4} U_{A.V}^{(\D_1-1,\D_2-1)}  }
\end{align} 
 A special case is $\D_1=\D_2=\frac{3}{2}$:
\begin{align} 
\label{eq:rel1}
U_{V}^{(\frac{3}{2},\frac{3}{2},\D=1))}   = \frac{1}{4}U_{A.V}^{(\frac{1}{2},\frac{1}{2})} 
\end{align} 
Eq.~\ref{eq:145} can be generalized:
 \begin{align}
 \label{eq:rel3prime}
\textbf{Identity\ 3':} \ \ \ \ \ \ \ \  \boxed{U^{(\D_1,\D_2, \D_1+\D_2-2-2N)}_V = \frac{1}{4}\sum_{n=0}^{N} a_{n,N} U^{(\D_1-1-n, \D_2-1-n)}_{A.V} }
\end{align}
where $N$ is an integer. We show this in Fig.~\ref{fig:bubblerelations3}. Solving for the coefficients gives the formula:
\begin{align}
 a_{n,N} = \frac{1}{(N+1)_{-n}(\D_1+\D_2-\frac{d}{2}-N-1)_{-n}}
\end{align}
where $(a)_b$ is the Pochhammer symbol. 

In \cite{DHoker:1999kzh,Dolan:2000ut} it was shown that certain 4-point exchange diagrams were equal to a finite sum of 4-point contact diagrams. Identity 3' implies that this arises from a more general property of a bulk vertex with 2 bulk-to-boundary and 1 bulk-to-bulk propagators attached to it. This identity has previously been derived in \cite{Goncalves:2019znr}, see footnote~\ref{footnote:56}. See also \cite{Jepsen:2019svc}.
 
\begin{figure}[t]
\centering
\includegraphics[clip,height=8.6cm]{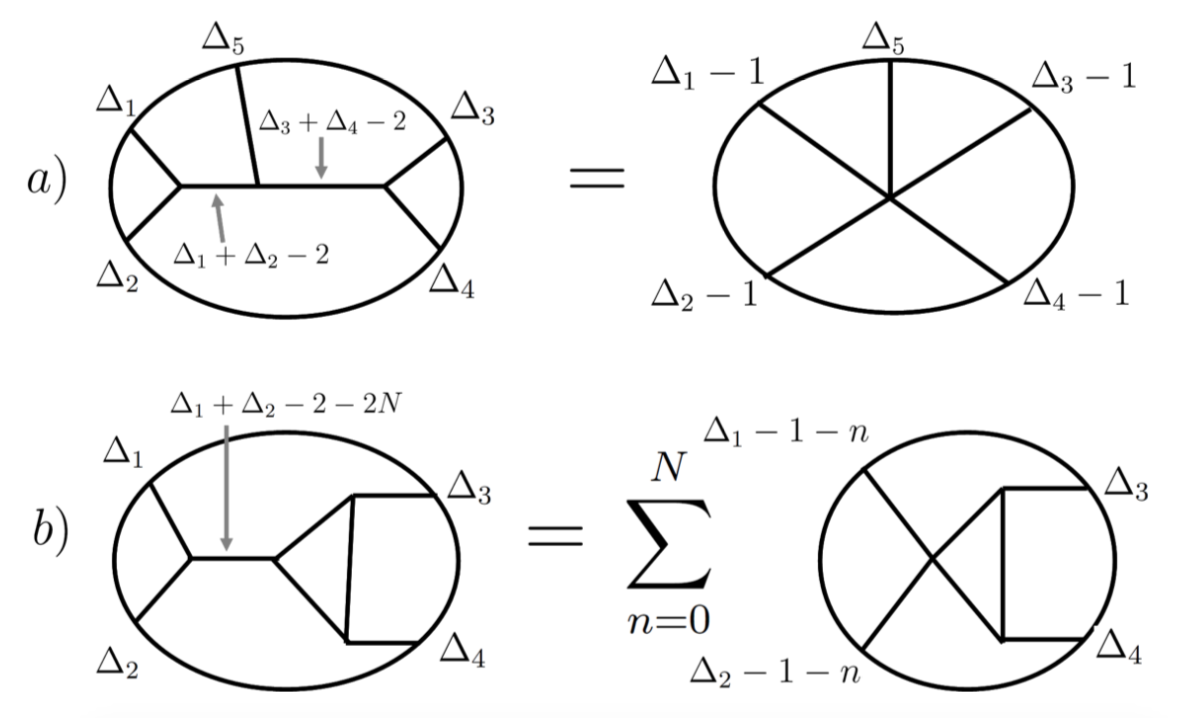}
\caption{Applications of Identities 3 and 3'. \textbf{a)} An example of a 5-point exchange diagram which equals the 5-point contact diagram on the right. \textbf{b)} Identity 3' applied to 4-point triangle diagram. We are suppressing numerical coefficients in these plots.}
\label{fig:bubblerelations6}
\end{figure}

Let us give 2 examples of applications of Identity 3:
\begin{itemize}
\item \underline{Example 1:} In Fig.~\ref{fig:bubblerelations6}a, we show a 5-point tree-level exchange diagram which from Identity 3 is equal to the 5-point contact diagram. See also \cite{Goncalves:2019znr}.

\item \underline{Example 2:} Identity 3' relates the two types of triangle diagrams shown in Fig.~\ref{fig:bubblerelations6}b.
\end{itemize}

\subsection{Identity 4}
\label{sec:bubblebulk}
\begin{figure}[t]
\centering
\includegraphics[clip,height=7.6cm]{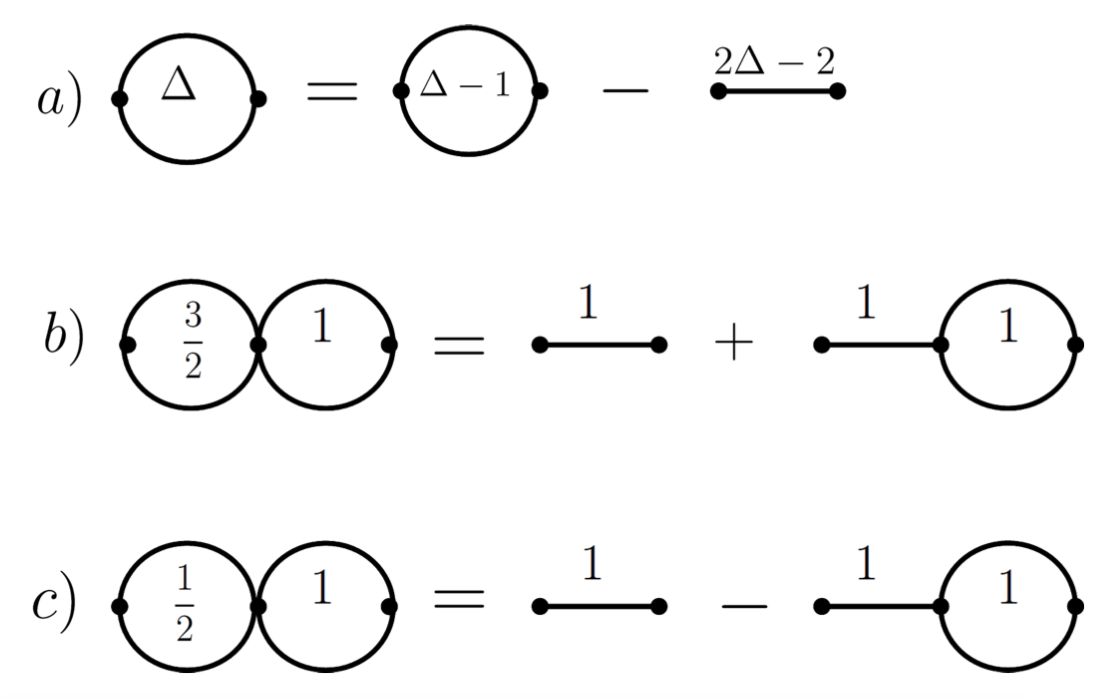}
\caption{\textbf{a)} Identity 4: The 1-loop bubble with scaling dimension $\D$ is simply related to the bubble with scaling dimension $\D-1$. \textbf{b)} Identity 5': The two-loop bubble on the left equals a 1-loop bubble plus a propagator.  \textbf{c)} Identity 5. Identities 5 and 5' enable to reduce an $M$ bubble diagram to an $M-1$ bubble diagram. We are suppressing numerical coefficients in these plots.}\label{fig:bubblerelations2}
\end{figure}
In this subsection we simply recall the relation for the 1-loop ($d=2$) bubble derived in Eq.~3.11 of \cite{Fitzpatrick:2010zm} or Eq.~53 of \cite{Fitzpatrick:2011hu}:
\begin{align} 
\label{eq:new453}
G^2_\D(x_1,x_2) = \frac{1}{2\pi} \sum^\infty_{n=0} G_{2n+2\D} (x_1,x_2) =   G^2_{\D-1} (x_1,x_2) -\frac{1}{2\pi} G_{2\D-2} (x_1,x_2)
\end{align} 
This relation in the spectral representation Eq.~\ref{eq:bulkspect4} is:
\begin{align} 
 \tilde{B}^{(\D)}_\n= \sum^\infty_{n=0} \frac{1}{\n^2+(2\D+2n-1)^2} 
\end{align} 
or,
\begin{align} 
\label{skdnen4}
\textbf{Identity\ 4:} \ \ \ \ \ \ \ \  \boxed{ \tilde{B}^{(\D)}_\n =  \tilde{B}^{(\D-1)}_\n - \frac{1}{2\pi} \frac{1}{\n^2+(2\D-3)^2} }
\end{align} 
See Fig.~\ref{fig:bubblerelations2}a.  Thus for example the bubble function for any integer scaling dimension $\D=$integer can be written in terms of the $\D=1$ bubble:
\begin{align} 
\label{eq:newrt4}
\tilde{B}^{(\D)}_\n =  \tilde{B}^{(\D-1)}_\n - \frac{1}{2\pi} \sum_{j=0}^{\D-2} \frac{1}{\n^2+(2j+2)^2} \ \ \ \ \  \ \ , \ \ \ \ \ \ \ \ \ \D=integer
\end{align} 

\subsection{Identity 5}
We consider mainly bubbles of scaling dimension $\D=1$ and $\D=\frac{3}{2}$, and from it one can get the bubble function for arbitrary integer (half-integer) $\D$ by using Eq.~\ref{skdnen4}.
Multiplying Eqs~\ref{eq:b1}, \ref{eq:b2}, and \ref{eq:b3}, we get a reduction of 2 bubbles in to a single bubble and a bulk propagator:
\begin{align} 
\label{eq:1457}
\textbf{Identity\ 5:} \ \ \ \ \ \ \ \ \boxed{\tilde{B}^{(\D=1)}_\n \tilde{B}^{(\D=\frac{1}{2})}_\n =   -\frac{ 1  }{64 ( i\n)^2}-\frac{1}{4\pi(i\n)^2}  \tilde{B}^{(\D=1)}_\nu  }
\end{align} 
and
\begin{align} 
\label{eq:1467}
\textbf{Identity\ 5':} \ \ \ \ \ \ \ \ \boxed{\tilde{B}^{(\D=1)}_\n \tilde{B}^{(\D=\frac{3}{2})}_\n =   -\frac{ 1  }{64 ( i\n)^2}+\frac{1}{4\pi(i\n)^2}  \tilde{B}^{(\D=1)}_\nu }
\end{align} 
These identities are shown in Fig.~\ref{fig:bubblerelations2}b and \ref{fig:bubblerelations2}c, and they enable to reduce $M$-loop bubble diagrams into a $M-1$-loop bubble diagrams.

\subsection{Identity 6}
\label{sec:relation4}
\begin{figure}[t]
\centering
\includegraphics[clip,height=4.3cm]{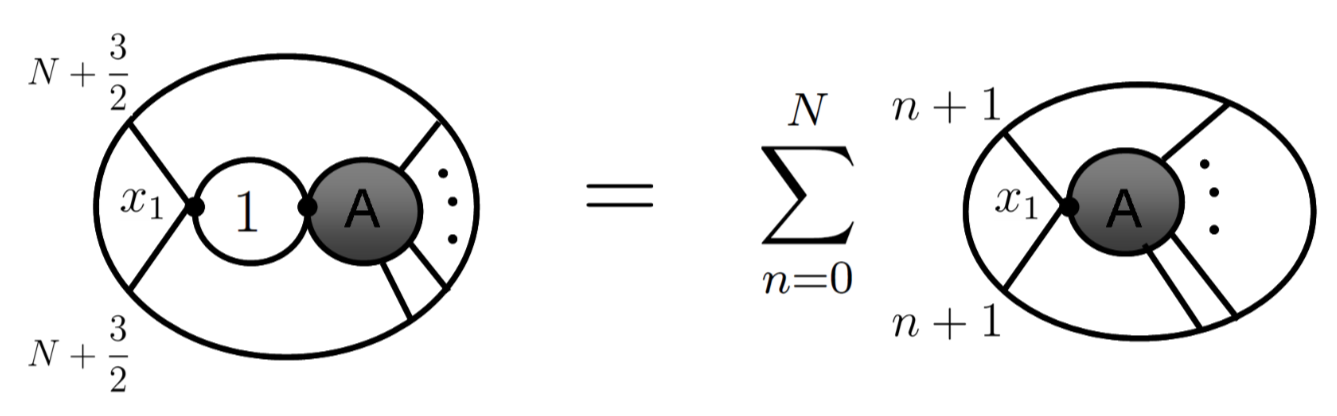}
\caption{ Identity 6'' gives a reduction of integer scaling dimension bubble diagrams, with half integer external legs. This identity is the 1-loop generalization of identity 3' shown in Fig.~\ref{fig:bubblerelations3}. The grey blobs are general. We are suppressing numerical coefficients in these plots, and the diagrams correspond to stripped correlators.}\label{fig:bubblerelations8}
\end{figure}
Consider the 1-loop bubble with scaling dimension $\D=1$. Using Eqs.~\ref{eq:AV}, \ref{eq:B}, and \ref{eq:b1}, and plugging $\D=1$ and $\frac{\D_1+\D_2}{2}=\frac{3}{2}$ in $U_B$, and $\frac{\D_1'+\D_2'}{2}=2$ in $U_{A.V}$, we get\footnote{In order for the full spectral representations to be equivalent, we need the conformal blocks to be equal between the two cases, thus we also impose that the differences be equal, i.e $\D_1'-\D_2' =\D_1-\D_2$.}:
\begin{align}
\label{eq:146}
\textbf{Identity\ 6:} \ \ \ \ \ \ \ \  \boxed{ U^{(\D_1,3-\D_1,\D=1)}_B = \frac{1}{16} U^{(\D_1-\frac{1}{2} , \frac{5}{2}-\D_1 )}_{A.V} } 
\end{align}
 A particular case is when $\D_2=\D_1$ in which:
 \begin{align}
 \label{eq:rel4}
U^{(\frac{3}{2},\frac{3}{2},\D=1)}_B = \frac{1}{16} U^{(1 ,1 )}_{A.V} 
\end{align}
We can derive another identity by multiplying Eq.~\ref{eq:145} by $B^{(\D=1)}_\n $ and then using Eq.~\ref{eq:146}
\begin{align}
\textbf{Identity\ 6':} \ \ \ \ \ \ \ \  \boxed{U^{(\D_1,1-\D_1,\D=1)}_B  =  \frac{1}{4}U_{V}^{(\D_1+\frac{1}{2},\frac{3}{2}-\D_1,\D=1)} }   
\end{align}
A particular case when $\D_2=\D_1$, gives:
\begin{align}
U^{(\frac{1}{2},\frac{1}{2},1)}_B  =  \frac{1}{4} U_{V}^{(1,1,1)}  
\end{align}
Eq.~\ref{eq:146} can be generalized to the case of arbitrary half-integer dimension $\frac{\D_1+\D_2}{2}=N+\frac{3}{2}$, where $N$ is an integer:
 \begin{align}
\label{eq:146a}
\textbf{Identity\ 6'':} \ \ \ \ \ \ \ \  \boxed{U^{(\D_1,2N+3-\D_1,\D=1)}_B = \sum_{n=0}^{N} a_n U^{(\D_1-N+n- \frac{1}{2},N+n+\frac{5}{2}-\D_1)}_{A.V} }
\end{align}
and it is simple algebra to solve for the coefficients $a_n$, which are just numbers. For the case of equal scaling dimensions $\D_1=\D_2$, Eq.~\ref{eq:146a} becomes:
 \begin{align}
 \label{eq:po0o0o0o}
U^{(N+\frac{3}{2},N+\frac{3}{2},1)}_B = \sum_{n=0}^{N} a_n U^{(n+1,n+1)}_{A.V}  
 \end{align}
This identity is shown in Fig.~\ref{fig:bubblerelations8}. For example, for $N=1$:
 \begin{align}
U^{(\frac{5}{2},\frac{5}{2},1)}_B = \frac{3}{64}U^{(1,1)}_{A.V}  +\frac{1}{16}U^{(2,2)}_{A.V} 
 \end{align}
In Appendix~\ref{sec:A} we derive 2 more identities, in addition to the 6 identities shown in this section.

\subsection{Obtaining loop diagrams using the Identities}
\label{sec:lkjhkl}
\begin{figure}[t]
\centering
\includegraphics[clip,height=8.1cm]{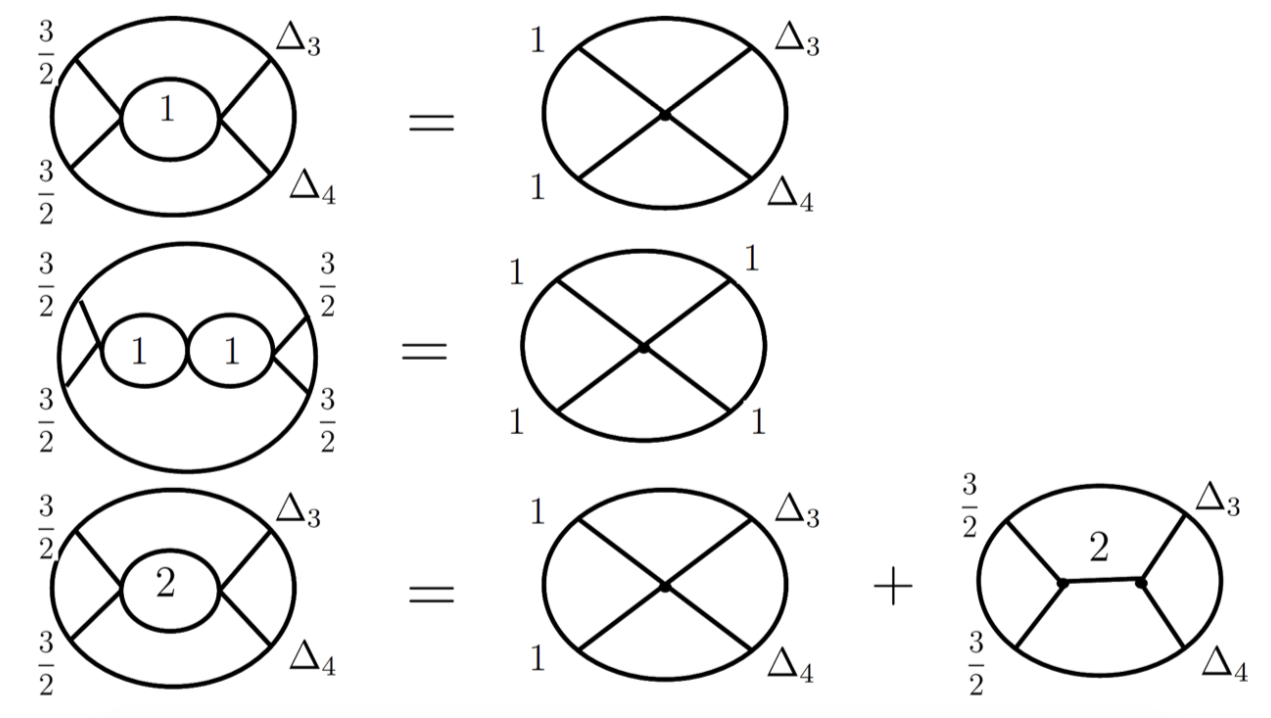}
\caption{Three examples of 1 and 2-loop diagrams written in terms of contact and exchange diagrams. This arises from Identity 6. The bottom figure also uses Identity 4. We are suppressing numerical coefficients in these plots.}
\label{fig:bubblerelations10}
\end{figure}

\begin{figure}[t]
\centering
\includegraphics[clip,height=7.4cm]{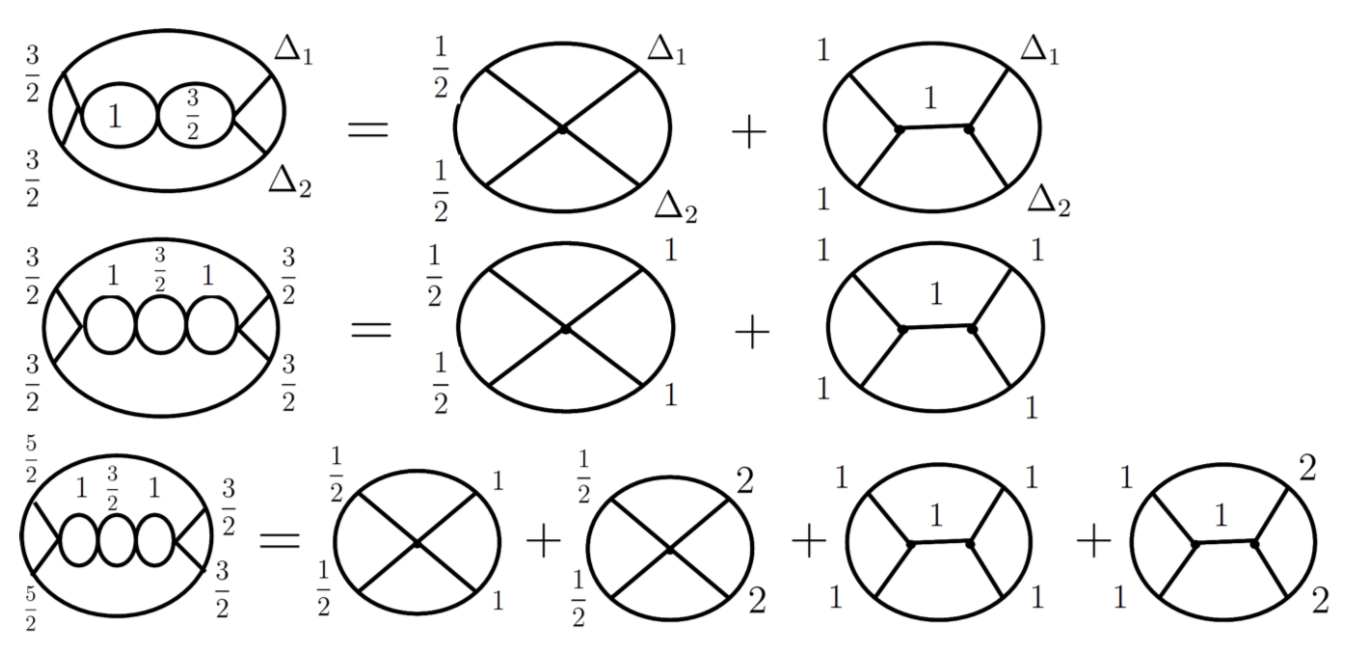}
\caption{Examples of alternating integer and half-integer bubble Witten diagrams which are equal to contact and exchange diagrams. Using Identities 5,6, and 7. The bottom diagram uses Eq.~\ref{eq:po0o0o0o}. We are suppressing numerical coefficients in these plots.}
\label{fig:bubblerelations11}
\end{figure}

\begin{figure}[t]
\centering
\includegraphics[clip,height=5.3cm]{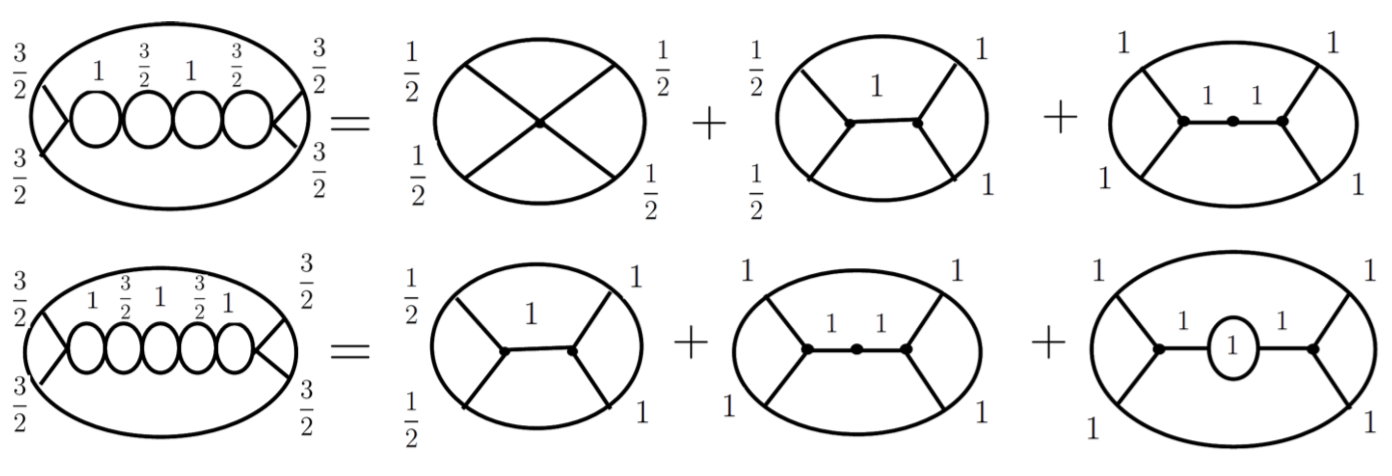}
\caption{Examples of alternating integer and half-integer bubble Witten diagrams. Using Identities 5,6, and 7. We are suppressing numerical coefficients in these plots. }
\label{fig:bubblerelations12}
\end{figure}

In this subsection we show applications of the bulk vertex/propagator identities to the reduction of loop bubble Witten diagrams in to lower loop diagrams, when the bubble has integer or half-integer scaling dimension.
\begin{itemize}
\item From Identity 6 Eq.~\ref{eq:rel4} one immediately sees that the 4-point 1-loop bubble of Fig~\ref{fig:bubblerelations10}-top is equal to a contact diagram.
\item Applying Identity 6 twice, the 2-loop bubble 4-point function bubble of Fig~\ref{fig:bubblerelations10}-middle is equal to the contact diagram on the right.
\item Combining Identity 6 with Identity 4 Eq.~\ref{skdnen4}, we get that the 4-point 1-loop bubble of Fig~\ref{fig:bubblerelations10}-bottom is equal to the contact diagram plus exchange diagram on the right.
 \item Examples of alternating integer and half-integer bubble Witten diagrams are shown in Fig~\ref{fig:bubblerelations11} and \ref{fig:bubblerelations12}. These relations can be obtained from Identities 5 and 6.  
\end{itemize}


\section{A finite coupling AdS 4-point function: The $O(N)$ model}
\label{sec:3}
\begin{figure}[t]
\centering
\includegraphics[clip,height=2.25cm]{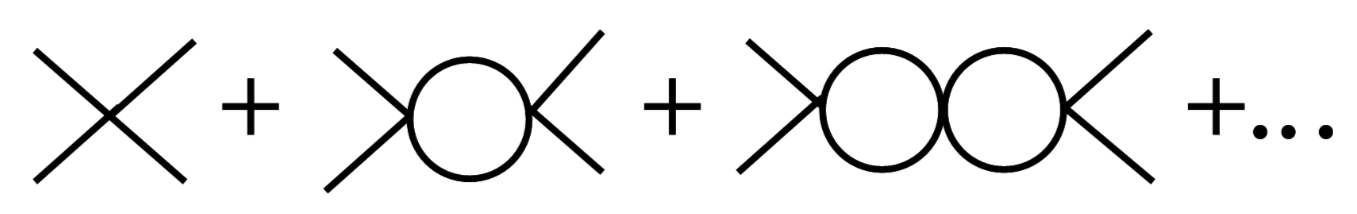}
\caption{The exact 4-point function in the large-$N$ $O(N)$ model on $AdS$ is given by an infinite sum of bubble diagrams.}\label{fig:b23}
\end{figure}
Now we focus on the case with all the scaling dimension equal $\D_1=\D_2=\D_3=\D_4\equiv \D$, and we will consider the 4-point function with resummed bubble diagrams, Fig.~\ref{fig:b23}. This finite coupling 4-point function is realized in the large-$N$ $O(N)$ model on $AdS_3$. The Lagrangian for the $O(N)$ model is:
\begin{align} 
\mm{L}= \frac{1}{2}(\pa \phi^i)^2 + \frac{m^2}{2} (\phi^i)^2 +\frac{\l}{2N}((\phi^i))^2
\end{align} 

As shown in \cite{Carmi:2018qzm}, the resummation of bubble diagrams is a geometric series in spectral space. The resummation gives the following spectral function $\tilde{F}_\n = \frac{1}{\l^{-1}+2\tilde{B}_\n}$. Thus from Eq.~\ref{eq:sdbsmd7}, or alternatively Eq.~4.18 of \cite{Carmi:2018qzm}, the 4-point function is:
\begin{align} 
\label{eq:solution2}
&g_{4}(z, \bar z)=   \int_{-\infty}^\infty  \frac{d\nu}{\l^{-1}+2\tilde{B}_\n}\frac{\Gamma_{\Delta-\frac{d+2i\nu}{4}}^2\G_{\Delta-\frac{d-2i\nu}{4}}^2\G_{\frac{d+2i\nu}{4}}^4}{ \G_{i\nu}\G_{\frac{d}{2}+i\nu}}\mathcal{K}_{\frac{d}{2}+i\nu}(z,\bar{z}) 
\nn
&=   \int_{-\infty}^\infty  \frac{d\nu}{\l^{-1}+2\tilde{B}_\n}  \G_{\Delta-\frac{1}{2}-\frac{i\nu}{2}}^2\G_{\Delta-\frac{1}{2}+\frac{i\nu}{2}}^2 4(i\n)Q_{\frac{i\n-1}{2}} (\hat z)  Q_{\frac{i\n-1}{2}} (\hat{\bar z})
\end{align}
where in second line we used the expression for the scalar conformal block in $d=2$:
\begin{align}
\label{eq:confblock}
\mathcal{K}_{\b}(z,\bar{z}) = k_\b(z) k_\b(\bar z) = 4 \frac{\G^2(\b)}{\G^4(\frac{\b}{2})} Q_{\frac{\b}{2}-1} (\hat z)  Q_{\frac{\b}{2}-1} (\hat{\bar z})
\end{align}
and where $Q_{\frac{\b}{2}-1} (\hat z)$ is the LegendreQ function, and the "hats" are defined as:
\begin{align}
\hat z \equiv \frac{2}{z} -1 \ \ \ \ \ \ , \ \ \ \ \ \ \ \ \hat{\bar z} \equiv \frac{2}{\bar z} -1 
\end{align}
In Eq.~\ref{eq:solution2} we close the contour in the $\n$-plane and pick up the poles. There are 2 towers of poles: double-trace poles coming from $\Gamma_{\Delta-\frac{d+2i\nu}{4}}^2$, and a tower of poles coming from the factor $\tilde{F}_\n = \frac{1}{\l^{-1}+2\tilde{B}_\n}$. We label these contributions $g_{d.t}$ and $g_B$:
\begin{align} 
\label{eq:gursoy}
g_4(z, \bar z) = g_{d.t}(z, \bar z)+g_B(z, \bar z)
\end{align} 
No we focus on the bulk conformal point (see section 5 of \cite{Carmi:2018qzm}): $\lambda \to \infty$ and $\D=1$. Plugging Eq.~\ref{eq:b1} inside Eq.~\ref{eq:solution2} gives:
\begin{align} 
\label{eq:app9}
g_{4}(z,\bar z )= 16\pi^2 \int_{-\infty}^\infty  \frac{ d\n }{\sin \frac{\pi}{2}(i\n)\sin \frac{\pi}{2}(i\n+1)  }     (i\n)^2
Q_{\frac{i\n}{2}-\frac{1}{2}} (\hat z)  Q_{\frac{i\n}{2}-\frac{1}{2}} (\hat{\bar z})
\end{align}
The residue theorem gives\footnote{ For general $\D$ we can write the double-trace sum from Eq.~\ref{eq:solution2}:
\begin{align}
 g_{d.t}(z, \bar z)= -64\pi^2 \sum_{n=0}^\infty \frac{(2n+2\D-1)^2\G^2(2\D+n-1)}{\G^2(n+1)} Q_{n+\D-1}(\hat z)Q_{n+\D-1}(\hat{\bar z}) 
\end{align} On the other hand, for the bubble poles we only know how to write the sum for the case of $\D=1$, as in Eq.~\ref{eq:lwnjdn5}. For general $\D$, one would need to analytically compute the zeros of the denominator $\l^{-1}+2\tilde{B}_\n=0$ of Eq.~\ref{eq:solution2}, and this is known only for $\D=1$ and $\l \to \infty$.}:
\begin{align} 
\label{eq:bvcxz}
g_{d.t}(z,\bar z )= -64\pi^2 \sum_{n=0}^\infty   (2n+1)^2 Q_n (\hat z)  Q_n (\hat{\bar z}) \ \ \ \ \ \ \ ,\ \ \  \ \ \ \ i\n = 2n+1
\end{align}
and
\begin{align} 
\label{eq:lwnjdn5}
g_{B}(z,\bar z )= 64\pi^2 \sum_{n=0}^\infty   (2n+2)^2 Q_{n+\frac{1}{2}} (\hat z)  Q_{n+\frac{1}{2}} (\hat{\bar z}) \ \ \ \ \ \ \ ,\ \ \  \ \ \ \ i\n = 2n+2
\end{align}
The two sums above are related to each other by $n \to n+\frac{1}{2}$.
These sums are complicated to compute analytically, and we have not found a reference which computed them. In Eq.~\ref{eq:sumb} of Appendix~\ref{sec:appb} we computed the sum $g_B$:
 \begin{align}
 \label{eq:exact6}
\boxed{
g_{B}(z,\bar z )= 64\pi^2  \Big(1+2D_{z,\bar z} \Big)  \Big[ \frac{\pi}{2^{\frac{3}{2}}} \frac{2\sqrt{c}}{\sqrt{a-c}} K\Big(\sqrt{\frac{b-c}{a-c}} \Big)  - \frac{1}{2}  Q_{-\frac{1}{2}}(\hat z)Q_{-\frac{1}{2}}(\hat{\bar z}) \Big]  }
 \end{align}
where $K$ is the elliptic integral function of the first kind, defined in Eq.~\ref{eq:elliptic78}. On the other hand, we were not able to directly compute the double-trace sum in Eq.~\ref{eq:bvcxz}. Note that the bubble sum is equal to the double-discontinuity:
 \begin{align}
 g_{B}(z,\bar z ) = dDisc_s  [ g_{4}(z,\bar z )]= dDisc_s  [ g_{B}(z,\bar z )]
 \end{align}
This can be seen from Eq.~\ref{eq:app9}, and the fact that taking the double-discontinuity cancels the double-trace poles. $g_B$ has the nice property that it's double-discontinuity is equal to itself.

\subsection{Identity 9}
\begin{figure}[t]
\centering
\includegraphics[clip,height=7.9cm]{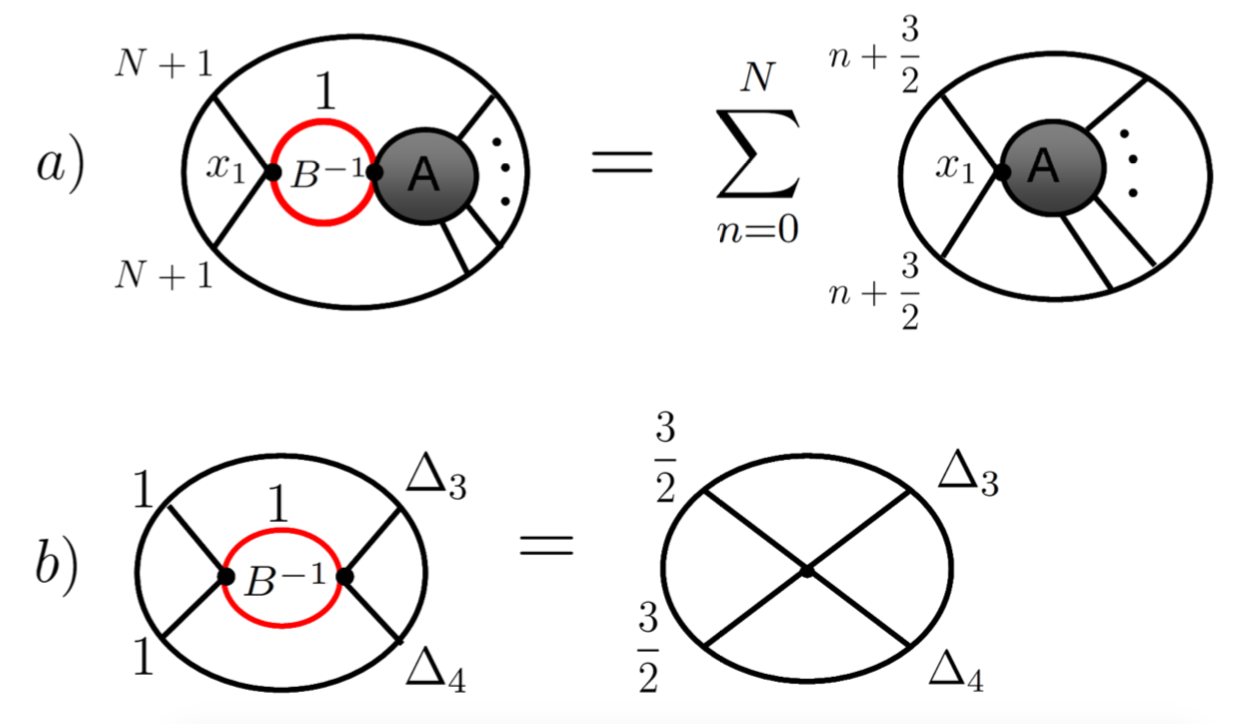}
\caption{\textbf{a)} Identity 9': reduces the resummed bubble diagrams with $\D=1$. The red loop is notation for the resummed bubble with scaling dimension $\D=1$, given on the right hand side of Eq.~\ref{eq:B1}. \textbf{b)} Applying Identity 9 immediately gives the result for the 4-point function in the conformal large-$N$ $O(N)$ model on $AdS_3$, in terms of the contact diagram on the right. $\D_3$ and $\D_4$ are arbitrary in this diagram. The grey blob is general, numerical coefficients are suppressed, and the diagrams correspond to stripped correlators.}
\label{fig:bubblerelations13}
\end{figure}

In this subsection we derive Identity 9, and show that it can be used to derive the 4-point function of the large $N$ conformal $O(N)$ that we attempted to compute in the previous subsection. 
We define
\begin{align}
\label{eq:B1}
U^{(\D_1,\D_2,\D)}_{B^{-1}} \equiv     \frac{1}{16} \G_{\frac{\Delta_1+\D_2}{2}-\frac{d}{4}+\frac{i \nu}{2}} \G_{\frac{\Delta_1+\D_2}{2}-\frac{d}{4}-\frac{i \nu}{2}} \times ( \tilde{B}^{(\D)}_\n  )^{-1}
\end{align}
Using Eqs.~\ref{eq:AV}, \ref{eq:B1}, and \ref{eq:b1}, and plugging $\D=1$ and $\frac{\D_1+\D_2}{2}=1$ in $U_{B^{-1}}$, and $\frac{\D_1'+\D_2'}{2}=\frac{3}{2}$ in $U_{A.V}$, we get:
\begin{align}
\label{eq:156}
\textbf{Identity\ 9:} \ \ \ \ \ \ \ \ \boxed{ U^{(\D_1,2-\D_1,\D=1)}_{B^{-1}} = U^{(\D_1+\frac{1}{2} , \frac{5}{2}-\D_1 )}_{A.V} } 
\end{align}
When $\D_2=\D_1$, this becomes:
\begin{align}
\label{eq:solution1}
 U^{(1,1,\D=1)}_{B^{-1}} =U^{( \frac{3}{2} , \frac{3}{2}  )}_{A.V} 
\end{align}
Eq.~\ref{eq:156} can be generalized to the case of arbitrary integer dimension $\frac{\D_1+\D_2}{2}=N+1$, where $N$ is an integer:
\begin{align}
\textbf{Identity\ 9':} \ \ \ \ \ \ \ \   \boxed{ U^{(\D_1,2N+2-\D_1,\D=1)}_{B^{-1}} = \sum_{n=0}^{N} a_n U^{(\D_1-N+n+ \frac{1}{2},N+n+\frac{5}{2}-\D_1)}_{A.V}  }
\end{align}
where one can easily find the coefficients $a_n$ by simple algebra. When $\D_2=\D_1$ this relation is:
 \begin{align}
U^{(N+1,N+1, \D=1)}_{B^{-1}} = \sum_{n=0}^{N} a_n U^{(n+\frac{3}{2},n+\frac{3}{2})}_{A.V}  
 \end{align}
This is shown in Fig.~\ref{fig:bubblerelations13}a.

Using Eq.~\ref{eq:solution2} and Identity 9 (more specifically Eq.~\ref{eq:solution1}), we get the finite coupling 4-point function of the conformal large $N$ $O(N)$ model in terms of a contact diagram!
 \begin{align}
\label{eq:exact7}
\boxed{
g_4 (z,\bar z)=  \frac{1}{2} g_4^{(contact)}(z,\bar z) }
 \end{align}
where $ g_4^{(contact)}(z,\bar z)$ is the contact diagram with $\D_i=(1,1,\frac{3}{2},\frac{3}{2})$, See Fig.~\ref{fig:bubblerelations13}b.\\

\underline{Aside:} 

We can reverse the logic, and use this section to compute the double-trace sum in Eq.~\ref{eq:bvcxz} in terms of $g_4^{(contact)}(z,\bar z)$ and elliptic integrals:
\begin{align} 
-64\pi^2 \sum_{n=0}^\infty   (2n+1)^2 Q_n (\hat z)  Q_n (\hat{\bar z}) = g_{d.t}(z,\bar z ) =  \frac{1}{2}g_4^{(contact)}(z,\bar z) -g_B(z, \bar z)
\end{align}
where $g_B$ is given by Eq.~\ref{eq:exact6}, and we used Eq.~\ref{eq:gursoy}. We are not aware of any reference which previously computed this sum.

\section{Scalar 4-point diagrams with $\D_{12}=\D_{34}=0$}
\label{sec:4}
\begin{figure}[t]
\centering
\includegraphics[clip,height=4.2cm]{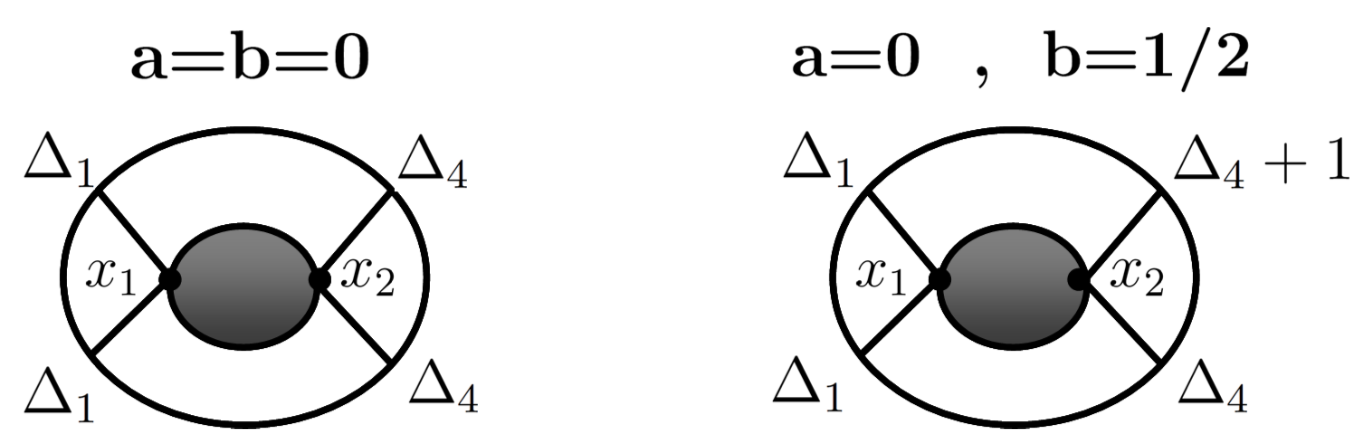}
\caption{\textbf{Left:} The 4-point function with $a=b=0$, discussed in section~\ref{sec:4}. \textbf{Right:} The 4-point function with $a=0$, $b=\frac{1}{2}$, discussed in section~\ref{sec:5}. The grey blob is general.}
\label{fig:bubblerelations14}
\end{figure}
\begin{figure}[t]
\centering
\includegraphics[clip,height=3.9cm]{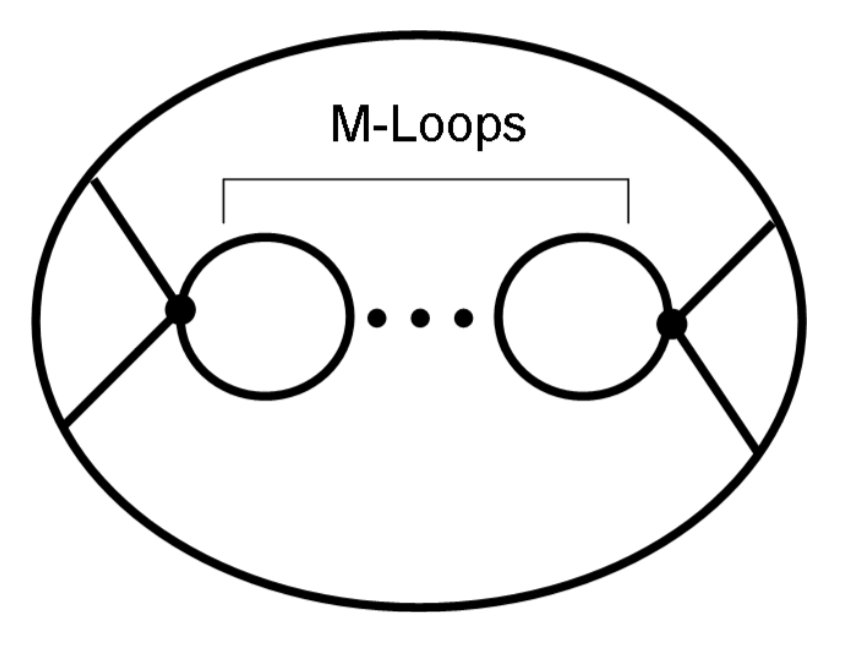}
\caption{The 4-point $M$-loop bubble diagram.}\label{fig:bb22}
\end{figure}
In this section we consider a scalar field in AdS with $\phi^4$ coupling, and compute perturbative 4-point functions with external scaling dimensions $\D_{12}=\D_{34}=0$, Fig.~\ref{fig:bubblerelations14}-Left. From Eq.~\ref{eq:sdbsmd7} the spectral representation of the 4-point function is given by:
\begin{align} 
g_{4}(z,\bar z ) =
 \int d\nu \tilde{F}_\n 
  \frac{ ( \G_{\D_1-\frac{d}{4}+\frac{i \nu}{2}} \G_{\D_1-\frac{d}{4}-\frac{i \nu}{2}} \G_{\D_4-\frac{d}{4}+\frac{i \nu}{2}} \G_{\D_4-\frac{d}{4}-\frac{i \nu}{2}})   \Gamma^4_{\frac{d}{4}+\frac{i\n}{2}} }{\G_{i\n }\Gamma_{\frac{d}{2}+i\n} } 
\mathcal{K}^{\D_i}_{\frac{d}{2}+i\nu}   (z,\bar z )
\end{align}
Plugging the $d=2$ conformal block Eq.~\ref{eq:confblock}, gives:
\begin{align} 
\label{eq:dsfm3md}
g_{4}(z,\bar z )= 4\pi^2\int d\nu \tilde{F}_\nu
\frac{\frac{\G_{\D_1-\frac{1}{2}+\frac{i \nu}{2}} }{\G_{-\D_1+\frac{3}{2}+\frac{i \nu}{2}}} \frac{\G_{\D_4-\frac{1}{2}+\frac{i \nu}{2}} }{\G_{-\D_4+\frac{3}{2}+\frac{i \nu}{2}}} }{\sin \pi(\D_1-\frac{1}{2}-\frac{i \nu}{2}) \sin \pi(\D_4-\frac{1}{2}-\frac{i \nu}{2}) }     (i\n)
Q_{\frac{i\n-1}{2}} (\hat z)  Q_{\frac{i\n-1}{2}} (\hat{\bar z})
\end{align}
where we also used a gamma function identity $\G_x\G_{1-x} = \frac{\pi}{\sin \pi x}$. Now we consider the diagram comprised of a sequence of $M$ bubbles, which has the spectral function $\tilde{F}_\n = \tilde{B}^M_{\n}$, Fig.~\ref{fig:bb22}.There are double-trace poles coming from the two sines in the denominator, and there are poles of order $M$ coming from the bubbles $\tilde{B}^M_\nu$. The poles can collide when the external $\D_1$ and $\D_4$ are separated by an integer, or when they are separated from the bubble poles by a integer. 

We want to obtain analytic expression for the 4-point function, by closing the $\n$ contour picking up the poles. Recall from Eq.~\ref{eq:bubble4} that in $d=2$ the bubble spectral representation is a combination of digamma functions. The poles of the bubble function are given in Eq.~\ref{eq:poles6}:
 \begin{align}
\label{eq:dsfm3md3}
\tilde{B}^{(d=2)}_\n \overset{1+i\n \, \sim  \,2\D+2n}{\sim} -\frac{1}{4\pi} \frac{1}{i\n-(2n+2\D-1)} \frac{1}{2\D+2n-1}
 \end{align}
Therefore we will have sums containing a product of digamma functions, LegendreQ's, and gamma functions. 

\subsection{Contact diagrams}
For contact diagrams we plug $\tilde{F}_\n=1$ in Eq.~\ref{eq:dsfm3md}, and the residue theorem gives:
\begin{align} 
\label{eq:lkjgfs1}
&g_{4}(z,\bar z )= \frac{16\pi^2}{\sin \pi(\D_4-\D_1)} \times
\nn
& \sum_{n=0}^\infty  \frac{\G_{2\D_1-1+n }}{\G_{n+1}} \frac{\G_{\D_4+\D_1-1+n} }{\G_{1-\D_4+\D_1+n }}   (2\D_1+2n-1)Q_{\D_1-1+n} (\hat z)  Q_{\D_1-1+n} (\hat{\bar z}) + \Big( \D_1 \leftrightarrow \D_4  \Big)
\end{align}
In the special case of $\D_4=\D_1+integer$, the poles collide. For example when $\D_4=\D_1$ we have:
\begin{align}
\label{eq:lkjgfs2}
g_{4}(z,\bar z )=  16\pi^2 \sum_{n=0}^\infty \frac{d}{dn} \bigg( \frac{\G^2_{2\D_1+n-1}}{\G^2_{n+1}} (2\D_1+2n-1) Q_{\D_1+n-1}(\hat z) Q_{\D_1+n-1}(\hat{\bar z})  \bigg) 
\end{align}
We don't know how to directly compute these sums in general\footnote{However, contact diagrams are known to give so called $\bar D$ functions. Thus, reversing the logic, we get that the complicated sums of Eqs.~\ref{eq:lkjgfs1} and \ref{eq:lkjgfs2} are given by $\bar D$ functions. As was shown in \cite{Dolan:2000ut}, the $\bar D$ functions can be written in terms of Appell functions.}, however there are special cases that we can do:

\begin{itemize}
\item \underline{Example 1: $\D_1=1$ and $\D_4=1$:}

From Eq.~\ref{eq:lkjgfs2} we have:
 \begin{align} 
 \label{eq:lkjgfs23}
&g_{4}(z,\bar z ) = 16\pi^2 \sum_{n=0}^\infty  \frac{d }{dh } \bigg( (h-1) Q_{\frac{h}{2}-1}(\hat z)Q_{\frac{h}{2}-1}(\hat{\bar z}) \bigg) \bigg|_{h=2n+2}  
\nn 
&= 16\pi^2 \sum_{n=0}^\infty \frac{d}{dn} \bigg( (2n+1) Q_{n}(\hat z) Q_{n}(\hat {\bar z})  \bigg)
\end{align}
where in the second equality we just wrote the sum in a more suggestive way, that explicitly shows the canonical coefficient $(2n+1)$ of Legendre functions. In order to compute the sum, consider the following identity of Legendre functions:
\begin{align}
\label{eq:polns7bbd}
(2\b+1)Q_{\b}(\hat z)Q_{\b}(\hat {\bar z})= \frac{(\b+1)\Big(Q_{\b}(\hat z)Q_{\b+1}(\hat {\bar z})-Q_{\b}(\hat {\bar z})Q_{\b+1}(\hat z) \Big)}{\hat {\bar z}-\hat z} - \Big(\b \to \b-1\Big)
\end{align}
where $\b$ is not necessarily an integer, and in order to show this, one uses the following recursion relation for Legendre functions:
\begin{align}
(2\b+1)yQ_{\b}(\hat {\bar z})= (\b+1)Q_{\b+1}(\hat {\bar z}) +\b Q_{\b-1}(\hat {\bar z})
\end{align}
Summing both sides of Eq.~\ref{eq:polns7bbd} clearly gives rise to a telescopic sum, thus we can write:
\begin{align}
\label{eq:yes}
&\sum_{n=0}^\infty (2(n+\a)+1) Q_{n+\a}(\hat z) Q_{n+\a}(\hat {\bar z}) = 
\nn
&(2\a+1)Q_\a (\hat z) Q_\a (\hat {\bar z})+\frac{(\a+1)\Big(Q_{\a}(\hat z)Q_{\a+1}(\hat {\bar z})-Q_{\a}(\hat {\bar z})Q_{\a+1}(\hat z) \Big)}{\hat z-\hat {\bar z}} 
\end{align}
Since $\a$ is an arbitrary real parameter, we can take derivatives with respect to it and obtain the result of the sum in Eq.~\ref{eq:lkjgfs23}:
\begin{align}
\label{eq:dknnns}
&g_{4}(z,\bar z ) = 16\pi^2 \sum_{n=0}^\infty \frac{d}{dn} \bigg( (2n+1) Q_{n}(\hat z) Q_{n}(\hat {\bar z})  \bigg)
 \nn
&= 16\pi^2 \frac{d}{d\a} \bigg[(2\a+1)Q_\a (\hat z) Q_\a (\hat {\bar z})+  \frac{(\a+1) \Big(Q_{\a}(\hat {\bar z})Q_{\a+1}(\hat z)-Q_{\a}(\hat z)Q_{\a+1}(\hat {\bar z}) \Big)}{\hat {\bar z} -\hat z} \bigg] \bigg|_{\a\to 0}
 \nn
 &= -4\pi^2 \frac{z \bar z}{z-\bar z} \Big( \log \Big(\frac{1-z}{1-\bar z}\Big)\log(z\bar z) +2 Li_2 (z) -2 Li_2 (\bar z) \Big)
\end{align}
This precisely matches the function $\bar D_{1111}$! \cite{Dolan:2000ut,Gary:2009ae}. The function inside the brackets is  the Bloch-Wigner dilogarithm.

\item \underline{Example 2: $\D_1=1,2$ and $\D_4=1,2$:}

We can derive these by applying the $\mm{D}^{(+)}_{z,\bar z}$ operator on Eq.~\ref{eq:dknnns}. This is just a special case of Identity 2, Eq.~\ref{eq:relation22} with $\D_1=\D_2=2$. That lowering/raising relations for contact diagrams are well known \cite{Dolan:2000ut}.

\item  \underline{Example 3: $\D_1=\frac{1}{2}$ and $\D_4=\frac{3}{2}$:}

In this case the sum over poles gives:
 \begin{align} 
g_4(z,\bar z ) = 16\pi^2 \sum_{n=0}^\infty  \frac{d }{dh } \bigg( (h-1) Q_{\frac{h}{2}-1}(\hat z)Q_{\frac{h}{2}-1}(\hat{\bar z}) \bigg) \bigg|_{h=2n+1}  
\end{align}
So this case is different from the previous three cases because the sum is over half integer values. The sum here is again a telescopic sum, and one can use Eq.~\ref{eq:yes} to compute it.

\item  \underline{Example 4: $\D_1=1$ and $\D_4=\frac{1}{2}$:}

In this case the sum over poles gives:
 \begin{align} 
g_4 (z,\bar z ) = -16\pi^2 \sum_{n=0}^\infty   Q_{n}(\hat z)Q_{n}(\hat{\bar z})   + 16\pi^2 \sum_{n=0}^\infty   Q_{n-\frac{1}{2}}(\hat z)Q_{n-\frac{1}{2}}(\hat{\bar z}) 
\nn
= -16\pi^2  \widehat{S}_{d.t}  +16\pi^2  \widehat{S}_{B} +16\pi^2 Q_{-\frac{1}{2}}(\hat z)Q_{-\frac{1}{2}}(\hat{\bar z})
\end{align}
where in the second line we simply noticed that the sums are the same as in Eqs.~\ref{eq:6hdls} and \ref{eq:sdt1}.

\end{itemize}

\subsection{Double-discontinuity of the 1-loop bubble} 
Let us consider taking the double-discontinuity of the 4-point function Eq.~\ref{eq:dsfm3md}. As explained around Eq.~\ref{eq:sdbsmd7q}, taking the double-discontinuity will cancel the double-trace poles. Considering the 1-loop bubble $M=1$ with scaling dimension $\D=1$, one has the poles shown in Eq.~\ref{eq:dsfm3md3}. Thus the residue theorem gives:
\begin{align}
\label{eq:dDisc} 
dDisc_s [g_4(z,\bar z ) ]=  -2\pi^2 \sum_{n=0}^\infty  
 \frac{\G_{\D_1-1+n+\D }}{\G_{-\D_1+1+n+\D}} \frac{\G_{\D_4-1+n+\D} }{\G_{-\D_4+1+n+\D }}     
Q_{n+\D-1} (\hat z)  Q_{n+\D-1} (\hat{\bar z})
\end{align}
We do not know in general how to compute these sums analytically, and therefore we try in the following to find simple cases:
\begin{itemize}
\item Plugging $\D_1=\D_4=1$ in Eq.~\ref{eq:dDisc} gives:
 \begin{align} 
dDisc_s [g_4(z,\bar z ) ]=  -2\pi^2 \sum_{n=0}^\infty  Q_{n+\D-1} (\hat z)  Q_{n+\D-1} (\hat{\bar z})
\end{align}
If $\D$ is an integer or half-integer then then this sum the same as the sums in Appendix~\ref{sec:appb}, except for a finite number of terms in the sum. This can also be seen from Identity 4 in Eq.~\ref{skdnen4}.

\item $M=1$ and $\D_1=\frac{1}{2}$ and $\D_4=\frac{3}{2}$:
 
 In this case we get exactly the same sum as in the previous example.
 \begin{align} 
 \label{eq:dento}
dDisc_s [g_4(z,\bar z ) ]= -2\pi^2 \sum_{n=0}^\infty   Q_{n+\D-1}(\hat z)Q_{n+\D-1}(\hat{\bar z})  
 \end{align} 
 This is just a special case of Identity 8, Eq.~\ref{eq:relation8k}.
 
\item $M=1$ and $\D_1=1$ and $\D_4=\frac{3}{2}$: 
 
 In this case we have:
\begin{align} 
dDisc_s [g_4(z,\bar z ) ]=-\pi^2 \sum_{n=0}^\infty   (2\D+2n-1) Q_{\D+n-1}(\hat z)Q_{\D+n-1}(\hat{\bar z}) 
\end{align}
This sum is a telescopic sum in the canonical Legendre form, and thus we can compute it for any $\D$! From Eq.~\ref{eq:yes} we get:
\begin{align} 
&\frac{-1}{\pi^2}dDisc_s [g_4(z,\bar z ) ]=  
\nn
&(2\D-1)Q_{\D-1}(\hat z)Q_{\D-1}(\hat{\bar z})+\frac{\D \Big(Q_{\D-1}(\hat z)Q_{\D}(\hat{\bar z})-Q_{\D-1}(\hat{\bar z})Q_{\D}(\hat z) \Big)}{ \hat z-\hat{\bar z}} 
\end{align}
Using Identity 2 Eq.~\ref{eq:relation22}, we can raise the external scaling dimensions by applying the differential operator $\mm{D}_{z,\bar z}^{(+)}$ on the result above, and obtain the whole family of integer/half-integer external scaling dimensions, with an arbitrary $\D$ in the bubble.
\end{itemize}

\subsection{A few solvable cases at 1 and 2-loops}
The first simplification is to put $\D=1$ in the internal bubble using Eq.~\ref{eq:b1}. Then Eq.~\ref{eq:dsfm3md} gives\footnote{Note that the bubble with $\D=1$ has the special property of having equally spaces zeros at $h=2n+3$ coming from the $\cot$ factor in Eq.~\ref{eq:b1}. Thus if we choose $\D_4$=half-integer, then the tower of double-trace poles coming from the $\sin \pi(\D_4-\frac{1}{2}-\frac{i \nu}{2})$ denominator in Eq.~\ref{eq:dsfm3md2} cancels with the zeros of the bubble from $\cot^M$. In fact, one can show that these zeros enable the fact that example 1 in this section is actually equal to a contact diagram!}:
\begin{align} 
\label{eq:dsfm3md2}
g_{4}(z,\bar z ) =4\pi^2 \int d\nu
\frac{\frac{\G_{\D_1-\frac{1}{2}+\frac{i \nu}{2}} }{\G_{-\D_1+\frac{3}{2}+\frac{i \nu}{2}}} \frac{\G_{\D_4-\frac{1}{2}+\frac{i \nu}{2}} }{\G_{-\D_4+\frac{3}{2}+\frac{i \nu}{2}}}  \Big( \frac{  \tan^M ( \frac{ \pi i\n }{2}) }{ 8^{M}(i\n)^{M-1}} \Big) }{\sin \pi(\D_1 -\frac{1+i \nu}{2}) \sin \pi(\D_4 -\frac{1+i \nu}{2}) }    Q_{\frac{i\n-1}{2}} (\hat z)  Q_{\frac{i\n-1}{2}} (\hat{\bar z})
\end{align}

\begin{itemize}
\item Example 1: $M=1$ and $\D_1=1$ and $\D_4=\frac{3}{2}$ .

The (double-trace) poles at $h=2n+3$ are canceled by the zeros of the bubble. There is a tower of poles at $h=2n+2$, which gives rise to the following sum:
 \begin{align} 
\label{eq:sno6nd}
&g_4(z,\bar z ) = \frac{\pi^3}{8} \sum_{n=0}^\infty  \frac{d }{dh } \bigg( (h-1) Q_{\frac{h}{2}-1}(\hat z)Q_{\frac{h}{2}-1}(\hat{\bar z}) \bigg) \bigg|_{h=2n+2} 
\nn
&  = \frac{\pi^3}{8} \sum_{n=0}^\infty  \frac{d }{dn } \bigg( (2n+1) Q_{n}(\hat z)Q_{n}(\hat{\bar z}) \bigg)  
\end{align}
We directly computed this in Eq.~\ref{eq:dknnns}, and showed that is matches the well known result $\bar D_{1,1,1,1}$. This also matches the result we got in section~\ref{sec:lkjhkl} and Fig.~\ref{fig:bubblerelations10}-top by using identity 6.

\item Example 2: $M=2$ and $\D_1=\frac{3}{2}$ and $\D_4=\frac{3}{2}$ :

Once again the (double-trace) poles at $2n+3$ cancel by the zeros of the bubble.
So need only consider the tower of poles at $h=2n+2$, which gives rise to:
\begin{align} 
g_4(z,\bar z ) = \frac{\pi^3}{16}\sum_{n=0}^\infty  \frac{d }{dn } \bigg( (2n+1) Q_{n}(\hat z)Q_{n}(\hat{\bar z}) \bigg)  
\end{align}
Notice that the sum is exactly the same as in Eq.~\ref{eq:sno6nd}, and this fact simply arises from Identity 6 Eq.~\ref{eq:rel4}. This is also shown in the top two diagrams of Fig.~\ref{fig:bubblerelations10}.
\end{itemize}


\section{Scalar 4-point diagrams with $\D_{12}=0$ and $\D_{34}=1$}
\label{sec:5}
In this section we consider the 4-point function with differences of external scaling dimensions tuned to $a\equiv \frac{\D_1-\D_2}{2} =0$ and $b\equiv \frac{\D_3-\D_4}{2} =\frac{1}{2}$, Fig.~\ref{fig:bubblerelations14}-Right. There is significant simplification in this case, which enables us to compute various bubble diagrams explicitly. The main simplification arises from the fact that the $d=2$ conformal block simplifies from a ${}_2 F_1$ to a power law:
\begin{align} 
\label{eq:cnns7}
\mathcal{K}^{\D_i}_{\b }(z,\bar z)   = z^{\frac{\b}{2}} {}_2 F_1 (\frac{\b}{2},\frac{\b+1}{2},\b,z) \times \bar{z}^{\frac{\b}{2}}  {}_2 F_1 (\frac{\b}{2},\frac{\b+1}{2},\b, \bar z)= \sqrt{\frac{u}{v}}   \big(4Z \big)^{\b-1}
\end{align} 
where we defined $Z\equiv \frac{\sqrt{z \bar z}}{(1+\sqrt{1-z})(1+\sqrt{1-\bar z})}$, and the cross-ratios are $u=z\bar z$ and $v=(1-z)(1-\bar z)$. We start from the spectral representation of a 4-point function Eq.~\ref{eq:sdbsmd7}, and notice that it simplifies when plugging $\D_2=\D_1$ and $\D_3=\D_4+1$:
\begin{align} 
 g_{4}(z,\bar z ) = \pi 2^{3-d} \int d\nu \tilde{F}_\n \frac{\Gamma_{\frac{d}{2}+i\n -1}  }{4^{i\n} \Gamma_{i\nu}  } ( \G_{\D_1-\frac{d}{4}+\frac{i \nu}{2}} \G_{\D_1-\frac{d}{4}-\frac{i \nu}{2}} \G_{\D_4+\frac{1}{2}-\frac{d}{4}+\frac{i \nu}{2}} \G_{\D_4+\frac{1}{2}-\frac{d}{4}-\frac{i \nu}{2}})  \mathcal{K}^{\D_i}_{\frac{d}{2}+i\n}
\end{align}
Putting $d=2$, and using Eq.~\ref{eq:cnns7} gives:
\begin{align} 
\label{eq:raquel}
g_{4}(z,\bar z )=2 \pi  \sqrt{\frac{u}{v}}  \int_{-\infty}^\infty d\nu \tilde{F}_\n( \G_{\D_1-\frac{1}{2}+\frac{i \nu}{2}} \G_{\D_1-\frac{1}{2}-\frac{i \nu}{2}} \G_{\D_4 +\frac{i \nu}{2}} \G_{\D_4 -\frac{i \nu}{2}})  Z^{i\n}
\end{align}
Or alternatively:
\begin{align} 
\label{eq:poldhhd}
g_{4}(z,\bar z ) =  \sqrt{\frac{u}{v}}  \int  d\nu  \frac{ 2\pi^3 \tilde{F}_\n   }{\sin \pi (\D_1-\frac{1+i \nu}{2}) \sin \pi (\D_4-\frac{i \nu}{2})}   \frac{\G_{\D_1-\frac{1}{2}+\frac{i \nu}{2}}  }{\G_{-\D_1+\frac{3}{2}+\frac{i \nu}{2}}}  \frac{ \G_{\D_4 +\frac{i \nu}{2}}  }{\G_{1-\D_4 +\frac{i \nu}{2}}} Z^{i\n}  
\end{align}
Now we close the contour to the right, and we have poles from $ \G_{\D_4-\frac{i \nu}{2}}$ at     $i \nu = 2\D_4+2m $ , and poles from $ \G_{\D_1-\frac{1}{2}-\frac{i \nu}{2}}$ at     $i \nu = 2\D_1-1+2m$:
\begin{align} 
\frac{8\pi^3 \sqrt{\frac{u}{v}}}{\cos \pi (\D_1-\D_4 )} \sum_{m=0}^\infty   \frac{\G_{\D_1+\D_4-\frac{1}{2} +m }}{\G_{\D_4-\D_1+\frac{3}{2} +m}} \frac{\G_{2\D_4+m} }{\G_{m+1}} \tilde{F}_{ 2\D_4+2m}   Z^{ 2\D_4+2m} + \Big( \D_1 \leftrightarrow \D_4+\frac{1}{2} \Big)
\end{align}
We will now use this to compute contact diagrams, exchange diagrams, and loop bubble diagrams.

\subsection{Contact diagram}
\label{sec:contdi}
To obtain the contact diagram, we put $\tilde{F}_\n =1$, and so we need to compute the following sum:
\begin{align} 
g_{4} (z,\bar z )= \frac{8\pi^3 \sqrt{\frac{u}{v}}}{\cos \pi (\D_1-\D_4 )} \sum_{m=0}^\infty      \frac{\G_{\D_1+\D_4-\frac{1}{2} +m }}{ \G_{\D_4-\D_1+\frac{3}{2} +m}} \frac{\G_{2\D_4+m} }{\G_{m+1}}  Z^{2\D_4+2m}  + \Big( \D_1 \leftrightarrow \D_4+\frac{1}{2} \Big)
\end{align}
These sums give ${}_2 F_1$'s:
\begin{align}
\label{eq:789s1}
&g_{4}(z,\bar z ) = 8\pi^3 \sqrt{\frac{u}{v}} \frac{ \G_{2\D_4}  \G_{\D_1+\D_4-\frac{1}{2}} Z^{2 \D_4}}{\cos \pi (\D_1-\D_4)} 
\ {}_2 F^{(reg)}_1 (2\D_4, \D_1+\D_4-\frac{1}{2},\D_4-\D_1+\frac{3}{2}, Z^{2})
\nn
&+ \Big( \D_1 \leftrightarrow \D_4+\frac{1}{2} \Big)
\end{align}
where ${}_2 F^{(reg)}_1(a,b,c,x) \equiv \frac{1}{\G_c}{}_2 F_1(a,b,c,x)$ is the regularized hypergeometric function. Eq.~\ref{eq:789s1} matches the result obtained from the Mellin representation Eq.~\ref{eq:kskkd}.
The hypergeometric functions above simplify when $\D_1$ and $\D_4$ are separated by an integer. For example,
for $\D_4=\D_1-1$:
\begin{align}
g_{4}(z,\bar z ) =8\pi^3 \sqrt{\frac{u}{v}} 2^{5 - 4 \D_1}  \G_{4\D_1-4}  Z^{2\D-2 } (1+Z)^{4 - 4 \D_1} 
\end{align}
and for $\D_4=\D_1$:
\begin{align}
\label{eq:789s2}
g_{4}(z,\bar z ) = -8\pi^3\sqrt{\frac{u}{v}} 2^{3 - 4 \D_1}  \G_{4\D_1-2}  Z^{1+2\D_1 } (1+Z)^{ - 4 \D_1}  (1+\frac{1}{Z})^2
\end{align}








\subsection{Exchange diagram}
For the exchange diagram with internal $\D$, we plug $\tilde{F}_\n = \frac{1}{\n^2+(\D-1)^2}$ in Eq.~\ref{eq:raquel}:
\begin{align} 
g_{4}(z,\bar z )= 2\pi  \sqrt{\frac{u}{v}}  \int_{-\infty}^\infty d\n  \frac{ \G_{\D_1-\frac{1}{2}+\frac{i \nu}{2}} \G_{\D_1-\frac{1}{2}-\frac{i \nu}{2}} \G_{\D_4 +\frac{i \nu}{2}} \G_{\D_4 -\frac{i \nu}{2}}  }{\n^2+(\D-1)^2}  Z^{i\n}
\end{align}
There is one pole from the exchange operator at $i\n=\D-1$, and a tower of double-trace poles:
\begin{align} 
g_4(z,\bar z )=  g_{exc}(z,\bar z )+g_{d.t} (z,\bar z )
\end{align} 
 The former gives the following contribution:
\begin{align} 
g_{exc}(z,\bar z )=  4\pi^2 \sqrt{\frac{u}{v}}   \frac{1}{2(\D-1)} ( \G_{\D_1-1+\frac{\D}{2}} \G_{\D_1-\frac{\D}{2}} \G_{\D_4 -\frac{1}{2} +\frac{\D}{2}} \G_{\D_4 +\frac{1}{2}-\frac{\D}{2}})  Z^{\D-1}
\end{align}
The contribution from the double-traces is:
\begin{align} 
\label{eq:slkjd3}
g_{d.t}(z,\bar z ) =\sqrt{\frac{u}{v}}\frac{8\pi^3}{\cos \pi (\D_1-\D_4 )} 
 \sum_{m=0}^\infty    \frac{   \frac{\G_{\D_1+\D_4-\frac{1}{2} +m }}{ \G_{\D_4-\D_1+\frac{3}{2} +m}} \frac{\G_{2\D_4+m} }{\G_{m+1}}  Z^{2\D_4+2m} }{-(2\D_4+2m)^2+(\D-1)^2} + \Big( \D_1 \leftrightarrow \D_4+\frac{1}{2} \Big)
\end{align}
The result of the sum is a ${}_3F_2$:
\begin{align} 
&g_{d.t}(z,\bar z ) =  8\pi^3  \sqrt{\frac{u}{v}} \frac{\G_{2\D_4}\G_{\D_1+\D_4-\frac{1}{2}}Z^{2\D_4}}{4(\D-1)\cos \pi(\D_1-\D_4)}
 \nn
&\times \bigg[  \G_{\D_4+\frac{\D}{2}-\frac{1}{2}}{}_3 F_2^{(reg)} (2\D_4,\D_4+\frac{\D}{2}-\frac{1}{2},\D_1+\D_4-\frac{1}{2}:\D_4+\frac{\D}{2}+\frac{1}{2},\frac{3}{2}+\D_4-\D_1,Z^{2})
\nn
& -\G_{\D_4-\frac{\D}{2}+\frac{1}{2}}{}_3 F_2^{(reg)} (2\D_4,\D_4-\frac{\D}{2}+\frac{1}{2},\D_1+\D_4-\frac{1}{2}:\frac{3}{2}-\frac{\D}{2}+\D_4,\frac{3}{2}+\D_4-\D_1,Z^{2}) \bigg]
\nn
&+ \Big( \D_1 \leftrightarrow \D_4+\frac{1}{2} \Big)
\end{align}
When $\D_4$ and $\D_1$ are integers, one can see that the sum in Eq.~\ref{eq:slkjd3} gives combinations of $\Phi(Z^{2},1,\frac{k \pm \D}{2})$, where $k$ is an integer. Where $\Phi$ is the Lerch transcendent function:
\begin{align}
\label{eq:lerchui}
\Phi(y,s,\a) \equiv \sum_{n=0}^\infty \frac{y^n}{(n+\a)^s}
\end{align}
For example when $\D_1=\D_4=1$:
\begin{align}
&g_{d.t}^{(\D_1=\D_4=1)} (z,\bar z ) =\pi^3 \sqrt{\frac{u}{v}} Z \Big[ \Phi(Z^{2},1,\frac{2- \D}{2})-Z\Phi(Z^{2},1,\frac{3- \D}{2})
\nn
&+\Phi(Z^{2},1,\frac{\D}{2})-Z\Phi(Z^{2},1,\frac{1+\D}{2}) \Big]
\end{align}

\subsection{1-loop diagram}
For the 1-loop bubble diagram we have poles from the bubble and double trace poles:
\begin{align}
g_4(z,\bar z )=g_{B}(z,\bar z )+g_{d.t}(z,\bar z )
\end{align}
\subsubsection*{Poles from the bubble}
Consider the 1-loop bubble with internal scaling dimension $\D$. The bubble poles are at $i \nu = 2m+2\D-1 $. Using Eq.~\ref{eq:raquel}:
\begin{align} 
g_{4}(z,\bar z )=2 \pi  \sqrt{\frac{u}{v}}  \int_{-\infty}^\infty d\nu \tilde{F}_\n( \G_{\D_1-\frac{1}{2}+\frac{i \nu}{2}} \G_{\D_1-\frac{1}{2}-\frac{i \nu}{2}} \G_{\D_4 +\frac{i \nu}{2}} \G_{\D_4 -\frac{i \nu}{2}})  Z^{i\n}
\end{align}
The poles from the bubble Eq.~\ref{eq:poles6} give:
\begin{align} 
&g_B(z,\bar z ) = -\pi \sqrt{\frac{u}{v}}\sum_{m=0}^\infty \frac{ \G_{\D_1+m+\D-1} \G_{\D_1-m-\D } \G_{\D_4+\D+m -\frac{1}{2}} \G_{\D_4-m-\D +\frac{1}{2}} }{ 2\D+2m-1 }   Z^{2m+2\D-1}
\nn
&= -\pi \sqrt{\frac{u}{v}}  \frac{\G_{\frac{1}{2} - \D + \D_4} \G_{-\frac{1}{2} +\D + \Delta_4} \G_{-\D + \D_1} \G_{-1 + \D + \D_1 }}{(2\D-1)} Z^{2\D-1}  \times
\nn
& {}_4 F_3 (1,\D-\frac{1}{2},\D_4+\D-\frac{1}{2},\D_1+\D-1: \frac{1}{2}+\D,\frac{1}{2}+\D-\D_4,1+\D-\D_1 ,Z^{2})
\end{align}
This holds when the bubble pole and double trace poles do not collide. When $\D=1$ we have:
\begin{align} 
&g^{(\D=1)}_B(z,\bar z ) = -\pi \sqrt{\frac{u}{v}} (  \G_{-\frac{1}{2}   + \D_4} \G_{\frac{1}{2}   + \Delta_4} \G_{-1 + \D_1} \G_{ \D_1 } )Z
\nn
&\times {}_4 F_3 (1, \frac{1}{2},\D_4+ \frac{1}{2},\D_1 : \frac{3}{2} ,\frac{3}{2} -\D_4,2-\D_1 ,Z^{2})
\end{align}

\subsubsection*{Double-trace poles}
For the bubble with $\D=1$, we have from Eq.~\ref{eq:b1}:
\begin{align} 
&g_{d.t}(z,\bar z ) = \frac{-\pi^3\sqrt{\frac{u}{v}}}{\cos \pi (\D_1-\D_4 )}  \sum_{m=0}^\infty    \frac{\cot (   \pi (\frac{1}{2}+m+\D_4) )}{ 2m+2\D_4}   \frac{\G_{\D_1+\D_4-\frac{1}{2} +m }}{ \G_{\D_4-\D_1+\frac{3}{2} +m}} \frac{\G_{2\D_4+m} }{\G_{m+1}}  Z^{ 2\D_4+2m}  
 \nn
&+ \Big( \D_1 \leftrightarrow \D_4+\frac{1}{2} \Big)
\end{align}
The result of the sum:
\begin{align}
 &g_{d.t}(z,\bar z ) = \frac{-\pi^3}{2} \sqrt{\frac{u}{v}} \frac{\G_{\D_1+\D_4-\frac{1}{2}} \G_{\D_1-\frac{1}{2}}  \G_{2\D_1-1} Z^{2\D_1-1}  }{\cos \pi (\D_4-\D_1)} \cot (\pi \D_1 )
 \nn
&\times \ {}_3 F_2^{(reg)} (\D_1-\frac{1}{2},\D_1+\D_4-\frac{1}{2} ,2\D_1-1:\D_1+\frac{1}{2},\D_1-\D_4+\frac{1}{2} , Z^{2})
\nn
& + \Big( \D_1 \leftrightarrow \D_4+\frac{1}{2} \Big)
\end{align}

\subsection{Higher loops}
\label{sec:higherloops}
Starting from Eq.~\ref{eq:poldhhd}, the $M$-loop bubble diagram Fig.~\ref{fig:bb22} gives:
\begin{align}
\label{eq:sjdo7} 
&g^{(M)}_{4}(z,\bar z )=  \sqrt{\frac{u}{v}}  \int  d\nu  \frac{ 2\pi^3 (\tilde{B}_\n)^M   }{\sin \pi (\D_1-\frac{1}{2}-\frac{i \nu}{2}) \sin \pi (\D_4-\frac{i \nu}{2})}   \frac{\G_{\D_1-\frac{1}{2}+\frac{i \nu}{2}}  }{\G_{-\D_1+\frac{3}{2}+\frac{i \nu}{2}}}  \frac{ \G_{\D_4 +\frac{i \nu}{2}}  }{\G_{1-\D_4 +\frac{i \nu}{2}}}  Z^{i\n}  
\nn
&=   \sqrt{\frac{u}{v}}  \frac{2\pi^3}{8^M} \int d\nu  \frac{ 1 }{\sin^{M+1} (\pi \frac{i \nu+1}{2})  }  \frac{  \sin^{M-1} ( \frac{ \pi i\n }{2})  }{(i\n)^{M}} \frac{\G_{\D_1-\frac{1}{2}+\frac{i \nu}{2}}  }{\G_{-\D_1+\frac{3}{2}+\frac{i \nu}{2}}}  \frac{ \G_{\D_4 +\frac{i \nu}{2}}  }{\G_{1-\D_4 +\frac{i \nu}{2}}}   Z^{i\n} 
\end{align}
where in the bottom line we put\footnote{One can also compute for $\D=$integer by using Eq.~\ref{skdnen4}.} $\D=1$ in the bubble. In order to simplify the gamma functions in Eq.~\ref{eq:sjdo7}, we put $\D_1$, $\D_4$= integer. The simplest case is $\D_1=\D_4=1$, which gives:
\begin{align}
&g^{(M)}_{4}(z,\bar z )=  \sqrt{\frac{u}{v}}  \frac{2\pi^4}{8^M}\frac{1}{\G_{M+1}}  \sum_{n=0}^\infty \frac{d^{M}}{dh^{M} } \bigg( \bigg[  \frac{ (h-(2n+2))^{M+1} }{\sin^{M+1}  (\pi \frac{h}{2})  }\bigg]  \frac{  \sin^{M-1} ( \frac{ \pi (h-1) }{2})  }{(h-1)^{M-1}}    Z^{h-1} \bigg)_{h\to 2n+2} 
\end{align}
For $M=2$, the sum gives:
\begin{align}
&g^{(2)}_{4}(z,\bar z )= - \sqrt{\frac{u}{v}} \frac{\pi Z}{32}  \Big[  \Phi \big(Z^{2},3,\frac{1}{2}\big)- 2\log(Z) \Phi \big(Z^{2},2,\frac{1}{2}\big)+ 2\log^2(Z) \Phi \big(Z^{2},1,\frac{1}{2} \big) \Big]
\end{align}
where the Lerch transcendent was defined in Eq.~\ref{eq:lerchui}. For $M=3$, it gives:
\begin{align}
&g^{(3)}_{4}(z,\bar z )=  \sqrt{\frac{u}{v}} \frac{Z}{768 }  \Big[  -6 \Phi \big(Z^{2},5,\frac{1}{2}\big)+9\log(Z) \Phi \big(Z^{2},4,\frac{1}{2}\big)
\nn
&+ \big(\pi^2-6 \log^2(Z) \big) \Phi \big(Z^{2},3,\frac{1}{2} \big)+ \big(-\pi^2+2 \log^3(Z) \big) \Phi \big(Z^{2},2,\frac{1}{2} \big) \Big]
\end{align}
More generally, for $\D_1 , \D_2=$integer and $\D=1$, we will have sums of the form:
\begin{align}
&[\log (Z)]^{k_2}  Z \sum_{n=0}^\infty \frac{n^{k}Z^{2n} }{(n+\frac{1}{2})^{k_1}}  = [\log (Z)]^{k_2} Z  (Z^{2}\pa_{Z^{2}})^k \sum_{n=0}^\infty \frac{Z^{2n} }{(n+\frac{1}{2})^{k_1}}
\nn
&=Z    [\log (Z)]^{k_2} (Z^{2}\pa_{Z^{2}})^k \  \Phi(Z^{2},k_1,\frac{1}{2})
\end{align}
where $k$, $k_1$, and $k_2$ are integers, and in the second equality we simply rewrote the $n^k$ factor in the sum by introducing the derivative operator $(Z^{2}\pa_{Z^{2}})^k$. Note that taking derivatives of the Lerch transcendent simply lowers it's order, as can be seen by the following identity:
\begin{align}
(\b+Z^{2}\pa_{Z^{2}}) \Phi(Z^{2},k_1,\b)= \Phi(Z^{2},k_1-1,\b) 
\end{align}
To summarize this section: we considered the 4-point function with external scaling dimensions obeying $\D_1-\D_2=0$ and $\D_3-\D_4=1$. We have computed the general contact and exchange diagrams in position space in terms of hypergeometric functions. We have computed the 1-loop bubble diagram with $\D=1$ in the bubble. We have shown that the $M$-loop bubble diagram for the case of $\D_1$, $\D_4$, $\D$=integer can be computed in terms of Lerch transcendent functions of increasing order.


\section{2-point bulk-to-bulk correlators}
\label{sec:6}
\begin{figure}[t]
\centering
\includegraphics[clip,height=4.3cm]{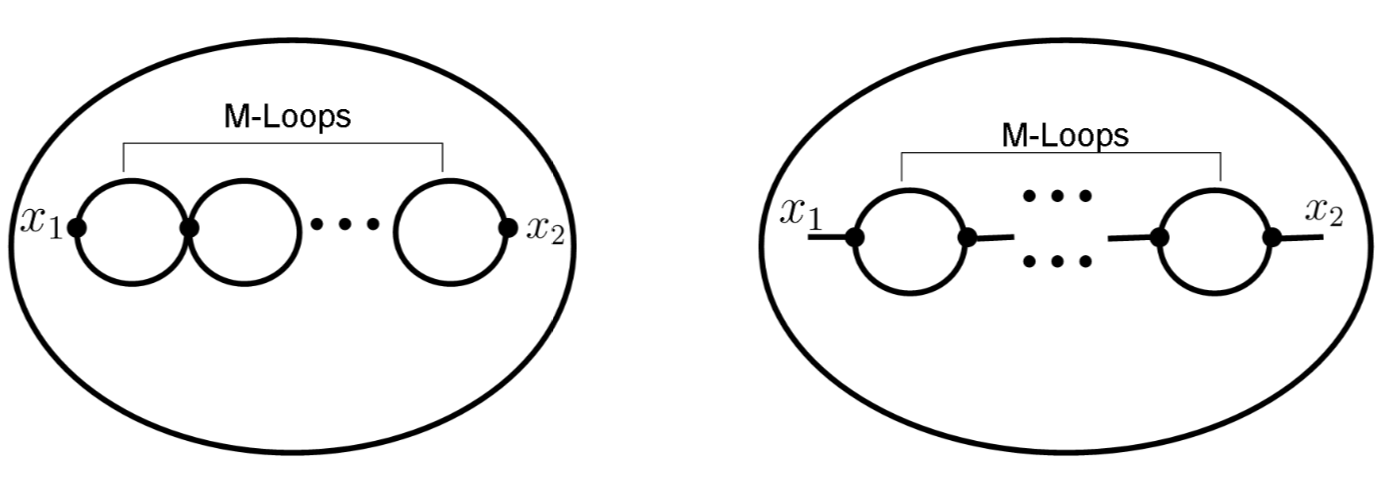}
\caption{The diagrams which we consider in section~\ref{sec:6}. \textbf{Left:} $\phi^4$ theory: The 2-point bulk-to-bulk correlator consisting of a sequence of $M$-bubbles.  \textbf{Right:} $\phi^3$ theory: The 2-point bulk-to-bulk correlator consisting of a sequence of $M$-bubbles.  }
\label{fig:bubblerelations24}
\end{figure}
In this section we compute bulk 2-point bubble diagrams in $AdS_3$, with the external points being $x_1$ and $x_2$ in the bulk, Fig~\ref{fig:bubblerelations24}. The position space  $AdS_{d+1}$ bulk-to-bulk propagator is:
\begin{align}
G_\D(x_1,x_2) =  \frac{\G_\D}{2\pi^{\frac{d}{2}} \G_{\D-\frac{d}{2}+1}} \zeta^{-\D} {}_2F_1 (\D,\D-\frac{d-1}{2},2\D-d+1,-4\zeta^{-1})
\end{align}
where $\zeta= \frac{(z_1-z_2)^2+(\vec{x}_1-\vec{x}_2)^2}{z_1z_2}$ is the square of the chordal distance between the two points $x_1$ and $x_2$. When $d=$even the bulk propagators above simplify. Interestingly, in $AdS_3$ the bulk-to-bulk propagator is a power law in $\D$:
\begin{align}
\label{eq:ksjnf23}
G_\D(x_1,x_2) =  \frac{1}{2\pi} \frac{1}{\sqrt{\zeta(\zeta+4)}}\Big( \frac{2}{\sqrt{\zeta}+\sqrt{\zeta+4}} \Big)^{2\D-2}
=   \frac{1}{2\pi} \frac{1}{\sqrt{\zeta(\zeta+4)}} \eta^{\D-1}
\end{align}
where for notational simplicity we introduced the variable $\eta \equiv \Big( \frac{2}{\sqrt{\zeta}+\sqrt{\zeta+4}} \Big)^2$. The range of the variable $\eta$ is $0 \leq \eta \leq1$.

\subsection{$\phi^4$ bubble diagrams}
\label{eq:nolegs}
Consider a $\phi^4$ scalar field in AdS. The  bubble diagrams Fig.~\ref{fig:bubblerelations24}-Left  gives a contribution to the 2-point function:
\begin{align}
\langle \phi^2(x_1) \phi^2(x_2) \rangle_{bulk}
\end{align}
The spectral representation of $M$ bubbles is just the $M$th power of a single bubble.
\begin{align} 
\label{eq:dkja45}
&g_2^{(M)}(\eta)=  \int_{-\infty}^\infty d\nu ( \tilde{B}_\n )^M  \Omega _{\nu}(x_1,x_2) = \frac{1}{\pi} \int_{-\infty}^\infty d\nu ( \tilde{B}_\n )^M  i \n G_{\frac{d}{2}+\nu}(\eta)
\end{align}
where $G_{\frac{d}{2}+\nu}(\eta)$ is the conformal block. We go to position space by closing the $\n$ contour and picking up the residues of the poles by using Cauchy's residue theorem. For a pole of order $M$ at $y=y_0$:
\begin{align}
Res(F(y))\Big|_{y=y_0} = \frac{1}{\G_M}  \frac{d^{M-1}}{dy^{M-1} } \Big[(y-y_0)^M F(y) \Big] \Big|_{y\to y_0}
\end{align}
Eq.~\ref{eq:dkja45} gives:
\begin{align} 
\label{eq:workwork}
&g_2^{(M)}(\eta)=   \frac{2}{\G_M}\sum^\infty_{n=0}  \frac{d}{d^{M-1} h} \bigg[ (h-(2n+2\D))^M ( \tilde{B}_\n )^M  (h-1) G_{h}(\eta) \bigg]_{h\to (2n+2\D)}
\end{align}
where $h\equiv \frac{d}{2}+i\n$. Now consider the 1-loop bubble $M=1$ in $AdS_3$ ($d=2$). From \ref{eq:ksjnf23}:
\begin{align} 
g_2^{(1)}(\eta)= \frac{1}{2\pi} \sum^\infty_{n=0} G_{2\D+2n}(\eta)
= \frac{1}{4\pi^2} \frac{1}{\sqrt{\zeta(\zeta+4)}}\sum^\infty_{n=0} \eta^{2\D+2n-1}= G^2_{\D}(\eta)
\end{align}
The sum was just a geometric sum, and it's result is the bulk propagator squared $G^2_\D(\eta)$. We just recovered the relation obtained in \cite{Fitzpatrick:2010zm,Fitzpatrick:2011hu}\footnote{One can also do the same computation in $AdS_2$ and match the relation of \cite{Fitzpatrick:2010zm,Fitzpatrick:2011hu}. In this case the sum will be over a LegendreQ function and not a geometric series. }, and which we wrote in Eq.~\ref{eq:new453}.

\subsubsection*{Bubble diagrams for $\D=$integer}
Let us start with the $M$-loop case for $\D=1$. Using Eq.~\ref{eq:b1} gives:
\begin{align} 
\label{eq:sdna3}
\widetilde{g}_2^{(M)}(\eta) =  \frac{1}{\pi8^M \G_M} \sum^\infty_{n=0}  \frac{d}{d^{M-1} h} \bigg[\Big(- (h-(2n+2 )) \cot \frac{\pi h}{2}\Big)^M  \times \frac{\eta^{h-1} }{(h-1)^{M-1}}  \bigg]_{h\to (2n+2)}
\end{align}
Where in Eq.~\ref{eq:sdna3} we absorbed a factor such that $g_2^{(M)}(\eta) = \frac{\widetilde{g}_2^{(M)}(\eta)}{\sqrt{\zeta(\zeta+4)} }$. We can compute these sums by employing the definition of the Lerch transcendent function:
\begin{align}
\label{eq:hurwitz}
\Phi(y,s,\a) \equiv \sum_{n=0}^\infty \frac{y^n}{(n+\a)^s}
\end{align}
For $M= 2, 3$ loops we have from \ref{eq:sdna3}:
\begin{align} 
\widetilde{g}_2^{(2)}(\eta)  = \frac{\eta}{64\pi^3}  \Big[2\Phi(\eta^2,1,\frac{1}{2}) \log(\eta) -\Phi(\eta^2,2,\frac{1}{2})  \Big]
\end{align}
and
\begin{align} 
\widetilde{g}_2^{(3)}(\eta)  =  \frac{\eta}{1024\pi^4 }  
\bigg(3 \Phi(\eta^2, 4,\frac{1}{2}) -4 \log(\eta) \Phi(\eta^2, 3,\frac{1}{2}) + (2\log^2(\eta) -\pi^2) \Phi(\eta^2, 2,\frac{1}{2}) \bigg)
\end{align}
For general $M$, one has sums of the form:
\begin{align}
[\log (\eta)]^{k_2}  \sum_{n=0}^\infty \frac{\eta^{2n+1} }{(2n+1)^{k_1}}  =  \frac{\eta }{2^{k_1}}  [\log (\eta)]^{k_2} \  \Phi(\eta^2,k_1,\frac{1}{2})
\end{align}
where $k_1$and $k_2$ are integers. More generally, if the bubbles have $\D$=integer, we can use the expression for the bubble in Eqs.~\ref{eq:newrt4} and \ref{eq:b1}:
\begin{align}
\tilde{B}^{(\D)}_\n = -\frac{1}{8} \frac{ \cot ( \frac{ \pi h }{2})  }{(h-1)} -\frac{1}{4\pi} \sum^{\D-2}_{j=0} \frac{1}{2j+1} \Big(\frac{1}{h+2j}-\frac{1}{h-2j-2} \Big) \ \ \ \ , \ \ \ \D=int
\end{align}
Plugging this back in Eq.~\ref{eq:workwork}, we see that we will have two types of terms:
\begin{align}
[\log (\eta)]^{k_2}  \sum_{n=0}^\infty \frac{ \eta^{2n+2\D-1} }{(n+\D-\frac{1}{2})^{k_1}}  =  \eta^{2\D-1}   [\log (\eta)]^{k_2} \  \Phi(\eta^2,k_1,\D-\frac{1}{2})
\end{align}
and additional sums of the form:
\begin{align}
[\log (\eta)]^{k_2}  \sum_{n=0}^\infty \frac{\eta^{2n+2\D-1} }{(n+k_3)^{k_1}}   =\eta^{2\D-1}   [\log (\eta)]^{k_2} \  \Phi(\eta^2,k_1,k_3)
\nn
= \eta^{2\D-1}   [\log (\eta)]^{k_2}  \frac{1}{\eta^{2(k_3-1)}}\Big[\frac{1}{\eta^2} Polylog(k_1,\eta^2)-  \sum_{j=0}^{k_3-2}\frac{\eta^{2j} }{ (j+1)^{k_1}} \Big]
\end{align}
where the last  equality holds because $k_3$ is an integer, and the Lerch transcendent function reduces to polylogs. All together, we see that for $M$ bubbles with $\D=$integer we generally have:
\begin{align} 
&\widetilde{g}_2^{(M)}(\eta)  = \eta^{2\D-1} \sum_{k_1=0}^{M-1}\sum_{k_2=0} a^{(M)}_{k_1,k_2} \Phi \Big(\eta^2,M-1+k_1,\D-\frac{1}{2} \Big) \log^{k_2}(\eta)
\nn
& + \eta^{2\D-1} \sum_{k_1=0}^{M-1}\sum_{k_2,k_3} a^{(M)}_{k_1,k_2} \Phi \Big(\eta^2,M-1+k_1,k_3 \Big) \log^{k_2}(\eta)
\end{align}




\subsubsection*{Bubble diagrams for $\D=$half-integer}
Let us first consider the case $\D=\frac{3}{2}$. From Eq.~\ref{eq:workwork} and \ref{eq:b3}
\begin{align} 
\widetilde{g}_2^{(M)}(\eta) =  \frac{8^{-M}}{\pi \G_M } \sum^\infty_{n=0}  \frac{d}{d^{M-1} h} \bigg[\Big((h-(2n+3)) \tan \frac{\pi h}{2}\Big)^M   \frac{(1+\frac{2\cot \frac{\pi h}{2} }{\pi(h-1)})^M \eta^{h-1} }{(h-1)^{M-1}}  \bigg]_{h\to (2n+3)}
\end{align}
For $M= 2, 3$ loops we have:
\begin{align} 
\widetilde{g}_2^{(2)}(\eta)  = \frac{1}{64\pi^3}  \Big(2 \log(\eta)\textrm{Li}_1(\eta^2) -3\textrm{Li}_2(\eta^2)   \Big)
\end{align}
and
\begin{align} 
\widetilde{g}_2^{(3)}(\eta)  =  \frac{1}{512\pi^4 }  
\bigg(\big[ \pi^2-2 \log^2(\eta)\big] \textrm{Li}_2(\eta^2) +10\log(\eta) \textrm{Li}_3(\eta^2) -15  \textrm{Li}_4(\eta^2) \bigg)
\end{align}
where $\textrm{Li}_n(\eta^2)$ is the polylogarithm function of order $n$. More generally, if the bubbles have $\D$=half-integer we see that we will have terms of the form:
\begin{align} 
\widetilde{g}_2^{(M)}(\eta)  = \sum_{k_1,k_2}a_{k_1,k_2} (\log(\eta) )^{k_1}\textrm{Li}_{k_2}(\eta^2) \ \ \ \ \ \ \ , \ \ \ \ \ \ \ \ \D =\textrm{half-integer}
\end{align} 

\subsubsection*{Resummed $\phi^4$ bubbles: conformal $O(N)$ model}
Consider again the large $N$ $O(N)$ model on $AdS_3$. The resummation of bubble diagrams \cite{Carmi:2018qzm} corresponds to the $\l$-exact bulk 2-point function of two Hubbard-Stratonovich fields $\s$:
\begin{align} 
g_2(\eta)\equiv   \langle  \s (x_1)\ \s(x_2) \rangle \ \ \ \ \ \ \  ,\ \ \ \ \text{where} \ \ \ \ \ \ \ \ \ \ \ \s = \frac{\l}{\sqrt{N}} (\phi^i)^2
\end{align}
The spectral representation of this 2-point function is (similarly to Eq.~\ref{eq:solution2}):
\begin{align} 
g_2(\eta)=  \int_{-\infty}^{\infty}  \frac{d\n}{\l^{-1}+2\tilde{B}_\n} \Omega _{\nu}(x_1,x_2) \,,
\end{align}
As we saw in section~\ref{sec:3}, at the bulk conformal point we have $\l \to \infty$ and $\D=1$, and $\tilde{B}_\n = -\frac{\cot \frac{\pi h}{2}}{8(h-1)} $. So we have:
\begin{align} 
&g_2(\eta) = \frac{64}{\pi}\sum_{n=0}^\infty  (n+1)^2 G_{2n+3}(x_1,x_2)
 =  \frac{32}{\pi^2} \frac{1}{\sqrt{\zeta(\zeta+4)}} \sum^\infty_{n=0} (n+1)^2 \eta^{2n+2} 
 \nn
&=  \frac{-32}{\pi^2} \frac{1}{\sqrt{\zeta(\zeta+4)}}   \frac{\eta^2(\eta^2+1)}{(\eta^2-1)^3}=  \frac{32}{\pi^2} \frac{2+\zeta}{\zeta^2(\zeta+4)^2}
\end{align}
The sum above is simply a geometric series, and we see that we get a very simple result, much simpler than the perturbative loop results, which gave Lerch transcendent functions.

\subsection{$\phi^3$ bubble diagrams}
Similarly to previous subsection, we can also compute bulk 2-point bubble diagrams for scalar $\mm{L} = \l \chi \phi^2$ on $AdS_3$, see Fig.~\ref{fig:bubblerelations24}-Right. Consider $\chi$ to have scaling dimension $\d$ and $\phi$ to have scaling dimension $\D$. The $M$-bubble diagram bulk spectral representation contributing to the 2-point function $\langle \chi(x_1) \chi(x_2)\rangle$ is:
\begin{align} 
g^{(M)}_2(\eta)= \int_{-\infty}^\infty d\nu   \frac{( \tilde{B}_\n )^M \Omega _{\nu}(x_1,x_2)}{\Big((-h-\d+2)( h-\d)\Big)^{M+1}} \equiv  g^{(1)}_{\chi}(\eta) + g^{(1)}_{\tilde{B}}(\eta)
\end{align}
where $g^{(1)}_{\chi}(\eta)$ comes from the pole (of order $M+1$) at $h=\d$:
\begin{align} 
\label{eq:sfdnnndd}
&g^{(M)}_{\chi}(\eta) =\frac{1}{\pi \G_{M+1} } \frac{1}{\sqrt{\zeta(\zeta+4)}} \frac{d^M}{dh^M}\Big[  \frac{(h-1) (\tilde{B}_\n)^M\eta^{h-1}}{(-h-\d+2)^{M+1}}\Big] \bigg|_{h =\d } 
\nn
&=\frac{1}{8\pi^2 \G_{M+1}}  \frac{1}{\sqrt{\zeta(\zeta+4)}} \frac{d^M}{dh^M}\Big[  \frac{\psi(\D-\frac{h}{2}) - \psi(\D+\frac{h}{2}-1)}{(-h-\d+2)^{M+1}(h-1)^M}  \eta^{h-1}\Big] \bigg|_{h =\d }
\end{align} 
where we used Eq.~\ref{eq:bubble4}, and $\psi$ are digamma functions\footnote{It may be useful to write the polygamma functions in Eq.~\ref{eq:sfdnnndd} in terms of the Lerch transcendent function $\Phi$: \begin{align}  \psi^{(M)} (x)= (-1)^{M+1}\G_{M+1} \Phi(1,M+1,x)  \end{align} }. We see that $g^{(M)}_{\chi}(\eta)$ is generally given by a combination of polygamma functions (which arise from the derivatives of digamma functions in Eq.~\ref{eq:sfdnnndd}). On the other hand, $g^{(1)}_{\tilde{B}}(\eta)$ comes from the tower of poles of the bubble $\tilde{B}_\n$ at $h=2n+2\D$. Let's compute it for the 1-loop case $M=1$, by using Eq.~\ref{eq:poles6}: 
\begin{align} 
&g^{(1)}_{\tilde{B}}(\eta) =  \frac{-1}{4\pi^2 }\frac{1}{\sqrt{\zeta(\zeta+4)}} \sum^\infty_{n=0}  \frac{\eta^{2\D+2n-1}}{\Big((2n+2\D+\d-2)( 2n+2\D-\d )\Big)^{2}}
 \nn
& =  \frac{-1}{4\pi^2  } \frac{1}{\sqrt{\zeta(\zeta+4)}}
\frac{\eta^{2\D-1}}{16  (\d-1)^3} \bigg[-2 \Phi( \eta^2, 1,\D-\frac{\d}{2} )+2\Phi( \eta^2, 1,\D+\frac{\d}{2}-1 )
\nn
&+(\d-1)  \Phi( \eta^2, 2,\D-\frac{\d}{2} )+(\d-1) \Phi( \eta^2, 2,\D+\frac{\d}{2}-1 ) \bigg]
\end{align} 
where $\Phi$ is the Lerch transcendent function defined in Eq.~\ref{eq:hurwitz}. One can go to higher loops when the bubble scaling dimensions are integer or half-integer, and as in the previous section we will get Lerch transcendent functions.

\subsection{Identity 10}
\begin{figure}[t]
\centering
\includegraphics[clip,height=3.35cm]{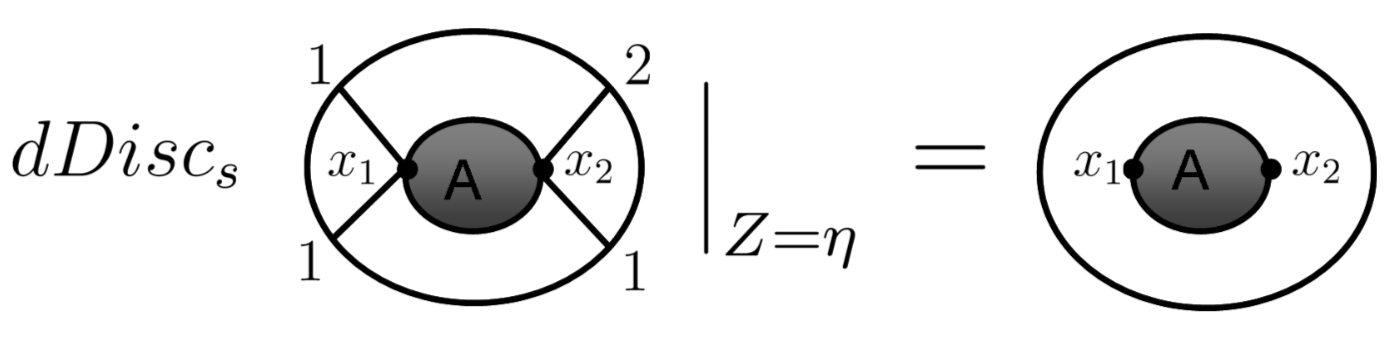}
\caption{Identity 10: relating the bulk 2-point function on the right, to the double-discontinuity of the boundary 4-point function on the left. This identity relates the computations of section~\ref{sec:5} and section~\ref{sec:6}. The blob 2-point function with end points $x_1$ and $x_2$ is general. The Witten diagrams correspond to stripped correlators.}
\label{fig:bubblerelations18}
\end{figure}

Using Eqs.~\ref{eq:nseeeeb}, \ref{eq:ksjnf2}, and \ref {eq:ksjnf23} a general bulk 2-point function has the following spectral representation:
\begin{align} 
\label{eq:guardo1}
&g_2(\eta)=  \int_{-\infty}^\infty d\nu \tilde{F}_\n \Omega _{\nu}(x_1,x_2) = \frac{1}{2\pi} \int_{-\infty}^\infty d\nu \tilde{F}_\n  i \n G_{1+ i \nu}(\eta)  
\nn
&= \frac{1}{(2\pi)^2} \frac{1}{\sqrt{\zeta(\zeta+4)}}\int_{-\infty}^\infty  d\nu  i\n  \tilde{F}_\n    \eta^{i \n} 
\end{align}
Now recall Eq.~\ref{eq:poldhhd}, and take the s-channel double-discontinuity of both sides, which cancels the $\sin$ denominators (see Eq.~\ref{eq:ddisc}):
\begin{align} 
\label{eq:guardo2}
dDisc_s \big[ g^{(\D_i)}_{4}(z,\bar z )\big]= 2\pi^3 \sqrt{\frac{u}{v}}  \int_{-\infty}^\infty d\nu  \tilde{F}_\n  \frac{\G_{\D_1-\frac{1}{2}+\frac{i \nu}{2}}  }{\G_{-\D_1+\frac{3}{2}+\frac{i \nu}{2}}}  \frac{ \G_{\D_4 +\frac{i \nu}{2}}  }{\G_{1-\D_4 +\frac{i \nu}{2}}}  Z^{i\n}  
\end{align}
Now plugging $\D_1=\D_4=1$, Eq.~\ref{eq:guardo2} becomes:
\begin{align} 
\label{eq:guardo3}
  dDisc_s \big[ g_4^{(1,1,2,1)}(z,\bar z ) \big]=   2\pi^3\sqrt{\frac{u}{v}}  \int_{-\infty}^\infty d\nu  \tilde{F}_\n  i\n  Z^{i\n}  
\end{align}
Comparing Eqs.~\ref{eq:guardo1} and \ref{eq:guardo3}, we get:
\begin{align} 
 \textbf{Identity\ 10:} \ \ \ \ \ \ \ \ \ \ \ \  \boxed{   dDisc_s \big[ g_4^{(1,1,2,1)} (z,\bar z ) \big] \Big|_{Z=\eta }= c\ g_2(\eta)  }
\end{align} 
where $c\equiv    8\pi^5  \sqrt{\frac{u}{v}}  \sqrt{\zeta(\zeta+4)}$. See Fig.~\ref{fig:bubblerelations18}.

\underline{Application:} Using Identity 10 we can immediately derive the results of section.~\ref{eq:nolegs} by plugging $\D_i=(1,1,2,1)$  and taking the $dDisc_s$ of the results of section~\ref{sec:higherloops}.

\section{Summary}
\label{sec:7}
Most of the results in this work were derived for $AdS_3$. For example, section~\ref{sec:5} and \ref{sec:6} relied on simplifications of the conformal block and bulk-to-bulk propagator that are special to $d=2$. Also, many results relied on the simplified $d=2$ form of the 1-loop bubble spectral representation, Eqs.~\ref{eq:bubble4}-\ref{eq:b3}. There are some results which hold in any dimension, e.g Identity 1, 2, 3, and 8 of sections~\ref{sec:relswitten1} and \ref{sec:rel88}.

It is possible to extend the results of this work in several directions \cite{Carmi:2021dsn}. Two such directions are to consider different space-time dimensions, i.e $AdS_{D\neq 3}$, and considering spinning particles in AdS. One complication that would arise in these cases is the need to deal with UV divergences. For example, it should be possible to extend the computations of section~\ref{sec:6} to $AdS_5$, by using the fact that the bulk-to bulk propagator simplifies to a power law. Likewise, it should be possible to extend some of the results of section~\ref{sec:5} to $AdS_2$ and $AdS_5$, because the conformal blocks simplify in these dimensions.

Regarding spinning particles in $AdS$, in \cite{Carmi:2018qzm} we computed the fermion bubble spectral representation in $AdS_3$ and $AdS_2$. Thus, it seems that many of the results of this paper can be extended to fermions with $\psi^4$ coupling or with a Yukawa coupling to a scalar field, \cite{Carmi:2021dsn}. In particular, the finite coupling resummed 4-point function of the large-$N$ $O(N)$ model of section~\ref{sec:3} can be extended to the large $N$ Gross-Neveu model on $AdS_3$.

The results of this paper were done for a special class of loop diagrams, i.e the bubble diagrams. It would be interesting to obtain analytic results for loop amplitudes which are not bubble diagrams. A relatively simple extension is to consider the sunset diagram instead of the bubble, as the basic building block. A more difficult extension is that of "generalized bubble functions", of which melonic diagrams in AdS are an example. Another class of diagrams are ladder diagrams in AdS, of which the 1-loop box diagram and triangle diagram are an example of. We leave these computations for future work, see \cite{Carmi:2021dsn}. Computations of amplitudes in flat space-time have led to the discovery of remarkable new structures and dualities. It is our hope, that computing loop amplitudes in $AdS$ would likewise uncover new interesting structures.\\ \\

\textbf{Acknowledgements:}
I especially thank Lorenzo Di Petro and Xinan Zhou for many discussions and comments. I thank Matthijs Hogervorst, Shota Komatsu, Dalimil Mazac, Marco Meineri, David Meltzer, Joao Penedones, Eric Perlmutter, Joao Silva, Allic Sivaramakrishnan, for discussions. D.C is supported by the European Research Council Starting Grant under grant no.\ 758903.


\appendix

\section{More identities for Witten diagrams}
\label{sec:A}
In this section we derive two more identities, adding to the eight identities shown in the main text.
\subsection{Identity 7}
\begin{figure}[t]
\centering
\includegraphics[clip,height=3.7cm]{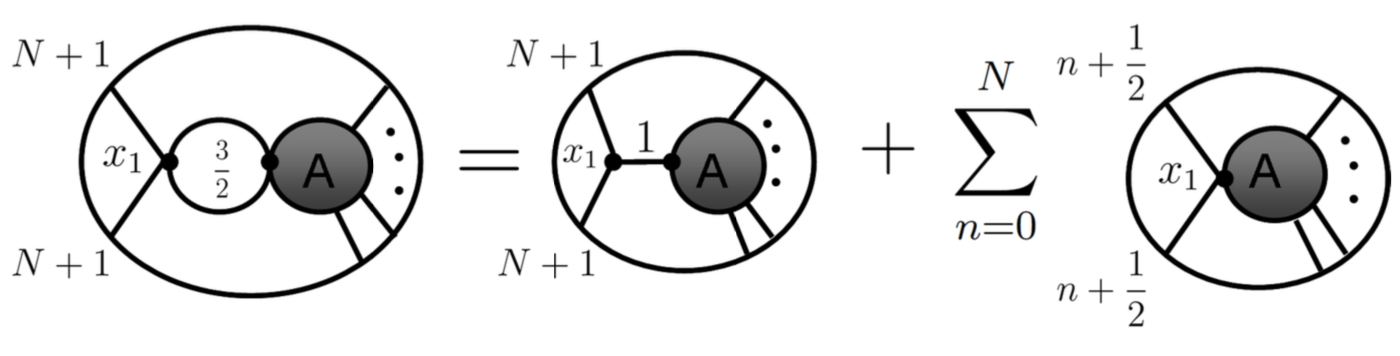}
\caption{Identity $7'$ gives a reduction of half-integer bubbles. The difference between this figure and Fig.~\ref{fig:bubblerelations8} is that here the external legs are integer scaling dimensions and the bubble is half-integer, whereas in Fig.~\ref{fig:bubblerelations8} it is the opposite. The grey blob is general, the numerical coefficients are suppressed, and the diagrams correspond to stripped correlators.}\label{fig:b9}
\end{figure}
Similarly to Identity~6 of subsection~\ref{sec:relation4}, here we derive an identity for the bubble of scaling dimension $\D=\frac{3}{2}$. Using Eqs.~\ref{eq:V}, \ref{eq:B}, and \ref{eq:b3} gives:
\begin{align}
\label{eq:fsgf43}
\textbf{Identity\ 7:} \ \ \ \ \ \ \ \  \boxed{U^{(\D_1,2-\D_1,\D=\frac{3}{2})}_{B} = \frac{1}{16}U^{(\D_1-\frac{1}{2} , \frac{3}{2}-\D_1 )}_{A.V} -\frac{1}{4\pi} U^{(\D_1  , 2-\D_1,\D=1 )}_{V} }
\end{align}
In the particular case when $\D_2=\D_1$, we have:
\begin{align}
U^{(1,1,\D=\frac{3}{2})}_B = \frac{1}{16}U^{(\frac{1}{2} , \frac{1}{2} )}_{A.V} -\frac{1}{4\pi} U^{(1 , 1,\D=1 )}_{V}
\end{align}
Eq.~\ref{eq:fsgf43} can be generalized to the case of arbitrary integer dimension $\frac{\D_1+\D_2}{2}=N+1$, where $N$ is an integer:
\begin{align}
\textbf{Identity\ 7':}  \ \ \ \    \boxed{U^{(\D_1,2N+2-\D_1,\D=\frac{3}{2})}_B = \sum_{n=0}^{N} \hat a_n U^{(\D_1-N+n- \frac{1}{2},N+n+\frac{3}{2}-\D_1)}_{A.V} -\frac{U^{(N+1  , N+1,\D=1 )}_{V} }{4\pi}   }
\end{align}
When $\D_2=\D_1$ this relation is:
\begin{align}
U^{(N+1,N+1,\D=\frac{3}{2})}_B = \sum_{n=0}^{N} a_n U^{(n+\frac{1}{2},n+\frac{1}{2})}_{A.V}   -\frac{1}{4\pi} U^{(N+1  , N+1,\D=1 )}_{V} 
\end{align}
This is shown in Fig.~\ref{fig:b9}.

\subsection{Identity 8}
\label{sec:rel88}
\begin{figure}[t]
\centering
\includegraphics[clip,height=4.1cm]{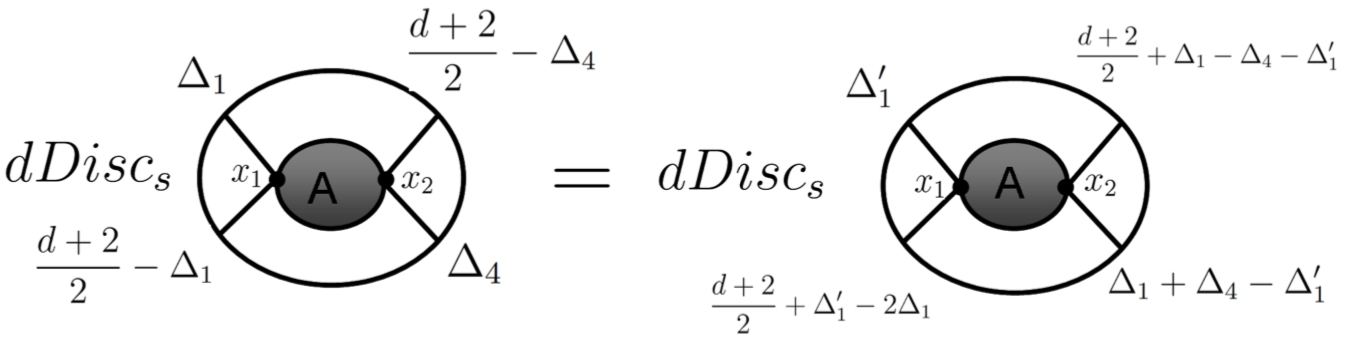}
\caption{Identity 8: a relation for the $dDisc _s$ of the 4-point function with different choices of the external scaling dimensions. $\D_1$, $\D_4$, and $\D_1'$ are arbitrary. The blob 2-point function with end points $x_1$ and $x_2$ is general, and the diagrams correspond to stripped correlators. }\label{fig:bubblerelations19}
\end{figure}

Consider the s-channel double-discontinuity of the 4-point function, written in Eq.~\ref{eq:sdbsmd7q}.
Using this equation we now derive an identity for $dDisc_s \big[ g_4^{(\D_i)} (z,\bar z ) \big]$ for two choices of external scaling dimensions: $\D_i$ and $\D_i'$. In order for the second line of Eq.~\ref{eq:sdbsmd7q} to be invariant, we must impose:
\begin{align} 
\label{eq:algebra1}
\D_1-\D_2= \D_1'-\D_2' \ \ \ \ \ \ \ \text{and} \ \ \  \ \ \ \D_4-\D_3= \D_4'-\D_3' 
\end{align} 
In order for the gamma functions in the square brackets of Eq.~\ref{eq:sdbsmd7q} to be invariant, we impose:
\begin{align} 
\label{eq:algebra2}
\frac{\D_1+\D_2}{2}-\frac{d}{4}= 1-\frac{\D_1+\D_2}{2}+\frac{d}{4} \ \ \ \ \ \ \ \text{and} \ \ \  \ \ \ \frac{\D_3+\D_4}{2}-\frac{d}{4}= 1-\frac{\D_3+\D_4}{2}+\frac{d}{4} 
\end{align} 
And
\begin{align}
\label{eq:algebra3} 
\frac{\D_1'+\D_2'}{2}-\frac{d}{4}= 1-\frac{\D_3'+\D_4'}{2}+\frac{d}{4} 
\end{align} 
These choices in fact make the square brackets of Eq.~\ref{eq:sdbsmd7q} equal to 1.
We solve Eqs.~\ref{eq:algebra1}-\ref{eq:algebra3}, with the result:
\begin{align}
\label{eq:rt3}
\textbf{Identity\ 8:} \ \ \ \ \ \ \ \  \boxed{dDisc_s \big[ g_4^{(\D_i)}(z,\bar z ) \big] = dDisc_s \big[ g_4^{(\D_i')}(z,\bar z ) \big] }
\end{align}
where we defined:
\begin{align} 
&\D_i \equiv \big( \D_1,\ 1+\frac{d}{2}-\D_1,\ 1+\frac{d}{2}-\D_4,\ \D_4 \big) 
\nn
&\D_i' = \big( \D_1',\ \D_1'-2\D_1+1+\frac{d}{2},\ 1+\frac{d}{2}-\D_1'+\D_1-\D_4,-\D_1'+\D_1+\D_4 \big)
\end{align} 
Where $\D_1$, $\D_4$, and $\D_1'$ are completely general. Identity 8 of Eq.~\ref{eq:rt3} is shown in Fig.~\ref{fig:bubblerelations19}. A special case of Identity 8 is when $d=2 $, $\D_1=\D_4=1$, and $\D_1'=\frac{3}{2}$, which gives:
\begin{align}
\label{eq:relation8k}
dDisc_s \big[ g_4^{(1,1,1,1)} (z,\bar z ) \big] = dDisc_s \big[ g_4^{(\frac{3}{2},\frac{3}{2},\frac{1}{2},\frac{1}{2})} (z,\bar z ) \big]
\end{align}
This identity was discussed below Eq.~\ref{eq:dento}.

\section{Computing a sum}
\label{sec:appb}
In this section we compute the sum in Eq.~\ref{eq:lwnjdn5}.

\subsection*{Computing the bubble sum}
Let us start with the sum of Eq.~\ref{eq:lwnjdn5}. 
 \begin{align}
 \label{eq:lkjhgfv}
S_B(z,\bar z)\equiv \sum_{n=0}^\infty (2n+2)^2Q_{n+\frac{1}{2}}(\hat z) Q_{n+\frac{1}{2}}(\hat{\bar z}) 
 \end{align}
 We will first compute a different sum:
 \begin{align}
 \label{eq:6hdls}
\widehat{S}_B(z,\bar z) \equiv \sum_{n=0}^\infty  Q_{n+\frac{1}{2}}(\hat z) Q_{n+\frac{1}{2}}(\hat{\bar z})	
 \end{align}
One can get the sum $S_B$ by applying the Casimir operator on the sum $\widehat{S}_B$:
 \begin{align}
\label{eq:lkj2}
S_B(z,\bar z) = \Big(1+2D_{z,\bar z} \Big) \widehat S_B  (z,\bar z)
 \end{align}
where $D_{z,\bar z}$ is given in Eq.~\ref{eq:diff56}. Now we show how to compute the sum $\widehat S_B$ analytically. We will use two results, given on pages 403 and 230 respectively of\footnote{Note that when $c=b$, one gets $d^2=  \frac{2b}{a-b}$ and $\hat z=\hat{\bar z} = \frac{\sqrt{a+b}}{\sqrt{2b}}$. In this case, Eq.~\ref{eq:ell1} matches with Eq.~3 of page 232 of \cite{prudnikov1}:
 \begin{align}
\int_a^\infty \frac{dx Q_{\n}(\frac{x}{b})}{\sqrt{(x-a)(x-b)} } = Q^2_\n \Big( \sqrt{\frac{a+b}{2b} } \Big) 
 \end{align}.} \cite{prudnikov1}:
 \begin{align}
 \label{eq:ell3}
\sum_{n=0}^\infty Q_{n+\frac{1}{2}} (\hat z)= \frac{\pi}{2^{\frac{3}{2}}} \frac{1}{\sqrt{\hat z-1}}- \frac{1}{2}Q_{-\frac{1}{2}}(\hat z) 
 \end{align}
and
 \begin{align}
 \label{eq:ell1}
\int_a^\infty \frac{dx Q_{\n}(\frac{x}{c})}{\sqrt{(x-a)(x-b)} } = Q_\n(\hat z)Q_\n(\hat{\bar z}) 
 \end{align}
where $a, b$, and $c$ are related to $\hat{ z}$ and $\hat{\bar z}$ via
 \begin{align}
\hat z = \frac{\sqrt{1+d^2}}{d} \ \ \ \ \  \ \ \ \ , \ \ \ \ \ \ \ \ \ \ \  \hat{\bar z}= \frac{\sqrt{4c^2+(a-b)^2d^2}}{2c}
 \end{align}
where $d$ is another parameter which obeys:
\begin{align}
\label{eq:ell2}
(1+d^2)(4c^2+d^2(a-b)^2) =(a+b)^2d^2
 \end{align}
Solving Eq.~\ref{eq:ell2} for $d^2$ gives:
 \begin{align}
d_{1,2}^2 =2 \frac{ab-c^2\pm \sqrt{c^4-c^2(a^2+b^2)+a^2b^2}}{(a-b)^2}
 \end{align}
We now compute our desired sum in $\widehat S_B$:
\begin{align}
\label{eq:bncks4}
&\widehat S_B(z,\bar z) = \sum_{n=0}^\infty Q_{n+\frac{1}{2}}(\hat z) Q_{n+\frac{1}{2}}(\hat{\bar z}) = \sum_{n=0}^\infty \int_a^\infty \frac{dx Q_{n+\frac{1}{2}}(\frac{x}{c})}{\sqrt{(x-a)(x-b)} } 
\nn 
&=  \int_a^\infty  \frac{dx}{\sqrt{(x-a)(x-b)} }   \sum_{n=0}^\infty  Q_{n+\frac{1}{2}}(\frac{x}{c})
\nn
&=  \frac{\pi}{2^{\frac{3}{2}}} \int_a^\infty  \frac{dx}{\sqrt{(x-a)(x-b)} } \frac{1}{\sqrt{\frac{x}{c}-1}} - \frac{1}{2} \int_a^\infty dx \frac{Q_{-\frac{1}{2}}(\frac{x}{c})}{\sqrt{(x-a)(x-b)} }  
\nn
&=  \frac{\pi}{2^{\frac{3}{2}}} \int_a^\infty  \frac{dx}{\sqrt{(x-a)(x-b)} } \frac{1}{\sqrt{\frac{x}{c}-1}} - \frac{1}{2}  Q_{-\frac{1}{2}}(\hat z)Q_{-\frac{1}{2}}(\hat{\bar z})
 \end{align}
 In the first line we used Eq.~\ref{eq:ell1}. In the second line we exchanged the summation and integration. To get the third line we used Eq.~\ref{eq:ell3}. To get the fourth line we used Eq.~\ref{eq:ell1} again.

We are only left with computing the integral in the last line of Eq.~\ref{eq:bncks4}, and it can be solved by mathematica:
 \begin{align}
\sqrt{c} \int_a^\infty  \frac{dx}{\sqrt{(x-a)(x-b)(x-c)} } =  \frac{2\sqrt{c}}{a-c} K(\sqrt{\frac{b-c}{a-c}}) 
 \end{align}
where $K$ is the elliptic integral of the 1st kind:
 \begin{align}
 \label{eq:elliptic78}
 K(x)= \frac{\pi}{2} {}_2 F_1 (\frac{1}{2},\frac{1}{2},1,x^2)= \frac{\pi}{2}+\frac{\pi}{8}x^2 + \ldots
 \end{align}
Thus, Eq.~\ref{eq:bncks4}	gives:
 \begin{align}
 \label{eq:ncsnn6}
\widehat S_B(z,\bar z)= \frac{\pi}{2^{\frac{3}{2}}} \frac{2\sqrt{c}}{\sqrt{a-c}} K(\sqrt{\frac{b-c}{a-c}})  - \frac{1}{2}  Q_{-\frac{1}{2}}(\hat z)Q_{-\frac{1}{2}}(\hat{\bar z})  
 \end{align}
The reader can easily check that this result is correct by plugging numerical values on both sides.  Using Eq.~\ref{eq:lkj2} we finally have:
 \begin{align}
 \label{eq:sumb}
\boxed{
S_B(z,\bar z)  = \Big(1+2D_{z,\bar z} \Big)  \Big[ \frac{\pi}{2^{\frac{3}{2}}} \frac{2\sqrt{c}}{\sqrt{a-c}} K(\sqrt{\frac{b-c}{a-c}})  - \frac{1}{2}  Q_{-\frac{1}{2}}(\hat z)Q_{-\frac{1}{2}}(\hat{\bar z}) \Big]  }
 \end{align}
We can alternatively write the elliptic integral above as $K(\sqrt{y}) = Q_{-\frac{1}{2}}(2y-1)$, and then both terms are Legendre functions:
 \begin{align}
S_B(z,\bar z)  = \Big(1+2D_{z,\bar z} \Big)  \Big[ \frac{\pi}{2^{\frac{3}{2}}} \frac{2\sqrt{c}}{\sqrt{a-c}} Q_{-\frac{1}{2}}\Big(2\frac{b-c}{a-c}-1\Big)  - \frac{1}{2}  Q_{-\frac{1}{2}}(\hat z)Q_{-\frac{1}{2}}(\hat{\bar z}) \Big] 
 \end{align}
 
\subsection*{The double-trace sum}
Consider the sum arising from the double-trace poles,  Eq.~\ref{eq:bvcxz}:
\begin{align}
S_{d.t}(z,\bar z) \equiv \sum_{n=0}^\infty (2n+1)^2 Q_n(\hat{  z}) Q_n (\hat{\bar z}) 
\end{align}
Now consider the simpler sum $\widehat{S}_{d.t}$:
 \begin{align}
 \label{eq:sdt1}
\widehat{S}_{d.t}(z,\bar z) =\sum_{n=0}^\infty Q_n(\hat{  z}) Q_n (\hat{\bar z})
 \end{align}
The two sums are related via:
 \begin{align}
S_{d.t}(z,\bar z) = \Big(1+2D_{z,\bar z} \Big) \widehat{S}_{d.t}  (z,\bar z)
 \end{align}
Using similar methods to those in this section, we attempted to compute this sum. However, in the final stage of the computation we encounter a complicated integral, and we did not manage to perform this integral analytically.

\section{Derivation of the 4-point function spectral representation }
\label{sec:appbb}
Consider the spectral representation of the 4-point function with scaling dimensions $\D_1,\D_2,\D_3, \D_4$. As we wrote in Eq.~\ref{eq:4pointspec}, the Witten diagram is given by:
\begin{align}
\label{eq:poland1}
\widehat{g}_{4} =\int d x_1 dx_2 F(x_1,x_2) K_{\Delta_1}(P_1,x_1)K_{\Delta_2}(P_2,x_1) K_{\Delta_3}(P_3,x_2)K_{\Delta_4}(P_4,x_2)
 \end{align}
where the integrals are over $AdS_{d+1}$ space. The spectral representation of the bulk 2-point function, Eq.~\ref{eq:nseeeeb}, is:
 \begin{align}
 \label{eq:poland2}
F(x_1,x_2)  =\int_{-\infty}^\infty d \n \tilde{F}_\n  \Omega_{\nu}(x_1,x_2)
 \end{align}
We write the split representation of the $AdS$ harmonic function $\Omega_{\nu}$:
\begin{align}
\label{eq:poland3}
\Omega_{\nu}(x_1,x_2)=\frac{\nu^2\sqrt{\mathcal{C}_{\frac{d}{2}+i\nu}\mathcal{C}_{\frac{d}{2}-i\nu}}}{\pi}\int dP_0\, K_{\frac{d}{2}+i\nu}(P_0,x_1)K_{\frac{d}{2}-i\nu}(P_0,x_2)\,.
\end{align}
where the $P_0$ integral is over the boundary, and where
\begin{align}\label{eq:CDeltaDef}
\mathcal{C}_{\Delta}\equiv \frac{\Gamma_\Delta}{2\pi ^{d/2}\Gamma_{\Delta-\frac{d}{2}+1}}\,.
\end{align}
Plugging Eqs.~\ref{eq:poland2} and \ref{eq:poland3} inside Eq.~\ref{eq:poland1} gives:
\begin{align}
\begin{aligned}
&\widehat{g}_{4} =\int d \n \tilde{F}_\n \frac{\nu^2\sqrt{\mathcal{C}_{\frac{d}{2}+i\nu}\mathcal{C}_{\frac{d}{2}-i\nu}}}{\pi}\int dP_0\int dx_1 \, K_{\Delta_1}(P_1,x_1)K_{\Delta_2}(P_2,x_1)K_{\frac{d}{2}+i\nu}(P_0,x_1)\\
&\times \int dx_2  K_{\Delta_3}(P_3,x_2)K_{\Delta_4}(P_4,x_2)K_{\frac{d}{2}-i\nu}(P_0,x_2)
\end{aligned}
\end{align}
We can now perform the bulk integrals over $x_1$ and $x_2$ since they are just three-point functions:
\begin{align}
\label{eq:3poointresult}
\int dx K_{a} (P_1, x) K_{b}(P_2,x)K_{c} (P_3,x)=\frac{B_{a,b,c}}{(P_{12})^{\frac{a+b-c}{2}}(P_{23})^{\frac{b+c-a}{2}}(P_{31})^{\frac{c+a-b}{2}}}\,,
\end{align} 
with
\begin{align}
\label{eq:Bdef}
B_{a,b,c}  \equiv \frac{\pi^{\frac{d}{2}}}{2}\Gamma_{-\frac{d}{2}+\frac{a+b+c}{2}}\sqrt{\mathcal{C}_a\mathcal{C}_b\mathcal{C}_c}\frac{\Gamma_{\frac{a+b-c}{2}}\Gamma_{\frac{a-b+c}{2}}\Gamma_{\frac{-a+b+c}{2}}}{\Gamma_{a}\Gamma_{b}\Gamma_{c}}\,.
\end{align}
We are left with the integral over $P_0$, which is just the shadow representation of the conformal partial wave, \cite{SimmonsDuffin:2012uy,Rosenhaus:2018zqn}:
\begin{align}
\label{eq:jsbbddd}
&\int dP_0 \frac{   \frac{1}{(P_{12})^{\frac{\D_1+\D_2-\frac{d}{2}-i\nu}{2}}(P_{34})^{\frac{\D_3+\D_4-\frac{d}{2}+i\nu}{2}}}  }{(P_{01})^{\frac{\D_1-\D_2+\frac{d}{2}+i\nu}{2}}(P_{20})^{\frac{\D_2-\D_1+\frac{d}{2}+i\nu}{2}}(P_{40})^{\frac{\D_4-\D_3+\frac{d}{2}-i\nu}{2}}(P_{03})^{\frac{\D_3-\D_4+\frac{d}{2}-i\nu}{2}}}
\nn
&= \frac{\Big( \frac{P^2_{14}}{P^2_{24}}\Big)^{\frac{\D_2-\D_1}{2}}\Big( \frac{P^2_{14}}{P^2_{13}}\Big)^{\frac{\D_3-\D_4}{2}}}{(P_{12})^{\frac{\Delta_1+\D_2}{2}}(P_{34})^{\frac{\Delta_3+\D_4}{2}}}\left(k^{\D_3, \D_4}_{\frac{d}{2}-i\nu}\mathcal{K}^{\D_i}_{\frac{d}{2}+i\nu}(z,\bar{z})+k^{\D_1,\D_2}_{\frac{d}{2}+i\nu}\mathcal{K}^{\D_i}_{\frac{d}{2}-i\nu}(z,\bar{z})\right)\,,
\end{align}
where $\mathcal{K}_{\Delta}(z,\bar z)$ is the $d$-dimensional scalar conformal block with exchanges scaling dimension $\Delta$, and $k_{a}$ is given in Eq 3.4 of \cite{Rosenhaus:2018zqn} 
\begin{align}
\label{eq:kdef1}
k^{\D_3, \D_4}_{\frac{d}{2}-i\n}=\frac{\pi^{\frac{d}{2}}\Gamma_{-i\n}\Gamma_{\frac{d}{4}+\frac{i\n}{2}-\frac{\D_{34}}{2}}\Gamma_{\frac{d}{4}+\frac{i\n}{2}+\frac{\D_{34}}{2}}}{\Gamma_{\frac{d}{2}+i\n}\Gamma_{\frac{d}{4}-\frac{i\n}{2}-\frac{\D_{34}}{2}} \Gamma_{\frac{d}{4}-\frac{i\n}{2}+\frac{\D_{34}}{2}} }\,.
\end{align}
Therefore we finally obtain the result of Eq.~\ref{eq:sdbsmd7}:
\begin{align} 
\label{eq:AAA1}
&\widehat{g}_4= A_{\D_i} \int  d\n \tilde{F}_\n 
\times \frac{  \G_{\frac{\Delta_1+\D_2}{2} +\frac{i \nu-\frac{d}{2}}{2}} \G_{\frac{\Delta_1+\D_2}{2} -\frac{i \nu+\frac{d}{2}}{2}} \G_{\frac{\Delta_3+\D_4}{2} +\frac{i \nu-\frac{d}{2}}{2}} \G_{\frac{\Delta_3+\D_4}{2} -\frac{i \nu+\frac{d}{2}}{2}}  }{\Gamma_{i\nu} \Gamma_{\frac{d}{2}+i\n} }
\nn
& \Big( \G_{\frac{\D_2-\Delta_1}{2}+\frac{i \nu+\frac{d}{2}}{2}} \G_{\frac{\Delta_1-\D_2}{2}+\frac{i \nu+\frac{d}{2}}{2}} \G_{\frac{\Delta_3-\D_4}{2}+\frac{i \nu+\frac{d}{2}}{2}} \G_{\frac{\D_4-\Delta_3 }{2}+\frac{i \nu+\frac{d}{2}}{2}} \Big) \mathcal{K}^{\D_i}_{\frac{d}{2}+i\nu} (z,\bar z)
\end{align}
where we defined:
\begin{align} 
\label{eq:prefactor}
A_{\D_i} \equiv c \frac{\frac{\Big( \frac{P^2_{14}}{P^2_{24}}\Big)^{\frac{\D_2-\D_1}{2}}\Big( \frac{P^2_{14}}{P^2_{13}}\Big)^{\frac{\D_3-\D_4}{2}}}{(P_{12})^{\frac{\Delta_1+\D_2}{2}}(P_{34})^{\frac{\Delta_3+\D_4}{2}}}}{\sqrt{\Gamma_{\Delta_1}\Gamma_{\Delta_2}\Gamma_{\Delta_3}\Gamma_{\Delta_4}\Gamma_{\Delta_1-\frac{d}{2}+1}\Gamma_{\Delta_2-\frac{d}{2}+1}\Gamma_{\Delta_3-\frac{d}{2}+1}\Gamma_{\Delta_4-\frac{d}{2}+1}} }
\end{align}
where $c$ is a numerical factor which will be unimportant for us. In this paper we will consider the stripped correlator $g_4(z, \bar z)$:
\begin{align}
g_4 (z, \bar z) \equiv \frac{1}{A_{\D_i}}\widehat{g}_4
\end{align}
$g_4$ is a function of the cross-ratios $(z, \bar z)$, and is written in the main text in Eq.~\ref{eq:sdbsmd7}.


\section{More general diagrams}
\label{sec:proofblob1}
We will show in this section that the (stripped) $N$ point function with a pair of bulk-to-boundary propagators originating from the same vertex $x_1$, depends on $\D_1+\D_2$ only through the factor $U^{(\D_1,\D_2)}_{A.V}$, defined in Eq.~\ref{eq:AV}. We will use this result in Eq.~\ref{eq:importantn}. Consider the diagram in Fig.~\ref{fig:bubblerelations20}, which is given by:
\begin{align} 
\label{eq:fmonnnf}
\widehat{g}_N(P_1,P_2,\{ P_i\}) = \int \prod^{i=N}_{i=2} d^d x_i (Blob )\times \int d^d x_1  G_{\D}(x_1,x_2) K_{\D_1}(P_1,x_1) K_{\D_2}(P_2,x_1)
\end{align} 
We focus on the dependence of this diagram on $\D_1$ and $\D_2$. The factor $(Blob)$ does not depend on $\D_{1}$ and $\D_2$. Let us look at the second factor:
\begin{align} 
\label{eq:nbvdsa}
&\int d^d x_1  G_{\D}(x_1,x_j) K_{\D_1}(P_1,x_1) K_{\D_2}(P_2,x_1)  
\nn
&=\int_{-\infty}^\infty \frac{ d\nu }{\n^2+(\D-\frac{d}{2})^2} \int d^d x_1 \Omega _{\nu}(x_1,x_2)   K_{\D_1}(P_1,x_1) K_{\D_2}(P_2,x_1)
\nn
&=\int_{-\infty}^\infty \frac{ d\nu \frac{\nu^2\sqrt{\mathcal{C}_{\frac{d}{2}+i\nu}\mathcal{C}_{\frac{d}{2}-i\nu}}}{\pi} }{\n^2+(\D-\frac{d}{2})^2}\int dP_0  K_{\frac{d}{2}-i\nu}(P_0,x_2) 
\nn
&\times \int d^d x_1  K_{\frac{d}{2}+i\nu}(P_0,x_1)   K_{\D_1}(P_1,x_1) K_{\D_2}(P_2,x_1)
\end{align} 
Let us look at the 3-point function on the last line. From Eq.~\ref{eq:3poointresult}:
\begin{align}
&\int d^d x_1  K_{\frac{d}{2}+i\nu}(P_0,x_1)   K_{\D_1}(P_1,x_1) K_{\D_2}(P_2,x_1)= A_0\times U^{(\D_1,\D_2)}_{A.V} (\n)\times U_2(P_0,\n)
\end{align}
where we separated the result into 3 functions:
\begin{align}
\label{eq:a0nbvf}
A_0 \equiv  \frac{\pi^{\frac{d}{2}}}{2} \frac{\sqrt{\mathcal{C}_{\D_1}\mathcal{C}_{\D_2}}}{\G_{\D_1}\Gamma_{\D_2}(P_{12})^{\frac{\D_1+\D_2}{2}}}\ \ \ \ \ \ \ , \ \ \ \ \ \ \ \ \ U^{(\D_1,\D_2)}_{A.V}(\n) \equiv \Gamma_{\frac{\D_1+\D_2}{2}-\frac{d}{4}+\frac{i\nu}{2}} \Gamma_{\frac{\D_1+\D_2}{2}-\frac{d}{4}-\frac{i\nu}{2}}
\end{align}
and
\begin{align}
\label{eq:u2p}
U_2(P_0,\n) \equiv  \frac{\frac{\sqrt{\mathcal{C}_{\frac{d}{2}+i\nu}}}{\Gamma_{\frac{d}{2}+i\nu}}  \Gamma_{\frac{\D_1-\D_2+\frac{d}{2}+i\nu}{2}}\Gamma_{\frac{-\D_1+\D_2+\frac{d}{2}+i\nu}{2}}}{(P_{12})^{\frac{-(\frac{d}{2}+i\nu)}{2}}(P_{20})^{\frac{\D_2-\D_1+\frac{d}{2}+i\nu}{2}}(P_{01})^{\frac{\frac{d}{2}+i\nu+\D_1-\D_2}{2}}} 
\end{align}
Plugging this back in Eq.~\ref{eq:nbvdsa} and \ref{eq:fmonnnf}, gives:
\begin{align} 
\label{eq:blobie1}
&\widehat{g}_N(P_1,P_2,\{ P_i\}) = \frac{A_0}{\pi}  \int \prod_{i=2} d^d x_i (Blob ) \int_{-\infty}^\infty d\nu  \frac{ \nu^2\sqrt{\mathcal{C}_{\frac{d}{2}+i\nu}\mathcal{C}_{\frac{d}{2}-i\nu}}  }{\n^2+(\D-\frac{d}{2})^2}
\nn
&\times \Gamma_{\frac{\D_1+\D_2}{2}-\frac{d}{4}+\frac{i\nu}{2}} \Gamma_{\frac{\D_1+\D_2}{2}-\frac{d}{4}-\frac{i\nu}{2}} \bigg[  \int dP_0  K_{\frac{d}{2}-i\nu}(P_0,x_2)\times U_2(P_0,\n) \bigg]
\end{align}
$A_0$ does not depend on $\n$ or on $P_0$, so it is a constant for our purpose, and therefore we can just strip off this factor. The $U_2$ factor depends on the difference $\D_1-\D_2$, but not on $\D_1+\D_2$. So the only factor which depends non-trivially on both $\n$ and $\D_1+\D_2$ are the gamma functions in the bottom line of Eq.~\ref{eq:blobie1}. This simplicity gives rise to many of the identities derived in this paper. We used Eq.~\ref{eq:blobie1} in Eq.~\ref{eq:importantn}.

\subsection{Example: the 5-point tree-level exchange diagram}
\begin{figure}[t]
\centering
\includegraphics[clip,height=4.4cm]{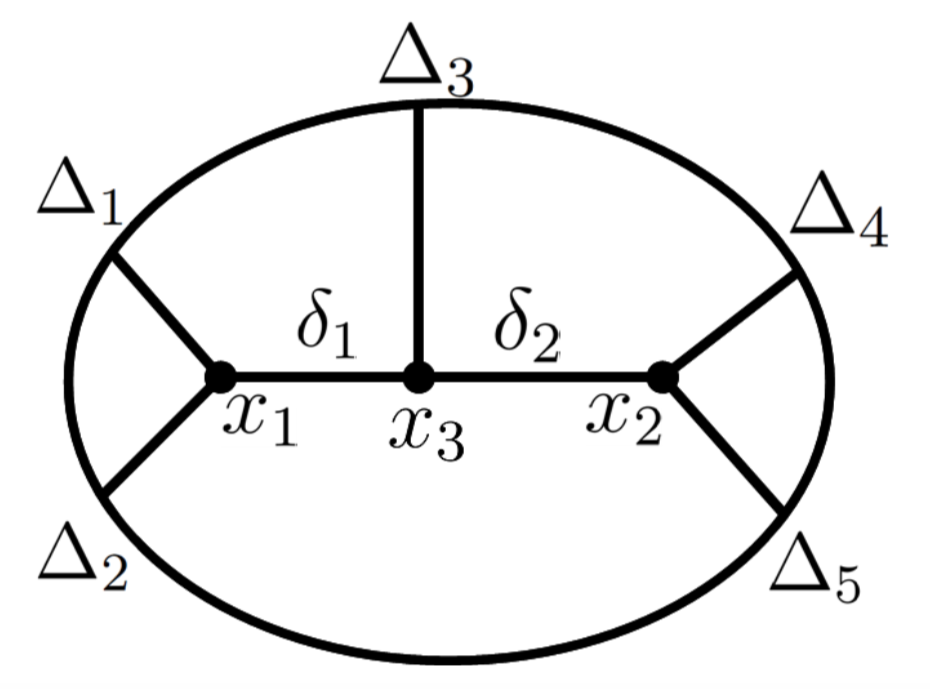}
\caption{Left: The 5-point tree-level exchange diagram.  }\label{fig:b21}
\end{figure}
Let us show an explicit example of the observations above, for the 5-point tree-level exchange diagram, see also \cite{Goncalves:2019znr,Jepsen:2019svc}. We start with the spectral representation of the 5-point tree level Witten diagram, Fig.~\ref{fig:b21}. This Witten diagram is given by:
\begin{align} 
&\widehat{g}_5(P_1,P_2,P_3,P_4,P_5) =\int d^d x_1 d^d x_2 d^d x_3 G_{\d_1}(x_1,x_3)G_{\d_2}(x_2,x_3)
\nn
&\times K_{\D_1}(P_1,x_1)K_{\D_2}(P_2,x_1)K_{\D_3}(P_3,x_4)K_{\D_4}(P_4,x_2) K_{\D_5}(P_5,x_2)
\end{align} 
where there are 2 exchanges of operators with scaling dimension $\d_1$ and $\d_2$. We will be concise in this section, because the computation is very similar to the 4-point function done in the previous section. We write the spectral representation of the exchanged bulk-to-bulk propagators as in Eq.~\ref{eq:bulkspect4}:
\begin{align} 
G_{\d_1}(x_1,x_3) = \int_{-\infty}^\infty \frac{ d\nu_1 \O _{\nu_1}(x_1,x_3)}{\n_1^2+(\d_1-\frac{d}{2})^2}  \ \ \  ,  \ \ \ G_{\d_2}(x_2,x_3) = \int_{-\infty}^\infty \frac{ d\nu_2 \O _{\nu_2}(x_2,x_3)}{\n_2^2+(\d_2-\frac{d}{2})^2} 
\end{align}
we can now write the split representations of $\O_{\n_1}$ and $\O_{\n_2}$, as in Eq.~\ref{eq:poland3}, and perform the conformal 3-point integrals, and get:
\begin{align} 
\label{eq:polu8}
&\widehat{g}_5 =  \frac{1}{\pi^2}\int_{-\infty}^\infty d \n_1d \n_2  \nu_1^2 \nu_2^2\sqrt{\mathcal{C}_{\frac{d}{2}+i\nu_1} \mathcal{C}_{\frac{d}{2}-i\nu_1} \mathcal{C}_{\frac{d}{2}+i\nu_2} \mathcal{C}_{\frac{d}{2}-i\nu_2}} 
\nn
&\frac{B_{\D_1,\D_2,\frac{d}{2}+i\n_1} B_{\D_4,\D_5,\frac{d}{2}+i\n_2}B_{\D_3,\frac{d}{2}-i\n_1,\frac{d}{2}-i\n_2}}{(\n_1^2+(\d_1-\frac{d}{2})^2)(\n_2^2+(\d_2-\frac{d}{2})^2)} \times  \Psi_{\frac{d}{2}+i\n_1, \frac{d}{2}+i\n_2}^{\D_i}(x_i)
\end{align} 
where $\mm{C}_\D$ and $B_{a,b,c}$ are written in Eqs.~\ref{eq:CDeltaDef} and \ref{eq:Bdef}, and $\Psi_{\D_a, \D_b}^{\D_i}$ is the 5-point conformal partial wave:
\begin{align} 
\Psi_{\D_a, \D_b}^{\D_i}(x_i) = \int d^dy_a d^dy_b \langle O_1 O_2 O_a \rangle \langle \tilde{O}_a O_3 O_b \rangle \langle \tilde{O}_b O_4 O_5 \rangle
\end{align} 
Let us write a few formulas from \cite{Rosenhaus:2018zqn,Gross:2017aos}, which computed these 5-point conformal blocks,
\begin{align} 
\Psi^{\D_i}_{\D_a,\D_b}(x_i) = 
\mm{P}^{\D_i}_{\D_a,\D_b} \mm{L}^{\D_i}_{\D_a,\D_b} \mm{M}^{\D_i}_{\D_a,\D_b}(w_i)
\end{align} 
where $w_i$ are cross-ratios, and
\begin{align} 
\label{eq:M564}
\mm{M}^{\D_i}_{\D_a,\D_b}(u_1,v_1,u_2,v_2,w) =  k^{\D_3,\D_b}_{\tilde{\D}_a} k^{\D_4\D_5}_{\tilde{\D}_b} g^{\D_i}_{\D_a,\D_b}(w_i)
\end{align} 
and $k$ is written in Eq.~\ref{eq:kdef1}, and
\begin{align} 
 \mm{L}^{\D_i}_{\D_a,\D_b} (x_i)= \frac{ \Big( \frac{x_{23}^2}{x_{13}^2} \Big)^{\frac{\D_{12}}{2}} \Big( \frac{x_{24}^2}{x_{23}^2} \Big)^{\frac{\D_{3}}{2}}\Big( \frac{x_{35}^2}{x_{34}^2} \Big)^{\frac{\D_{45}}{2}} }{(x_{12}^2)^{\frac{\D_1+\D_2}{2}}(x_{34}^2)^{\frac{\D_3}{2}}(x_{45}^2)^{\frac{\D_4+\D_5}{2}}}
\end{align} 
and
\begin{align} 
 \mm{P}^{\D_i}_{\D_a,\D_b} = \frac{ \pi^d }{\G(\frac{\D_a+\D_{12}}{2})\G(\frac{\D_a-\D_{12}}{2})\G(\frac{\tilde{\D}_b+\D_{45}}{2})\G(\frac{\tilde{\D}_b-\D_{45}}{2})\G(\frac{\tilde{\D}_a+\D_b-\D_{3}}{2})\G(\frac{\D_a-\tilde{\D}_b+\D_{3}}{2}) }
\end{align} 
The conformal block $g^{\D_i}_{\D_a,\D_b}(w_i)$ in Eq.~\ref{eq:M564} was computed \cite{Rosenhaus:2018zqn,Gross:2017aos} in $d=1,2$ in terms Appell$F_2$ functions. 

Similarly to the 4-point case, the 5-point conformal partial wave $\Psi_{\D_a, \D_b}^{\D_i}$ depends on $\D_1+\D_2$ and $\D_4+\D_5$ only through the overall factor $\mm{L}^{\D_i}_{\D_a,\D_b} (x_i)$, which we strip off. Looking at Eq.~\ref{eq:polu8}, this means that all of the non-trivial dependence of the spectral representation of $g_5$ on $\D_1+\D_2$ and $\D_4+\D_5$ comes from the factor:
\begin{align} 
B_{\D_1,\D_2,\frac{d}{2}+i\n_1} B_{\D_4,\D_5,\frac{d}{2}+i\n_2} \ni  
 \G_{\frac{\Delta_1+\D_2}{2} +\frac{i \nu_1-\frac{d}{2}}{2}} \G_{\frac{\Delta_1+\D_2}{2} -\frac{i \nu_1+\frac{d}{2}}{2}} \G_{\frac{\Delta_4+\D_5}{2} +\frac{i \nu_2-\frac{d}{2}}{2}} \G_{\frac{\Delta_4+\D_5}{2} -\frac{i \nu_2+\frac{d}{2}}{2}}  
\end{align}
The gamma functions on the right hand side are the factor $U^{(\D_1,\D_2)}_{A.V}\times U^{(\D_3,\D_4)}_{A.V}$, defined in Eq.~\ref{eq:AV}. We see that this very similar to the 4-point case of section~\ref{sec:appbb}, and in particular it is clear that Identity 3 and 3' of Eqs.~\ref{eq:145} and \ref{eq:rel3prime} will hold for the 5-point case of this section. An application of this identity is that one can write certain 5-point exchange diagrams as a finite sum of 5-point contact diagrams, Fig.~\ref{fig:bubblerelations6}a.


\section{The 4-point $\D_{12}=0$, $\D_{34}= 1$ contact diagram from Mellin space}
\label{sec:Mellinp}
In this note subsection we compute, starting from the Mellin representation, the contact diagram in the case when $a\equiv \frac{\D_1-\D_2}{2} =0$ and $b\equiv \frac{\D_3-\D_4}{2} = \frac{1}{2}$. As we will see, the contact diagram (i.e $\bar{D}_{\D_1,\D_1,\D_4+1,\D_4}$) will be given by a  hypergeometric ${}_2F_1$. This matches the result of section~\ref{sec:contdi}, obtained from the spectral representation.

The Mellin representation of the 4-point amplitude, see e.g \cite{Costa:2012cb}:
\begin{align}
g_4(u,v) = \int dt ds M(s,t) u^{\frac{t}{2}}v^{-\frac{s+t}{2}} ( \G_{\frac{\D_1+\D_2-t}{2}} \G_{\frac{\D_3+\D_4-t}{2}})
(\G_{\frac{\D_{34}-s}{2}} \G_{\frac{-\D_{12}-s}{2}} )
(\G_{\frac{t+s}{2}} \G_{\frac{t+s+\D_{12}-\D_{34}}{2}}) 
\end{align}
Putting $\D_{12}=0$ and $\D_{34}=1$ and $M(s,t) =c=const$ for contact diagrams:
\begin{align}
g_4(u,v) =c \int dt ds \  u^{\frac{t}{2}}v^{-\frac{s+t}{2}} ( \G_{\D_1+\frac{-t}{2}} \G_{\D_4 +\frac{1}{2}+\frac{ -t}{2}} )
(\G_{\frac{1-s}{2}} \G_{\frac{-s}{2}} )(\G_{\frac{t+s}{2}} \G_{\frac{t+s-1}{2}}) 
\end{align}
Using a gamma function identity $\G_z \G_{z+\frac{1}{2}}= \sqrt{\pi} 2^{1-2z}\Gamma_{2z}$, the last two gamma function pairs can be simplified:
\begin{align}
g_4(u,v) = 8\pi c \int dt ds \   2^{-t} u^{\frac{t}{2}}v^{-\frac{s+t}{2}} ( \G_{\D_1-\frac{t}{2}} \G_{\D_4 +\frac{1}{2}-\frac{ t}{2}} )
 \G_{-s} \G_{t+s-1}
\end{align}
For concreteness let us choose $v$ such that we can close the $s$ contour integral to the right. There are poles only from $\G_{-s}$ at $s_{poles}=m$, where $m=0,1,2, \ldots$. So we need to perform the sum:
\begin{align}
\int ds  v^{-\frac{s}{2}}  \G_{-s} \G_{t+s-1} \ \     \to \ \   2\pi \sum_{m=0}^\infty   \frac{\G_{t+m-1}}{\G_{m+1}} (-1)^mv^{-\frac{m}{2}} = 2\pi \G_{t-1} \Big(1+\frac{1}{\sqrt{v}}\Big)^{1-t} 
\end{align}
We are left with the following $t$ integral:
\begin{align}
g_4(u,v) = 16 \pi^2 c \int dt u^{\frac{t}{2}}v^{-\frac{t}{2}}  2^{-t}\Big(1+\frac{1}{\sqrt{v}}\Big)^{1-t}( \G_{\D_1-\frac{t}{2}} \G_{\D_4 +\frac{1}{2}-\frac{ t}{2}} ) \G_{t-1}
\end{align}
Let us assume that $u$ and $v$ are in the range that we can close the $t$-contour to the left. There are only poles from $\G_{t-1}$, at $t_{poles}=1-m$, where $m=0,1,2, \ldots$. From the residue theorem we get the following sum:
\begin{align}
g_4(u,v) = 16\pi^3 c \sum_{m=0}^\infty u^{\frac{1-m}{2}}v^{-\frac{1-m}{2}}  2^{m}\Big(1+\frac{1}{\sqrt{v}}\Big)^{m}\frac{\G_{\D_1+\frac{m-1}{2}} \G_{\D_4 +\frac{ m}{2}} }{\G_{m+1}} 
\end{align}
The result of this sum:
\begin{align}
\label{eq:kskkd}
&g_4(u,v) = 16\pi^3 c\sqrt{\frac{u}{v}} \bigg[ \G_{\D_4}\G_{\D_1-\frac{1}{2}} \ {}_2F_1 \Big(\D_4, \D_1-\frac{1}{2},\frac{1}{2}, \frac{(\sqrt{v}+1)^2}{u} \Big)
\nn 
&- 2 \frac{1+\sqrt{v}}{\sqrt{u}}   \G_{\D_1}\G_{\D_4+\frac{1}{2}} \ {}_2F_1 \Big(\D_4+\frac{1}{2}, \D_1 ,\frac{3}{2}, \frac{(\sqrt{v}+1)^2}{u} \Big) \bigg]  
\end{align}
This matches the result Eq.~\ref{eq:789s1}, obtained from the spectral representation. Now if we further take $\D_4=\D_1$, the ${}_2 F_1$ simplifies, an we get:
\begin{align}
g_4(u,v) =  16\pi^3 c \G_{\D_1}\G_{\D_1-\frac{1}{2}}  \sqrt{\frac{u}{v}}  \Big(1+ \frac{1+\sqrt{v}}{\sqrt u}\Big)^{1-2\D_1} 
\end{align}
As in Eq.~\ref{eq:789s2}. Note that $Z$ from section~\ref{sec:5} is related to the cross ratios $u$ and $v$ as $\frac{(Z+1)^2}{2Z}=1+\frac{1+\sqrt{v}}{\sqrt{u}}$.

\bibliographystyle{utphys}
\bibliography{bulkdiagrams}

\end{document}